\documentclass[reprint,superscriptaddress,amsmath,amssymb,aps, tightenlines,]{revtex4-2}
\usepackage{graphicx}
\usepackage{dcolumn}
\usepackage{bm}
\usepackage{bbding} 
\usepackage{fdsymbol}
\usepackage{color}
\usepackage[colorinlistoftodos]{todonotes}
\usepackage{longtable} 
\usepackage{rotating}
\usepackage{tcolorbox}
\usepackage{hyperref}
\usepackage{capt-of}
\usepackage{color}
\usepackage{mathrsfs} 
\newcommand{\Nabla}{\triangledown} 

\begin{document}

\title{Creation of an isolated turbulent blob fed by vortex rings} 

\author{Takumi Matsuzawa}
\affiliation{James Franck Institute and Department of Physics, University of Chicago, Chicago, IL 60637, USA}

\author{Noah P. Mitchell}
\affiliation{James Franck Institute and Department of Physics, University of Chicago, Chicago, IL 60637, USA}

\author{St\'{e}phane Perrard}
\affiliation{James Franck Institute and Department of Physics, University of Chicago, Chicago, IL 60637, USA}

\author{William T.M. Irvine}
\affiliation{James Franck Institute and Department of Physics, University of Chicago, Chicago, IL 60637, USA}
\affiliation{Enrico Fermi Institute, University of Chicago, Chicago, IL 60637, USA}

\date{\today}

\begin{abstract}
Turbulence is hard to control.
A plethora of experimental methods have been developed to generate  this ephemeral state of matter, leading to fundamental insights into its statistical and structural features as well as its onset at ever higher Reynolds numbers. 
In all cases however, the central role played by the material boundaries of the apparatus poses a challenge on understanding what the turbulence has been fed, and how it would freely evolve. 
Here, we build and control a confined state of turbulence using only elemental building blocks: vortex rings. 
We create a stationary and isolated blob of turbulence ($Re_{\lambda}=50-300$) in a quiescent environment, initiated and sustained solely by vortex rings. 
We assemble a full picture of its three-dimensional structure, onset, energy budget and tunability.
Crucially, the incoming vortex rings can be endowed with conserved quantities, such as helicity, which can then be controllably transferred to the turbulent state.
Our `one eddy at a time' approach paves the way for sculpting turbulent flows much as a state of matter,  `printing' it at a targeted position, localizing it, and ultimately harnessing it. 
Our work paves the way to gaining a complete picture of this ephemeral state of flow. 
\end{abstract}

\maketitle

Vorticity, which measures the local rotation rate of a fluid, is the building block of flow. 
In its absence, any fine structure in an incompressible flow decays rapidly with distance from material boundaries.
Conversely, injection of vorticity can power complex bulk flows~\cite{van1982album, von1963aerodynamics}, the quintessential example being  the iconic multi-scale liveliness of turbulence. 
Canonical methods of generating turbulence rely on the spontaneous shedding of vorticity from boundaries
~\cite{smits2011high, adrian_hairpin_2007, christensen_statistical_2001, zhou1999mechanisms,acarlar1987studyone, acarlar1987studytwo, saddoughi_veeravalli_1994, theodorsen1952mechanism},
be it of pipes~\cite{reynolds1883xxix, mullin2011experimental, eckhardt2007turbulence, hof2003scaling}, grids~\cite{hurst2007scalings, kistler1966grid, comte1971simple, bodenschatz2014variable}, or spinning plates~\cite{de2007slow, volk2011dynamics, labbe1996study}. 
This makes it hard to control, or have detailed knowledge of, the fabric of the injected vorticity. 
It also often couples the turbulence to boundaries, posing a challenge to study its unconstrained evolution. 
Yet, our most basic models of turbulence are cast in terms of vorticity alone, with no reference to walls. 

Knowing the structure of the vorticity that feeds turbulence is fundamental to a full understanding of turbulence because  it determines the inviscid invariants including the amount of energy, helicity, linear impulse, and angular impulse that are injected into turbulence. 
The balance of the latter two invariants, for example, might lead to different types of turbulence 
in the large scales~\cite{krogstad2010grid}, and have been proposed to rule the decay of turbulence~\cite{davidson2015turbulence, davidson2009role, saffman1967note, landau2013fluid}. 
To make it possible to address these fundamental questions, we set out to build and sustain an isolated region of turbulence far away from boundaries, while controlling the injection of inviscid conserved quantities and fully observing its free evolution.

\begin{figure}[htbp]
\centering
\includegraphics[width=0.9\columnwidth]{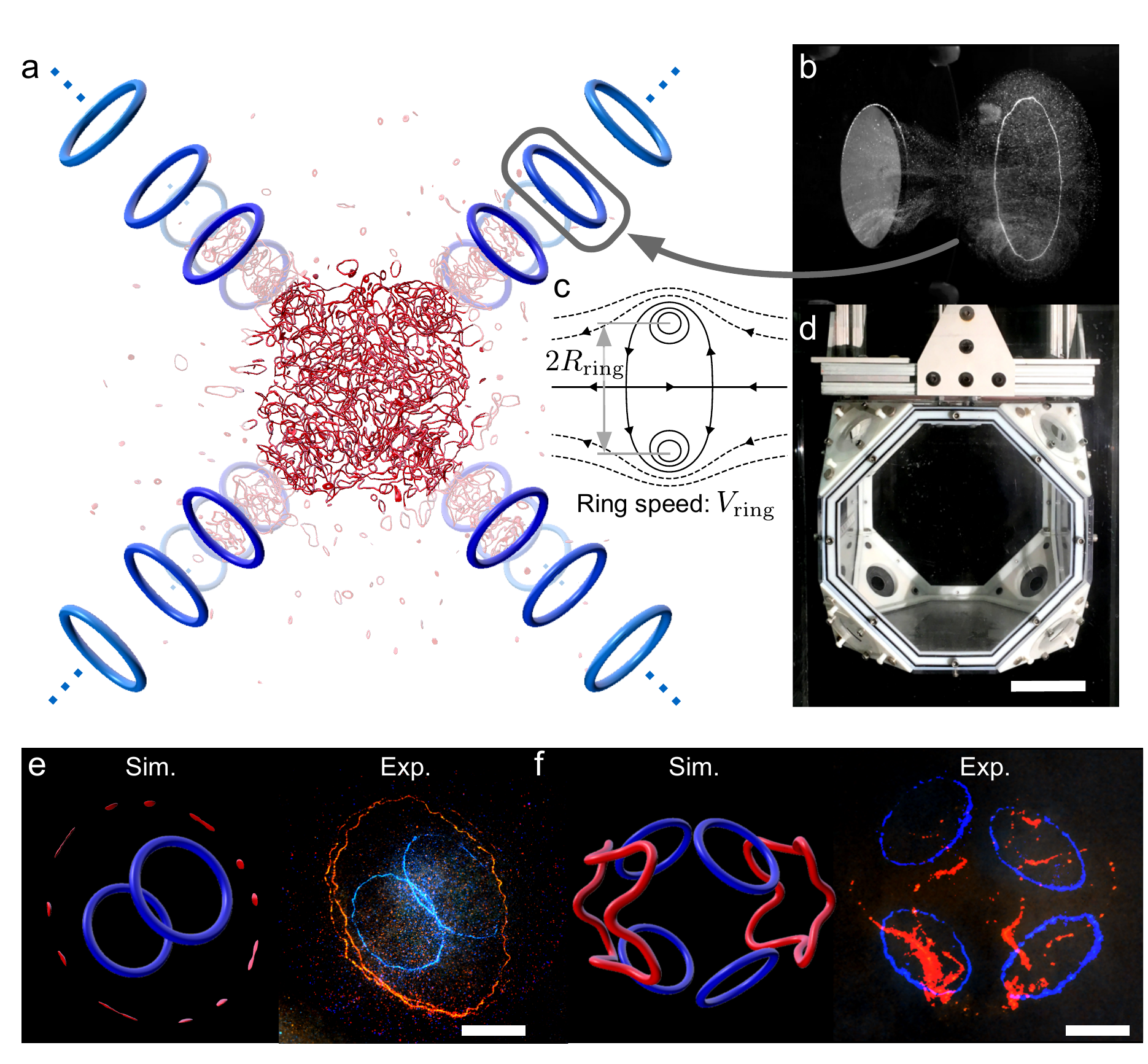}
\caption{\textbf{Generation of turbulence using vortex rings and their resistance to confinement.} (\textbf{a}) We envisage colliding vortex rings creates turbulence at a target location far from boundaries with a controlled injection rate of energy. (\textbf{b}) Extruding fluids through an orifice generates a vortex ring that is visualized by bubbles. (\textbf{c}) Streamline of a vortex ring in the co-moving frame. (\textbf{d}) A photograph of the experimental chamber where eight vortex rings are generated at its truncated faces every forcing cycle. The scale bar represents 100mm. (\textbf{e-f}) Voriticty resists confinement. (\textbf{e}) A head-on collision of two identical vortex
rings (blue) generates numerous secondary rings after reconnections occur (red). The left panel shows the result of the Gross-Pitaevskii simulation, whereas the right panel shows the experiment, visualized by bubbles. The scale bar represents 30mm along the semimajor axis of the red ellipse. The red ellipse is a projection of a circle. (\textbf{f}) Same as (e) but the centers of the four identical rings are initially set in a square configuration (blue), ejecting two vortex loops after reconnections (red). The scale bar is the same as (e).}
\label{fig:vorticity_escape}
\end{figure}

Vortex loops are a natural candidate to this end.
A vortex ring is readily generated by impulsively drawing water through an orifice in a tank  (Figure \ref{fig:vorticity_escape}b). 
Seeding the water with bubbles  reveals the coherent motion of the ring as it travels across the tank carrying its `atmosphere' as it propagates (Figure \ref{fig:vorticity_escape}c). 
Such a ring can, in an ideal fluid,  travel infinitely far away from the boundaries. In real fluids, vortex rings  eventually decay via viscous processes, or break down due to instabilities~\cite{crow1970stability, tsai1976stability, le2002theoretical}.  Nonetheless, they coherently carry their vorticity, and associated inviscid invariants, far from the boundaries that gave rise to them. 

\begin{figure}[htbp]
\centering
\includegraphics[width=0.9\textwidth]{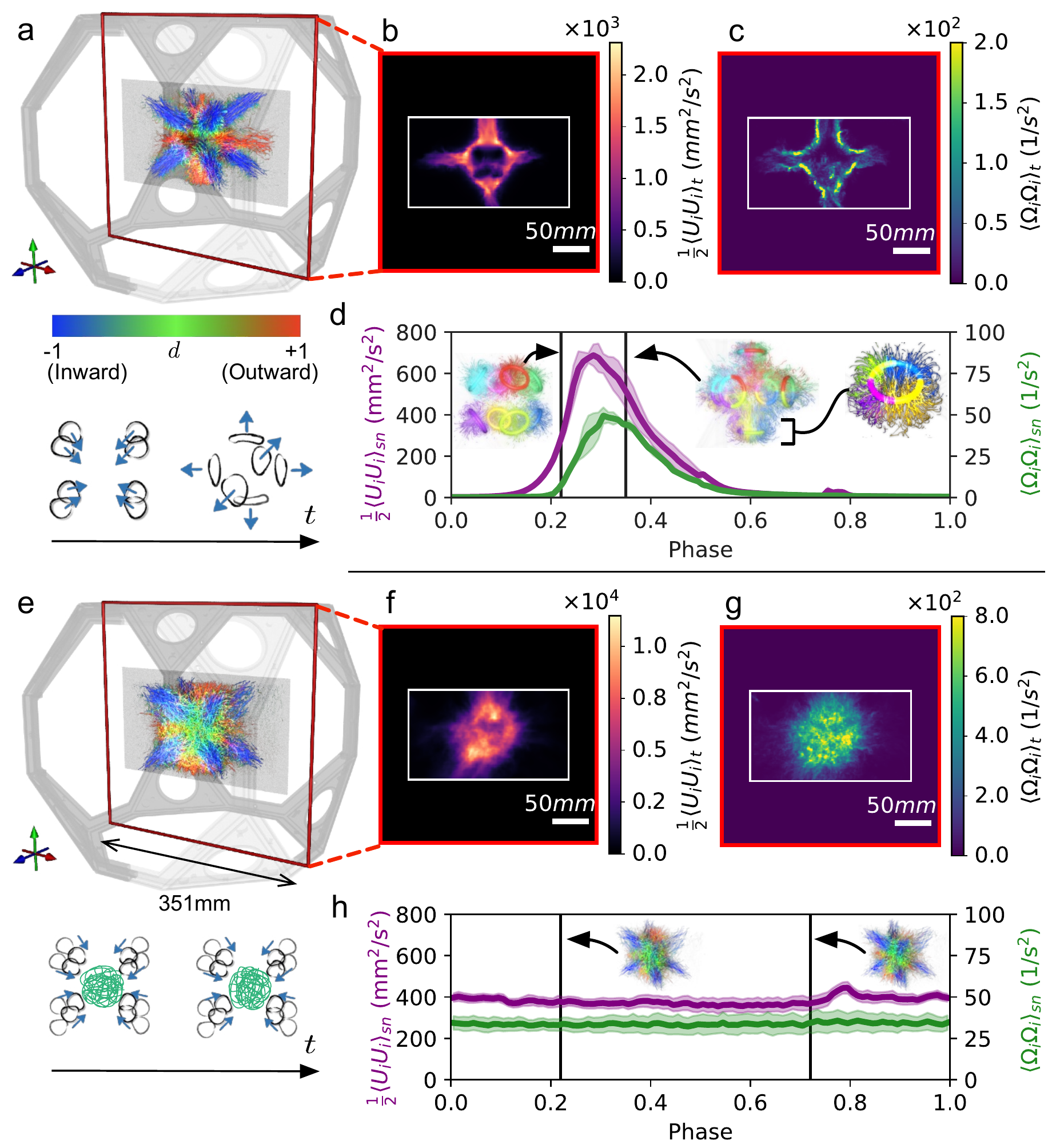}
\caption{\textbf{Two phases emerge as eight vortex rings repeatedly collide: coherent reconnections and a confined turbulence.} (\textbf{a}) Coherent vortex reconnections of eight vortex rings (Blue) result in six secondary rings (Red), visualzied by 3D Lagrangian trajectories throughout the experiments (t=[0, 6] in s). The color is the radial component of instantaneous Lagrangian speed, $d=(\vec{U}_{Lag} \cdot \hat{r}) /  |\vec{U}_{Lag}|$. Opacity is weighted by speed to highlight the fast dynamics ($\alpha=\exp{[-5(1-|\vec{U}_{Lag}|/U_{Lag, 0})]}$ for $|\vec{U}_{Lag}|<U_{Lag, 0}=200mm/s$). $(V_{\rm ring}/ R_{\rm ring}, f)=(20\text{Hz}, 0.2\text{Hz})$ (\textbf{b-c}) Time-averaged energy/enstrophy on the central slice shows the trace of the secondary rings after vortex reconnections. (\textbf{d}) Average energy and enstrophy on the measured plane indicates the entry of the eight vortex rings, and the propagation of the secondary rings. The inlets show the 3D Lagrangian trajectories before and after the reconnections.  (\textbf{e}) Lagrangian trajectories around a turbulent blob display uniform, nearly isotropic outflow from the core. Duration: t=[0, 6] in s. $(V_{\rm ring}/ R_{\rm ring}, f)=(20\text{Hz}, 5\text{Hz})$ (\textbf{f-g}) Time-averaged energy/enstrophy on the central slice shows an isolated region with high energy/enstrophy. (\textbf{h}) Average energy and enstrophy on the measured plane indicates the steadiness of the state with weak dependence on the periodic forcing. The piston draws the fluids into the chamber at phase=0.77.}
\label{fig:collision_comparison2}
\end{figure}

We set out to combine vortex loops  like LEGO blocks, firing them together to `print' a stationary region of turbulence in the center of our tank (Figure \ref{fig:vorticity_escape}a).
As demonstrated in iconic vortex collision experiments~\cite{oshima_head-collision_1978, lim_instability_1992, mckeown_turbulence_2020}, recently revisited as a minimal means to understand the inertial cascade in real space~\cite{brenner2016potential, mckeown2018cascade, mckeown_turbulence_2020, ostilla2021cascades}, two vortex rings fired together can multiply into a series of smaller rings, giving rise to turbulence. 
Figure \ref{fig:vorticity_escape}e shows a version of this experiment using a pair of vortices created by drawing fluid into our tank through  opposing orifices (Figure \ref{fig:vorticity_escape}d). 
The rings, visualized using bubbles, approach each other, stretch and recombine into smaller, outwardly  propagating rings. 
This example however, also highlights the tendency of colliding vortices to divide and redirect, escaping confinement. 
The situation is unchanged in the case of four vortices (Figure \ref{fig:vorticity_escape}f) or eight (Figure~\ref{fig:collision_comparison2}a-d). 
This generic behavior of vortices colliding, reconnecting, and escaping challenges the idea that a blob of turbulence can be printed and confined at a target position.

In a na\"ive attempt to hold the escaping vorticity in place, we fired subsequent  sets of eight vortex rings at repeating intervals so that the outgoing vortices would interact with the in-going vortices.
To image the flow we use a combination of Particle Imaging Velocimetry (PIV), seeded bubble tracking, and 3D Particle Tracking Velocimetry (PTV). (See SI section II for a detailed description of the acquisition and visualization processes.) 
At a low frequency ($f=0.2$Hz), we observe a simple repetition of the single-shot reconnection dynamics (Figure \ref{fig:collision_comparison2}a-d); coherent vorticity comes in and leaves.
However, when the frequency is sufficiently high to enable the outgoing vortex rings to interact with the incoming rings, a novel state with a remarkably different vorticity distribution emerges (Figure \ref{fig:collision_comparison2}e-g, $f=4$Hz). 

In this new state, vorticity is confined, and is evenly distributed within an approximately spherical region.
The developed flow inside the blob is in stark contrast to its surroundings which remain relatively quiescent.
The blob is sustained as long as the vortex rings are injected. Both energy and enstrophy averaged over the measured plane indicate the comparative steadiness of the state (Figure \ref{fig:collision_comparison2}h) with weak dependence on the periodic forcing.

\begin{figure}[htbp]
\centering
\includegraphics[width=\textwidth]{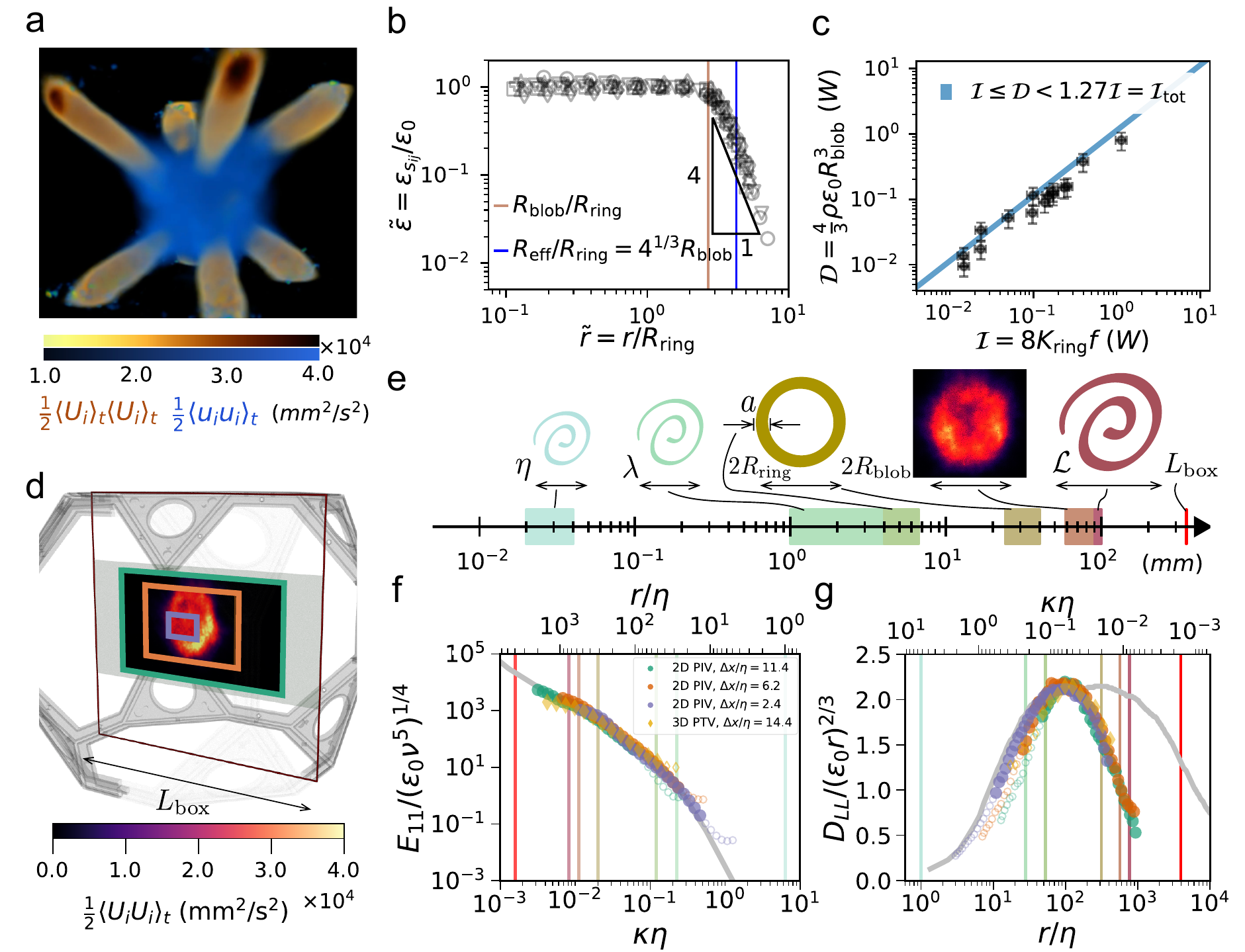}
\caption{\textbf{Turbulent dissipation inside the blob matches the power injected by vortex rings.} (\textbf{a})Reynolds decomposition $U_i = \langle U_i \rangle + u_i$ reveals that the blob consists of a turbulent core with a mean flow originated from the incoming vortex rings. Yellow: the mean flow energy $\langle U_i \rangle_t \langle U_i \rangle_t/2$. Blue: the mean turbulent energy $\langle u_i u_i \rangle_t/2$. ($i=1, 2, 3$)(\textbf{b}) Radial profile of the dissipation rate on the central plane reveals an homogeneous region up to $R=R_{\rm blob}\approx \sqrt{6}R_{\rm ring}$, and a tail that rapidly decays. The profile is universal over rings with different radii. (Piston stroke ratio, Piston effective stroke velocity, frequency)=$(L/D, v_{eff} \text{ in mm/s}, f\text{ in Hz})$- $\circ:(1.5, 196, 5), \triangle:(2.0, 418, 5), \triangledown:(3.0, 443, 5), \square:(3.0, 443, 7), \Diamond:(3.5, 318, 5), +:(3.5, 594, 5), \times:(3.5, 594, 7), \medwhitestar:(3.5, 594, 8) $  (\textbf{c}) Dissipated power in the sphere of radius $R_{\rm blob}$ linearly scales with the power injected into the blob by vortex rings. (\textbf{d}) Three-scale 2D PIV measurements are conducted on the plane cutting the center of the blob.($i=1, 2$) (\textbf{e}) Turbulence lengthscales with respect to the relevant geometries (blob radius $R_{\rm blob}$, ring radius $R_{\rm ring}$, and core diameter $a$). (\textbf{f}) Rescaled one-dimensional spectra are computed in the homogeneous region ($r \le R_{\rm blob}$). ($\epsilon_0 =6.0\times 10^4$ mm$^2/$s$^3$, $\nu=1.004$mm$^2/$s, $Re_{\lambda}=200$). The gray master curve is collected by multiple experiments with walls, taken from ~\cite{saddoughi_veeravalli_1994}~($Re_{\lambda}\approx 600$) as a reference. The attenuated signal due to PIV is addressed by hollow data points. (\textbf{g}) Rescaled second-order structure functions of the same data as (\textbf{f}) are shown with a reference~\cite{saddoughi_veeravalli_1994}~($Re_{\lambda}\approx 600$).}
\label{fig:turbulence_signal}
\end{figure}

In Figure~\ref{fig:turbulence_signal}a we show a Reynolds
decomposition of this chaotic flow  into mean and fluctuating components: $U_i = \langle U_i \rangle + u_i$. 
The blue cloud  represents the average energy associated with the fluctuations and occupies the central region alone, whereas the yellow clouds represent the mean flow energy, associated with the paths along which the vortex rings are fed.
We find that the flow inside the blob is dominated by fluctuations ($\langle u^2 \rangle / \langle U \rangle^2 \approx  10^{1.5}-10^{3}$)  whereas the flow outside the blob is dominated by coherent flow. 
Furthermore, the  velocity fluctuations inside the blob are only weakly dependent of the forcing phase, whereas the coherent flow outside the blob is phase-dependent, reflecting the laminar motion of the vortex rings. The temporally-and-radially averaged profiles of both fluctuating energy and enstrophy are approximately constant up to a radius $R_{\rm blob}$ (see SI section VI A), and decay rapidly for $r>R_{\rm blob}$, and can be parameterized approximately as a power law with an exponent $r^{-4}$. 
The local dissipation rate $\epsilon_{s_{ij}}(r)=2\nu \langle s_{ij} s_{ij} \rangle$ also possesses the same radial profile as energy and enstrophy (Figure \ref{fig:turbulence_signal}b) where $s_{ij}=(\partial_j u_i + \partial_i u_j)/2$. 

To investigate the character of the flow inside the blob, we compute the fluctuating energy spectrum and the second-order structure function. 
PIV measurements are inherently limited at small scales by image resolution and at large scales by the finite field of view. To span the full range of scales in our turbulent flow, we performed 2D PIV measurements at three levels of magnification (Figure\ref{fig:turbulence_signal}d, spatial resolutions: $\Delta_x=$0.5, 1.3, 2.4mm=$2.4\eta, 6.2\eta, 11.4\eta$) and stitched the results together by taking into account the spectral leakage and low-pass filtering effects of PIV.
With 3D PTV, we measure a one-dimensional energy spectrum on the slice that cuts the middle of the turbulent blob with a spatial resolution of $\Delta_x=3.0$mm$=14.4\eta$. In addition, the resulting 4D velocity field offers a direct computation of a three-dimensional energy spectrum without the assumption of isotropy (see SI section VII). 
The resulting one-dimensional energy spectrum $E_{11}(\kappa_1)$ and the second-order structure function $D_{LL}$ of the fluctuating component of the flow are shown in Figure \ref{fig:turbulence_signal}e-f. Our measurements at the three levels of magnification agree where their ranges of validity overlap. 
The rescaled spectrum is in excellent agreement with the universal curve obtained by grid turbulence and turbulent boundary layer experiments~\cite{saddoughi_veeravalli_1994}. Similarly, the second-order longitudinal structure function when rescaled by the  2/3 power law in the inertial sub-range is consistent with that of homogeneous isotropic turbulence. 

Our spectra and structure function support the notion that the flow inside the blob is turbulent and therefore that its statistical properties can be captured by a dissipation rate $\epsilon_0$ and an integral scale $\mathcal{L}$~\cite{kolmogorov_local_1941, kolmogorov_refinement_1962}, together with the fluid viscosity $\nu$. 
While the value of $\epsilon_0$ is notoriously challenging to measure~\cite{de_jong_dissipation_2008, xu_accurate_2013}, it can be inferred from the local strain rate measurements, from fitting the measured spectrum to the universal curve, or by fitting the peak value in the scaled second order structure function~\cite{de_jong_dissipation_2008}. As discussed in SI section VI,  we find all three methods are in agreement when computed on our median-filtered, spatio-temporally resolved velocity fields.
The corresponding value of the Kolmogorov length  $\eta=(\nu^3/\epsilon_0)^{1/4}$ is shown in Figure \ref{fig:turbulence_signal}e.
A measurement of the turbulent r.m.s. velocity $u'=\sqrt{\langle u_i u_i \rangle / 3}$ in turn provides the estimate of the integral length scale  $\mathcal{L}= u'^3/ \epsilon_0$.

How are the properties of this turbulent blob controlled by the incoming vortex rings?
As shown in Figure~\ref{fig:turbulence_signal}e, we find the value of the integral length-scale to be close to that of the blob diameter $2 R_{\rm blob}$, suggesting both blob radius and integral length-scales are determined by the largest scale in the incoming vortex rings.
This observation is supported by repetitions of our experiment in which we varied  frequency of injection of the incoming vortex rings and found no change in either $\mathcal{L}$ or $R_{\rm blob}$.
A repetition of our experiment in which the incoming vortex ring radius was halved, resulted instead in a halving of both $\mathcal{L}$ or $R_{\rm blob}$ (see SI section X).

The smallest (Kolmogorov) length-scale of the turbulent blob $\eta$ (Figure~\ref{fig:turbulence_signal}e) has by contrast little relation to the vortex ring radius, and is instead strongly affected by the incoming vortex ring energy and frequency of injection. 
This is consistent with the notion that at the smallest length-scales turbulence `forgets' about the large-scale forcing that gave rise to it and the velocity field depends only on energy flux $\epsilon_0$ and viscosity $\nu$. 
We thus turn our attention to the balance of energy in our system. 

Because the flow is at dynamical equilibrium, the dissipated power  must match the power injected  by the vortex rings. 
If we neglect  any residual dissipation due to the mean flow,
the energy balance is
\begin{align}
4\pi \rho \int \epsilon(r) r^2 dr \approx 8 K_{\rm ring} f,
\end{align}
where $\epsilon(r)=\epsilon_0$ if $r\leq R_{\rm blob}$, $\epsilon_0 (R_{\rm blob}/r)^4$ otherwise. 
$K_{\rm ring}$ is the kinetic energy inside the vortex atmosphere of any one of the incoming vortices.

When integrated over all space, the L.H.S. evaluates to $16/3 \pi \epsilon_0 R_{\rm blob}^3=4/3 \pi \epsilon_0 R_{\rm eff}^3$, whereas if integrated up to $R_{\rm blob}$ we have $4/3 \pi \epsilon_0 R_{\rm blob}^3$. The R.H.S. requires knowledge of $K_{\rm ring}$. When there is a vortex ring in a flow, the energy over all space $K$ is generally the sum of the energy inside the vortex atmosphere $K_{\rm ring}$ and the energy of the added mass associated with the potential flow that surrounds the atmosphere $K_{\rm added}$. $K_{\rm ring}$ can be further decomposed into the translational kinetic energy of the vortex atmosphere $K_{\rm rect}=4/3\pi R_{\rm atmosphere}^3 V_{\rm ring}^2$, and the energy associated with the rotational motion within the vortex atmosphere $K_{\rm int}$. While the exact partitioning varies by the vortex model, the variation for $K_{\rm ring}$ is small ($<$3.3\%) for realistic  vortex ring models. 
We directly measured the energy of our vortex rings, and found $K_{\rm ring}= (2.0\pm 0.4)K_{\rm rect}$ similar to $23/14K_{\rm rect}\approx 1.6K_{\rm rect}$ of Hill's spherical vortex (see SI section III C).

In Figure~\ref{fig:turbulence_signal}c we compare the measured dissipated vs injected power for a collection of  blobs that are created by altering ring size, speed, and frequency of injection.  
The dissipated power  scales linearly  with the injected power,  with a slope of approximately 1. 
A more granular accounting, e.g. including only the  energy contained within the vortex ring atmosphere  and computing  $\epsilon_0$ only  within $R_{\rm blob}$, yields a  linear relationship with lower proportionality constants: 1 (total energy, total turbulent dissipation), 0.68 (energy within the incoming vortex ring atmospheres, turbulent dissipation within a sphere of radius $R_{\rm eff}$,) and 0.33 (energy within the incoming vortex ring atmospheres, turbulent dissipation within a sphere of radius $R_{\rm blob}$).

Crucially, increasing the velocity or frequency of injection increases the rate of energy dissipation while keeping the integral length-scale fixed, thereby increasing the separation of scales $\mathcal{L}/\eta$. Thus ring radius and energy injection provide independent control knobs for producing turbulence of a desired intensity  localized to a given region. 

The picture is in stark contrast to the single-collision experiment (Figure \ref{fig:collision_comparison2}a) in which vortices come in, reconnect, and go out. 
At these low forcing frequencies the conversion from coherent vortex motion to turbulence is far less efficient. Even though in practice  reconnections trigger energy loss within the outgoing vortices, in the limit of a single coherent collision with large separation of scales, the fraction of advected energy can be in principle 100\%.

\begin{figure}[htbp]
\centering
\includegraphics[width=\textwidth]{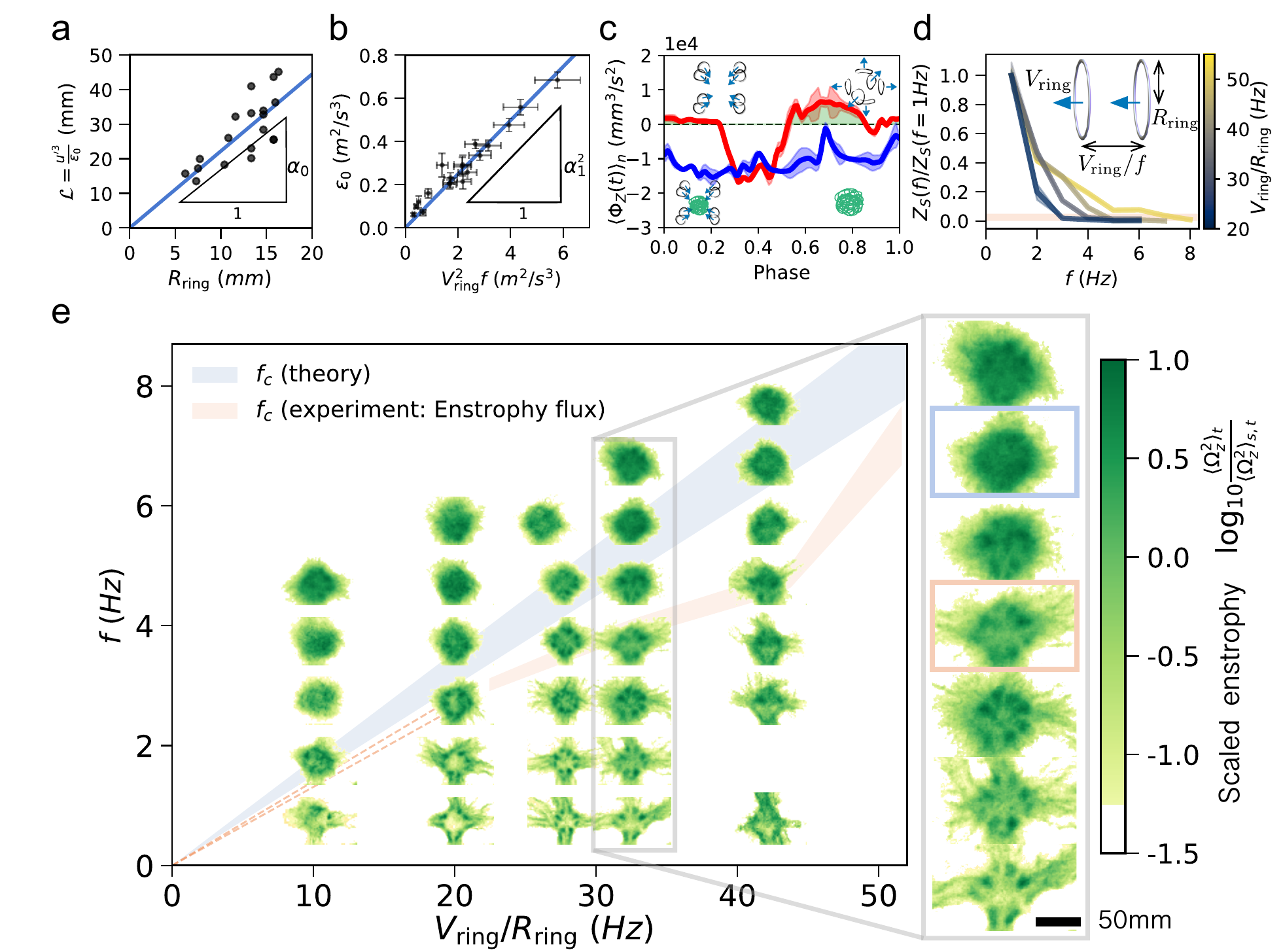}
\caption{\textbf{$V_{\rm ring}/(R_{\rm ring}f)$ governs a smooth transition from coherent reconnections to turbulence.} (\textbf{a}) Our theory predicts the transition frequency as $(\alpha_0/\alpha_1)(V_{\rm ring} / R_{\rm ring})$. The integral length scale $\mathcal{L}$ is linear to the radius of the injected vortex ring $R_{\rm ring}$ with $\alpha_0$ being a proportionality constant. (\textbf{b}) The second parameter $\alpha_1$ is a proportionality constant between $\epsilon_0$ inside the turbulent core and the power injected by vortex ring $\sim V_{\rm ring}^2 f$. (\textbf{c}) Phase-averaged enstrophy flux through a sphere (radius $R'$ enclosing a turbulent blob captures the degree of confinement ($R'=1.25R_{\rm eff}\approx 1.9R_{\rm blob}$). Below the transition frequency (Red, $f=1$Hz, $V_{\rm ring} / R_{\rm ring}=20$Hz), the enstrophy flux takes positive values when the outflow is greater than inflow. Above the transition (Red, $f=6$Hz, $V_{\rm ring} / R_{\rm ring}=20$Hz), the flux is always negative throughout a cycle because the outflow is weaker than inflow (confinement). Integrating the positive regions (a green shaded region) yields escaping enstrophy per cycle $Z_S(f)$. (\textbf{d}) Scaled escaping enstrophy per cycle decreases as the injection frequency increases, indicating a smooth transition to a blob. The orange band corresponds to $<$5\%. (\textbf{e}) Formation of a turbulent blob depends on the injection frequency and the ratio between the ring velocity and the radius. Scaled time-averaged enstrophy fields show agreements with the transition frequencies expected from our theory and measurements on the enstrophy flux. The orange band represents the frequencies that the escaping enstrophy becomes $<$5\% of the values at $f=1$Hz.}
\label{fig:blob_transition}
\end{figure}

What governs the transition to a blob state?
The most basic criterion is suggested by geometry:  the outgoing rings will collide with the incoming rings for $f \gtrsim V_{\rm ring}/R_{\rm ring}$.
A completely different conceptual approach is to seek to `match' the incoming vortex `eddies' to the turbulent state. 
A central idea in a turbulent cascade is that energy from each scale $\ell$ is transported to the next in a time $\tau_\ell \sim (\ell^2 / \epsilon_0)^{1/3}$.
If we demand that the time between incoming vortices $1/f$ match the timescale for the largest eddy to transfer energy down the cascade we have $1/f>\tau_\mathcal{L}$.
For our fully developed turbulent blob we have $\mathcal{L}\propto R_{\rm ring}$ (Figure \ref{fig:blob_transition}a, $\alpha_0=2.17\pm0.13$) and $\epsilon_0 \propto V_{\rm ring}^2 f$ (Figure \ref{fig:blob_transition}b, $\alpha_1=0.35\pm0.02$) with proportionality constants determined in experiment. We then obtain a criterion for transition: $f_c \sim V_{\rm ring}/R_{\rm ring}$ with the proportionality constant  determined by the independent measurements of $\mathcal{L}(R_{\rm ring}$) and $\epsilon_0(V_{\rm ring}$) in the fully developed turbulent state.

Figure \ref{fig:blob_transition}e shows the time-averaged enstrophy field for experiments in which we varied both $f$ and $V_{\rm ring}/R_{\rm ring}$. 
The transitional range predicted by matching vortex arrival intervals with  the largest eddy turnover time is shown by the blue band for comparison. 
The relationship between $f_c$ and $V_{\rm ring}/R_{\rm ring}$, consistent with predictions,  is qualitatively visible from the change in shape as frequency is increased.
A second criterion to classify whether a given flow is in a blob state is to compute the enstrophy flux through a sphere that encloses the blob (Figure \ref{fig:blob_transition}c). For the barotropic, incompressible fluids the integrated enstrophy flux is given by
\begin{align}
   \Phi_Z(t) = \int_{\partial \mathcal{V}} \Omega^2 U_i n_i dS,
    \label{eq:enstrophy_evo}
\end{align}
where $\Omega_i = \epsilon_{ijk} \partial_j U_k$ is vorticity.
The phase-averaged (integrated) flux $\langle \Phi_Z \rangle_n$ is shown in Figure \ref{fig:blob_transition}c for an experiment with $f<f_c$ (red) and one with $f>f_c$ (blue).
The red curve shows a trough (influx $>$ outflux), followed by a crest (influx $<$ outflux) as the secondary rings transport enstrophy away from the considered volume.
The blue curve, by contrast, shows little to no outflux. 
As $f$ is increased for a given $V_{\rm ring}/R_{\rm ring}$,  the escaping enstrophy per cycle (shaded green region in Figure \ref{fig:blob_transition}c) smoothly decreases as the frequency is increased (Figure \ref{fig:blob_transition}d.). This  corresponds to the suppression of coherent reconnections and development of turbulence. 
Placing a threshold ($<$5\% relative to the values at $f=1$Hz) on the escaping enstrophy reveals that the transitional frequency depends on $V_{\rm ring}/R_{\rm ring}$ in a linear fashion (Figure \ref{fig:blob_transition}e,  orange band). 
The upper limit of the orange band in Figure~\ref{fig:blob_transition}e lies within 10-20\% of the predicted transition frequency. 


For $f>f_c$, the energy and enstrophy are completely transferred to the blob, in sharp contrast to the single shot experiment ($f\ll f_c$) in which neither are left behind nor in fact penetrate the central region. 
By contrast, the mass associated with the vortex atmospheres must flow in and out in equal amounts and cannot be left behind (see SI section IX).
We find it interesting that the blob state can occur in the first place given this fact. 
Our work raises the question of whether vortex ring trains are in some sense optimally suited to confining and `feeding' turbulence. For example, if jets were used instead of ring trains~\cite{chang2012experimental, zimmermann2010lagrangian}, the input would have a higher ratio of mass to enstrophy and energy, and no blob state has been reported in this case to our knowledge~\cite{hwang2004creating, goepfert2010characterization}. 
More generally, what types of flow `input' lead to a maximally localized blob states vs delocalized states in which the necessary outward advection destabilizes confinement?

Finally, we explore the tunability of our turbulent blob through control of the vortex rings. 
The ring radius $R_{\rm ring}$ tunes the integral scale $\mathcal{L}$ and the blob radius $R_{\rm blob}$. Meanwhile, the energy balance sets the smallest scale of turbulence (Kolmogorov scale $\eta$) as it leads to $\epsilon_0=\alpha_1^2 V_{\rm ring}^2 f\sim \Gamma^2_{\rm ring}f / R_{\rm ring}^2$. Hence, the separation of scales is given by $\mathcal{L} / \eta  \sim ( \Gamma_{\rm ring} /  \nu )^{3/4} (R_{\rm ring}f / V_{\rm ring})^{1/4}$, consistent with the usual relation $\mathcal{L} / \eta \sim Re_{\mathcal{L}}^{3/4} \sim (u'\mathcal{L} / \nu)^{3/4}$ for general turbulence~\cite{pope_turbulent_2000}. Notice that it is expressed solely by the variables of the injecting vortex rings and thus can be completely controlled by tuning their properties. 

\begin{figure}[htbp]
\centering
\includegraphics[width=0.7\textwidth]{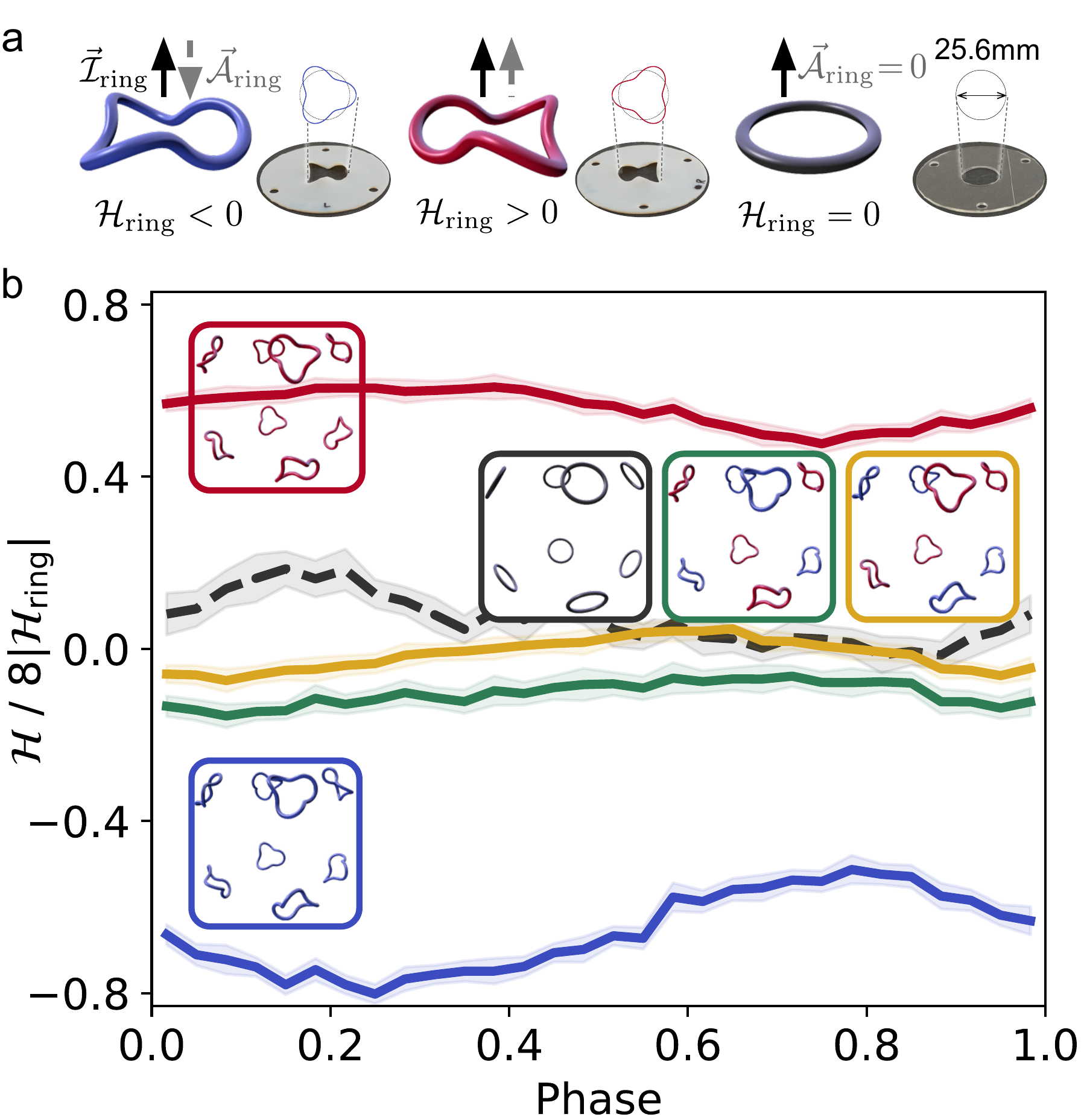}
\caption{\textbf{Repeated collision of helical rings transfers helicity to turbulence in a controlled fashion.} (a) Puffing fluids through helical and circular masks generate helical and planar rings respectively.  Handedness is defined by the relative orientation of linear impulse $\vec{\mathcal{I}}_{\rm ring}$ to angular impulse $\vec{\mathcal{A}}_{\rm ring}$, and is tuned from antiparallel (blue) to parallel (red). (b) 
Helicity is transferred from helical rings to a blob of turbulence. Different configurations allow injection of helicity with different handedness. The five configurations of the measurements are illustrated in the inlet figures. (Net heliciy per cycle, net angular impulse per cycle) = $(8 \mathcal{H}_{\rm ring}, 0)$ [Red: 8 Right], $(0, 0)$ [Green: 4 Right + 4 Left], $(0, 0)$ [Yellow: 4 Right + 4 Left], $(0, 0)$ [Black: 8 Planar], $(-8 \mathcal{H}_{\rm ring}, 0)$ [Blue: 8 Left]. A vortex ring of $(V_{\rm ring}/R_{\rm ring}, f) = (40 \text{Hz}, 5 \text{Hz})$) was used, and the graph shows the helicity integrated over a sphere of radius 60 mm$\approx R_{\rm blob}$. The shade represents the standard error of the mean.}
\label{fig:helical_blob}
\end{figure}

Can our approach to building a turbulent blob be harnessed to endow the turbulence with additional properties? Beyond energy,  natural candidates include the inviscid invariants of impulse, angular impulse and helicity. 
In Figure \ref{fig:helical_blob} we show measurements of the total helicity in a blob created by colliding helical vortex loops in combinations that inject a total helicity of $+8\mathcal{H}_{\rm ring}$, $-8\mathcal{H}_{\rm ring}$ and $0\mathcal{H}_{\rm ring}$, while injecting zero angular impulse and zero linear impulse. Although the vorticity field is not completely resolved, clearly the answer is affirmative. 

We have discovered that a collection of vortex rings periodically fired together leads to a self-confining turbulent blob. 
This bottom-up approach to turbulence provides unique design principles to position, localize and control turbulence as a state of flow.
In the canonical picture of the Richardson cascade, injection and dissipation go hand in hand at dynamical equilibrium. Nevertheless, their connection often remains elusive due to the uncontrolled injection and evolution of vortical structures. 
The use of coherent, controllable vortex rings overcomes this issue, enabling us to inject fully controlled arbitrary ratios of inviscid conserved quantities. 
Enabled by the self-confinement effect we discovered, our experiment provides a unique control of injection and dissipation in turbulence. 
The turbulent blob, which can   be measured in its entirety and is free to evolve in isolation, offers a playground for fundamental studies on inhomogeneous turbulence such as transfer at the turbulent/non-turbulent interface, decay of turbulence without the boundary effects, response of turbulence to a periodic drive~\cite{cekli2010resonant, verschoof2018periodically}, and the role of inviscid invariants such as helicity~\cite{kraichnan1973helical, alexakis_helically_2017} and angular impulse~\cite{saffman1967note, davidson2009role} in turbulence. 
Our work views turbulence as a state of  matter that can be controlled and manipulated coherently.

\begin{center}
{\textbf{METHODS}}\\
{\textbf{A: Experimental chamber and actuation}}
\end{center}
The experimental chamber is fabricated by a commercial 3D printer using
ultraviolet-cured polymers (Objet VeroWhite and VeroBlack, Objet Connex 350, Stratasys), and is primarily used throughout the experiments. A windowed chamber made of acrylic with a similar geometry is also used to measure properties of vortex rings. See Supplementary information for the design and the exact dimensions. An electric linear actuator (Copley Controls) controls the motion of the acrylic piston through signals output from a DAQ board (National Instruments.) As the piston attached to the top surface lifts up, fluid gets pulled into the chamber through the orifices, creating eight vortex rings travelling towards the center.
A rubber flap attached to the top surface around the piston's entry point prevents unwanted flow in or out of the chamber near the piston. 
We use a transmissive optical encoder (EM2, US Digital) to track the motion of the piston with sub-millimeter precision.
With the tracking data, we extract two important parameters regarding the properties of the vortex rings: the formation number $L/D$~\cite{gharib_rambod_shariff_1998} (stroke ratio normalized by the orifice diameter) and the effective velocity of the piston $v_{eff}$. (See Supplementary information Section II). The former governs the radius and the stability of the generated vortex ring, and is a function of the diameter of the orifices $D_o$ and that of the piston $D_p$. The latter controls the speed of the vortex ring. In order to generate two sizes of the vortex rings ($R_{\rm ring} \approx$ 15, 25mm), we used two settings $(D_p, D_o)$=(160.0mm, 25.6mm), and (56.7mm, 12.8mm). The first setting offers a large blob of turbulence, suited for turbulent analysis through the 2D particle image velocimetry. The second setting offers a blob of turbulence small enough with respect the illuminted volume for 3D particle tracking velocimetry to conduct the 3D flux measurements for Figure \ref{fig:blob_transition} without clipping the blob. 

We tested effects of the different thickness of the orifices on the generated rings but found that it did not qualitatively affect the dynamics. 

\begin{center}
{\textbf{B: Velocity field extraction}}
\end{center}
In order to characterize the flow, we illuminated fluorescent polyethylene microspheres ($d=100\mu$m, $\rho=1.090$g/cc, Cospheric) with a Nd:YLF single cavity diode pumped solid state laser ($<$40mJ/pulse, 527nm). A high-speed camera captured the beads' motion (Phantom v2515 or Phantom VEO4k, Vision Research) on a thin laser sheet (thickness: 1mm) for 2D Particle Image Velocimetry (PIV). We varied the frame rate of the cameras, depending on the speed of the vortex ring, ranging from 250 to 3000 fps while a ``quarter rule''~\cite{raffel_particle_2018} was always satisfied for the largest displacement observed.  We extracted the velocity fields with a software called DaVis (LaVision, Inc.). There, we used the pyramid algorithm~\cite{sciacchitano_multi-frame_2012} to generate a velocity field for turbulent analysis (energy spectrum, structure function, dissipation rate, and turbulence length scales) as it was shown to extract the small-scale motion more accurately than the standard cross-correlational algorithm (WIDIM~\cite{scarano2001iterative}).

For 4D measurements, we set up an array of three to four cameras to capture the motion of the same beads illuminated in a volume of $120$mm$\times 100$mm$\times 80$mm, created by two cylindrical lenses. The 3D particle tracking algorithm called ``Shake-the-box''~\cite{schanz_shake--box_2016} detected $O(10^{5})$ particles from the images of the different perspectives, and reconstructed their trajectories. Binning the Lagrangian velocities gave the underlying Eulerian velocity field at that frame. On average, 3-7 trajectories were present in each voxel with a width of 2.9mm. For the voxels with no trajectories, the velocity field was interpolated linearly or filled with the neighboring values.

\begin{center}
{\textbf{C: 3D visualization}}
\end{center}
The Lagrangian trajectories obtained by 3D PTV were first characterized by their lifespans, traveled distances, average speeds, and positions when they were first detected. 
We used this information to identify the particles transported by the vortex rings. 
The selected particles were then visualized as pathlines using a rendering software called Houdini (SideFX). The SI movies show the pathlines combined of the four recordings for each experiment (coherent reconnections and a turbulent state.)

As for the visualization of the mean flow energy and the mean turbulent energy, we used a software called Dragonfly (Object Research Systems).

\onecolumngrid

\newpage
\clearpage

\newpage
\begin{center}
\large
\textbf{Creation of an isolated turbulent blob fed by the injection of vortex loops\\Supplementary material}\\
\mbox{} \\
\large
Takumi Matsuzawa, Noah P. Mitchell, and St\'ephane Perrard\\
\mbox{} \\
{\it{James Franck Institute and Department of Physics,\\
University of Chicago, Chicago, IL 60637, USA}}\\
\mbox{} \\
William T.M. Irvine\\
\mbox{} \\
{\it{James Franck Institute, Enrico Fermi Institute and Department of Physics,\\
University of Chicago, Chicago, IL 60637, USA}}\\
\end{center}

\newpage

\tableofcontents
\clearpage

\section{Symbols and conventions} \label{sect:nomenclature}
In this section we summarize the symbols and conventions used in the main text and SI. 

For a vector field $U_i(x_j, t)$,
\begin{align}
    \langle U_i \rangle(x_j, t) = \int f(x_j, t) U_i(x_j, t) dx_j dt
    \label{eq:def_ensemble_avg}
\end{align}
refers to an ensemble average with the probability function $f(x_j, t)$. In experiments, $f(x_j, t)$ is unknown a priori. Instead of ensemble averaging, we frequently perform temporal averaging of $U_i(x_j, t)$:
\begin{align}
    \langle U_i \rangle_t(x_j) = \frac{1}{\tau} \int_0^\tau  U_i(x_j, t) dt.
    \label{eq:def_temporal_avg}
\end{align}
The ensemble averaging and the temporal averaging are said to be identical for a statistically stationary system (ergodicity).

The spatial average is denoted by a subscript $s$.
\begin{align}
    \langle U_i \rangle_s(t) = \frac{1}{\mathcal{V}} \int_\mathcal{V}  U_i(x_j, t) dV
    \label{eq:def_spatial_avg}
\end{align}
Similarly, the average over a polar angle is denoted by $\theta$. 
\begin{align}
    \langle U_i \rangle_\theta(r, \phi, t) = \frac{1}{2\pi} \int_0^{2\pi}  U_i(r, \theta, \phi, t) rd\theta
    \label{eq:def_polar_avg}
\end{align}
To study the phase dependence of the flow, we use the following notion to denote the phase $\Theta \in [0, 1)$ throughout the paper. 
\begin{align}
    t = n T + \Theta
    \label{eq:def_phase}
\end{align}
Here, $n\in \mathbb{Z}$ is a number of cycles, and $T$ is a forcing period. The phase-locked (or phase-averaged) quantity is denoted by a subscript $n$, and is given by
\begin{align}
    \langle U_i \rangle_n(x_j, \Theta) = \frac{T}{\tau}\int_0^\tau  U_i(x_j, t) \delta[ (t  \text{ mod } T)  - \Theta]dt
    \label{eq:def_phase_avg}
\end{align}
for $\tau \text{ mod } T=0$. These averaging procedures can be combined. For example, we presented the energy averaged over space and cycles (Eq.~\ref{eq:def_spatial_avg} and \ref{eq:def_phase_avg}) in Figure 2d.
\begin{align}
    \frac{1}{2}\langle U_i U_i \rangle_{sn}(\Theta) = \frac{T}{\tau} \frac{1}{\mathcal{V}}\int_0^\tau \int_\mathcal{V} \frac{1}{2} U_i(x_j, t) U_i(x_j, t)  \delta[(t \text{ mod } T)  - \Theta] dV dt
    \label{eq:def_spatial_phase_avg}
\end{align}

\begin{longtable}[c]{ c  l }
    \caption{Nomenclature}\\
      & \bf{Basics} \\ \hline
     $p$ & pressure\\
     $Re$ & Reynolds number, $Re=(u' \mathcal{L}) / \nu$\\
     $t$ & Time, $t=(n+\Theta) T; n \in \mathbb{Z}$\\
     $U_i$ & the $i$-th component of a velocity field \\
     $u_i$ & the $i$-th component of a fluctuating velocity field, or a velocity field in a general context \\
     $\Omega_i$ & the $i$-th component of a vorticity field   \\
     $\omega_i$ & the $i$-th component of a vorticity of a fluctuating velocity field, or vorticity of a general velocity field  \\
     $\Theta$ & Phase with respect to a forcing period, $\Theta \in [0, 1)$\\
     $\rho$ & Density of a fluid\\
     $\nu$ & Kinematic viscosity\\
     $x, y, z$ & Cartesian coordinates\\
     $\sigma, \varphi, z$ & Cylindrical coordinates\\
     $r, \theta, \phi$ & Spherical coordinates\\
     $\vec{\kappa}, \kappa_i$ & Wavenumber\\
     $\delta$ & Kronecker delta\\
    \\
    
      & \bf{Symbols} \\ \hline
     $\langle A \rangle$ & Ensemble average of $A$ \\
     $\langle A \rangle_n$ & Phase-locked average of $A(t)$ (Average over cycles with frequency $f$), \\
     & $\langle A \rangle_n (\Theta) = \frac{1}{fT}        \int_0^T A(t) \delta((ft-\Theta) {~\rm{mod}~} 1))dt$\\
     $\langle A \rangle_s$ & Spatial average of $A(x_i)$, $  \langle A \rangle_s = \frac{1}{V} \int_\mathcal{V} A(x_i) dV  $ \\
     $\langle A \rangle_t$ & Temporal average of $A(t)$, $  \langle A \rangle_t = \frac{1}{T} \int_0^T A(t) dT  $ \\
     $\langle A \rangle_\phi$ & Azimuthal average of $A(r, \theta, \phi)$ in spherical coordinates, $  \langle A \rangle_\phi = \frac{1}{2\pi} \int_0^{2\pi} A(r, \theta, \phi) r^2\sin{\theta} d\phi$  \\
     $\langle A \rangle_\theta$ & Polar average of $A(r, \theta, \phi)$ in spherical coordinates, $\langle A \rangle_\theta = \frac{1}{\pi} \int_0^{\pi} A(r, \theta, \phi) r^2\sin{\theta}  d\theta$  \\
     $\breve{u}_i$ & Discrete Fourier transform of a fluctuating velocity field $u_i$ \\
     $\tilde{u}_i$ & Continuous Fourier transform of a fluctuating velocity field $u_i$ \\
    \\
    
     & \bf{Section II: Experimental methods} \\ \hline
     $f$ & Forcing frequency\\
     $v_{eff}$ & Effective velocity of a piston, $v_{eff}=P\langle V_p \rangle$ \\
     $D_o$ & Diameter of an orifice in the experimental setup\\
     $D_p$ & Diameter of a piston in the experimental setup\\
     $L_p$ & Stroke length\\
     $L_{\rm box}$ & Length of the experimental chamber\\
     $N$ & Number of orifices in the experiments\\
     $R_p$ & Radius of a piston; $R_p = D_p / 2$\\
     $T$ & Forcing period, $T=1/f$\\
     $V_{\rm  p}$ & Velocity of a piston  \\
     \\
     
     & \bf{Section III: Vortex rings (Inviscid invariants)} \\ \hline
     $\vec{A}$ or $A_i$ & Angular impulse over an entire space\\
     $H$ & Helicity over an entire space\\
     $\vec{I}$ or $I_i$ & Hydrodynamic impulse over an entire space\\
     $K$ & Kinetic energy over an entire space\\
     $\vec{L}$ & Angular momentum\\
     $\vec{P}$ & Linear momentum\\
     $\Gamma$ & Circulation  \\
    \\
    
    & \bf{Section III: Vortex rings (Canonical models and Norbury's family of vortex rings)} \\ \hline
     $a$ & Radius of a vortex core\\
     $c$ & Proportionality constant related to energy partition of a vortex ring, $K_{\rm atmosphere} = c K_{\rm ring}$ \\
     $c_1$ & Parameter related to vorticity distribution of experimentally vortex rings Eq. \ref{eq: vort_dist}\\
     $A_{\rm sph}$ & Radius of Hill's spherical vortex\\
     $H_{\rm ring}$ & Helicity of a vortex ring/loop\\
     $I_{\rm ring}$ & Hydrodynamic impulse of a vortex ring\\
     $K_{\rm ring}$ &  Kinetic energy inside the vortex atmosphere with a presence of a vortex ring; $K_{\rm ring} = c K$\\
     $R_{\rm ring}$ & Radius of a vortex ring  (radial distance to the first moment of vorticity)\\
     $V_{\rm ring}$ & Self-induced velocity of a vortex ring  \\
     $\Gamma_{\rm ring}$ & Circulation of a vortex ring  \\
     $\Omega_{\rm ring}$ & Volume of a vortex atmosphere \\
     $\alpha$ & Constant in a vortex ring model\\
     $\beta$ & Constant in a vortex ring model\\
    
     $\psi$ & Streamfunction\\
     $K_1$ & The first kind of the elliptic integrals\\
     $K_2$ & The second kind of the elliptic integrals\\
     
     $\alpha_{\rm Norbury}$ & Shape parameter of a vortex family in ~\cite{norbury_family_1973}\\
     $K_{\rm added}$ & Kinetic energy outside the vortex atmosphere with a presence of a vortex ring \\ 
     $K_{\rm rect}$ & Kinetic energy associated to a rectilinear motion of a vortex ring \\
     $K_{\rm int}$ & Kinetic energy associated to a internal motion of a vortex ring \\
     $K_\psi$ & Kinetic energy inside a volume bounded by an isocontour of a streamfunction $\psi$ \\
     $M_{ij}$ & Added mass tensor\\
    \\
     & \bf{Section III: Vortex rings (Production and measurement)} \\ \hline
     ${\vec{d}}$ & Unit vector of propagation of a vortex ring\\
     $h$ & Semi-major radius of a vortex ring\\
     $l$ & Semi-minor radius of a vortex ring\\
     $L_*, L/D$ & Formation number\\
     $K_{\rm exp}$ & Kinetic energy inside the vortex atmosphere, measured by the PIV experiments\\
     $P$ & Velocity program factor, $P=\langle V_p^2 \rangle_t / \langle V_p \rangle_t^2$\\
     $R_{\rm slug}$ & Reynolds number of a slug of fluid\\
     $V_{\rm slug}$ & Velocity of a slug of fluid\\
     $\Gamma_{\rm c}$ & Circulation of a vortex ring within a core \\
    \\
     & \bf{Section III: Vortex rings (Helicity attenuation)} \\ \hline
     $a'$& Clipping length\\
     $R_{\rm major}$& Semi-major length of a twisted vortex loop\\
     $R_{\rm minor}$& Semi-minor length of a twisted vortex loop\\
     $Tw$& Twist of a vortex loop\\
     $Wr$& Writhe of a vortex loop\\
     $\psi$ & Complex scalar function to construct a twisted vortex loop\\
     $\chi$ & Parameter associated to a core size of a vortex loop\\
    \\
     & \bf{Section V-VI: Turbulent blob and Turbulent statistics} \\ \hline
     $k$ & Turbulent kinetic energy, $k=\frac{1}{2} \langle u_i u_i \rangle$ \\
     $u'$ & Characteristic fluctuating speed, $u'=\sqrt{u_i u_i} / N; i=1-N$\\
     $C$ & Kolmogorov constant related to $E(\kappa)$, $E(\kappa)=C\epsilon_0^{2/3}\kappa^{-5/3}$ in the inertial subrange \\
     $C_1$ & Kolmogorov constant related to $E_{11}(\kappa)$, $E_{11}(r) = C_1 \epsilon_0^{2/3}\kappa_1^{-5/3}$ in the inertial subrange\\
     $C_2$ & Kolmogorov constant related to $D_{LL}(r)$, $D_{LL}(r) = C_2 (\epsilon_0 r)^{2/3}$ in the inertial subrange\\
     $\mathcal{E}_0$ & Average kinetic energy inside a turbulent blob\\
     $\mathscr{E}$ & Kinetic energy per unit mass $\mathscr{E} = U_i U_i/ 2$ \\ 
     $\mathcal{L}$ & Large eddy turnover time, $\mathcal{L} = u'^3/\epsilon_0$ \\
     $R_{\rm blob}$ & Radius of a turbulent blob\\
     $R_{\rm eff}$ & Effective radius of a turbulent blob, $R_{\rm eff}=4^{1/3}R_{\rm blob}$\\
     $\delta u$& Different of fluctuating velocites between two points, $\delta u =u(\vec{x} + \vec{r}) - u(\vec{x}) $ \\\
     $\epsilon$ & Dissipation rate\\
     $\epsilon_0$ & Dissipation rate inside a homogeneous turbulent blob\\
     
     $\eta$ & Kolmogorov length scale $\eta = (\frac{\nu^3}{\epsilon_0})^{\frac{1}{4}}$\\
     $\lambda$ & Transverse Taylor microscale\\
     $\overline{\Omega}_i$ & the $i$-th component of vorticity of a temporally averaged flow $\langle U_j \rangle_t$\\
    \\
     & \bf{Section VII: Energy spectrum and structure function} \\ \hline
     $f(r)$ & longitudinal velocity correlation function\\
     $g(r)$ & transverse velocity correlation function\\
     $k$ & Average fluctuating kinetic energy\\
     $k_{\rm local}$ & Average fluctuating kinetic energy in the measurement domain\\
     $w(x_i)$ & Window function\\
     $D_{ij}$ & Second-order structure function\\
     $D_{LL}$ & Second-order longitudinal structure function\\
     $E_{ij}(\kappa_1)$ & One-dimensional turbulent energy spectrum along $x_1$\\
     $E(\kappa)$ & Turbulent energy spectrum\\
     $\mathcal{F}_{x_i}$ & Fourier operator along the direction $\hat{x}_i$\\
     $J$ & Jacobian\\
     $R_{ij}$ & Velocity two-point correlation function\\
     $W_x$ & Interrogation window size along the $x$ direction \\
     $W_y$ & Interrogation window size along the $y$ direction \\
     $\zeta$ & Correction factor due to windowing \\
     $\kappa_\theta$ & Polar component of a wavenumber vector\\
     $\kappa_\phi$ & Azimuthal component of a wavenumber vector\\
     $\kappa_{Ny}$ & Nyquist wavenumber\\
     $\kappa_r$ & Radial component of a wavenumber vector, $\kappa_r = \sqrt{\kappa_i \kappa_i}$\\
     $\Phi_{ij}$ & Velocity spectrum tensor \\
     \\
     & \bf{Section VIII: Energy balance of a blob} \\ \hline
     $p(r/R_{\rm blob})$ & Fraction of dissipation (energy, or enstrophy) inside a sphere of radius $r$ to the total Eq. \ref{eq:dissipation_fraction_model}\\
     $s_{ij}$ & rate-of-strain tensor of a fluctuating velocity,  $s_{ij} = \frac{1}{2} (\partial_j \langle u_i \rangle + \partial_i  \langle u_j \rangle)$\\
     $\mathcal{D}$ & Dissipated power, $\mathcal{D} = \rho \int_\mathcal{V} \epsilon(\vec{r}) dV$\\
     $\mathcal{D}_{\rm tot}$ & Total dissipated power inside the experimental chamber\\
     $\mathcal{I}$ & Injected power, $\mathcal{I}=8K_{\rm ring} f$\\
     $\mathcal{I}_{\rm tot}$ & Total injected power inside the experimental chamber\\
     $S_{ij}$ & Rate-of-strain tensor, $S_{ij}=\frac{1}{2} (\partial_j U_i + \partial_i U_j)$ \\ 
     $\overline{S_{ij}}$ & Rate-of-strain tensor of mean flow, $S_{ij}=\frac{1}{2} (\partial_j \langle U_i \rangle + \partial_i  \langle U_j \rangle)$ \\
     $\epsilon_{s_{ij}}$ & Dissipation rate that is computed via a rate-of-strain tensor $s_{ij}$\\
     $\epsilon_{D_{LL}}$ & Dissipation rate that is computed via a second-order structure function $D_{LL}$\\
     $\epsilon_{E_{11}}$ & Dissipation rate that is computed via a one-dimensional energy spectrum $E_{11}$\\
     $\tau_\mathcal{L}$ & Large eddy turnover time, $\tau_\mathcal{L} = \mathcal{L} / u'$\\
     \\
     & \bf{Section IX: Enstrophy flux and confinement transition} \\ \hline
     $f_c$ & Transitional forcing frequency from coherent vortex reconnections to formation of a turbulent blob\\
     $m_{ij}$ & Image moment of order (i, j) (Eq. \ref{eq:imagemoments_raw})\\
     $\vec{J}_m$ & Mass current $\vec{J}_m = \rho \vec{U}$\\
     $\vec{J}_{\Omega^2}$ & Enstrophy current, $\vec{J}_{\Omega^2} =  \Omega^2 \vec{U}$\\
     $\alpha_0$ & Parameter0 in Figure 4; $\mathcal{L}=\alpha_0 R_{\rm ring}$\\
     $\alpha_1$ & Parameter1 in Figure 4; $\mathcal{\epsilon_0}=\alpha_1 V_{\rm ring}^2 f$\\
     $\nu_{ij}$ & Second-order image moments\\
     $I_{\rm img}$ & Intensity distribution of an image\\
     $\Phi_m$ & Integrated mass flux \\
     $\Phi_Z$ & Integrated enstrophy flux \\
     \\
     & \bf{Section X: Gross-Pitaevskii simulation} \\ \hline
     $\Psi$ & Wavefunction\\
     $\Phi$ & Phase\\
     $\rho$ & Density\\
     $\xi$ & Healing length\\
     \\
    \label{tab:nomenclature}
 \end{longtable}
\newpage

\section{Experimental Methods}

In this section we describe the experimental flow chamber geometry, the flow actuation and our imaging methods. 

\subsection{Flow chambers}

\begin{figure}[!htb]
\centering
\includegraphics[width=0.99\textwidth]{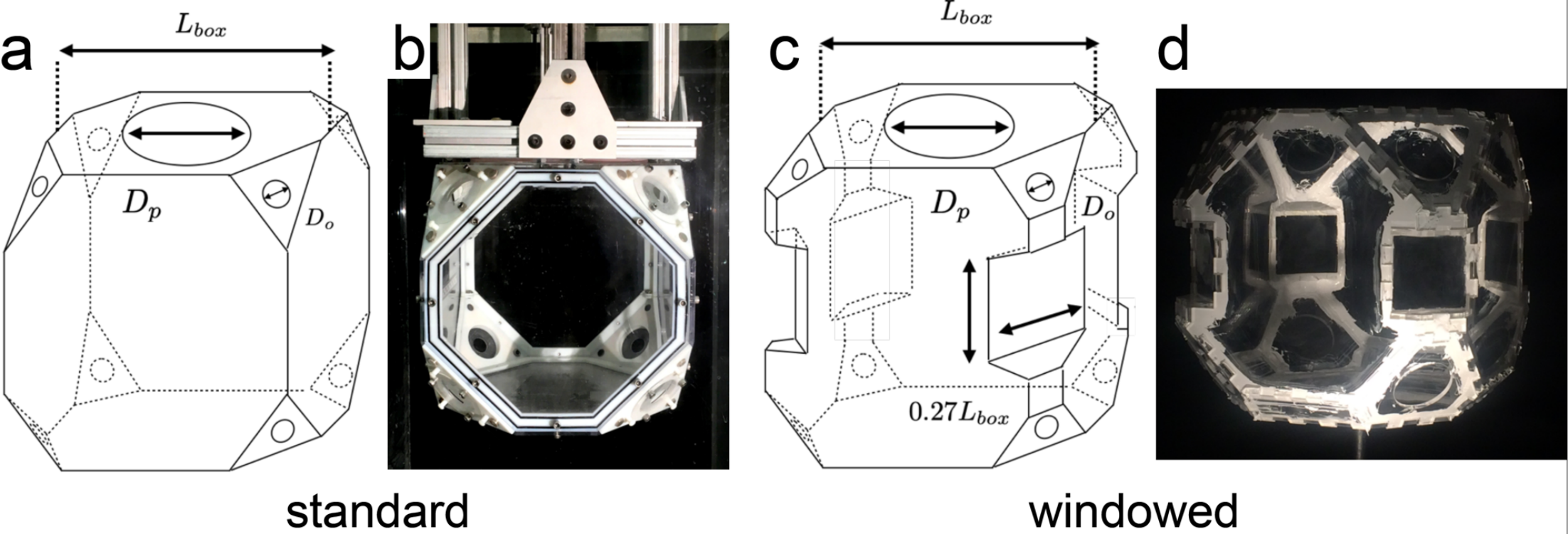}
\caption{\textbf{Geometry of the flow chambers.} (a-b) The ``standard chamber'' consists of a 3D-printed frame and acrylic faces. (c-d) The ``windowed chamber'', made of acrylics, has four additional windows to measure vortex ring properties by particle image velocimetry.}
\label{fig:FlowChmaberGeometry}
\end{figure}

Our experiments were performed in either of the two flow chambers depicted in Figure~\ref{fig:FlowChmaberGeometry}. Both chambers have the geometry of a cube with  corners truncated by triangular plates, each  containing an orifice through which  vortex rings are generated.  
The linear dimension of the cube is  $L=351mm$ in both cases. 
The chambers differ in their method of construction and their optical accessibility.

The chamber hereby referred to as the `standard chamber' consists of a 3D-printed frame onto which acrylic sheets are bolted, whereas the 'windowed chamber' was assembled by gluing acrylic sheets together and features four windows inserted long the vertical edges of the cube. These windows allow `edge-on' visualization of the flow, typically by laser sheet in a diagonal plane which contains the axis along which the vortex rings travel. 

For both chambers, the top plate of the chamber contains a circular hole with diameter of $200mm$. An additional acrylic sheet with a hole of diameter $D_p$ is bolted on top face of the chamber, through which a piston is inserted. This allows forcing with different size of the piston. The piston is an acrylic cylinder with diameter $D_p$, and is attached to the actuator by stainless steel threads. Additionally, we attach a silicone rubber on the top face to reduce any excess flow dragged by the motion of the piston. At each truncated triangular plate, we magnetically attach a holster bolted with a 1/16in acrylic sheet with a hole of diameter $D_o$, through which a vortex ring is generated. In this study, we use two primary settings of $D_p$ and $D_o$ as summarized in Table \ref{tab:exp_dimensions}. 

Unlike the standard chamber, the windowed chamber is fully made of acrylic and calking materials. The other difference is that it does not have holsters at the triangular plates, instead a circular acrylic piece with a hole of diameter $D_p$ is bolted to the plate directly.

\begin{figure}[!htb]
\centering
\includegraphics[width=0.8\textwidth]{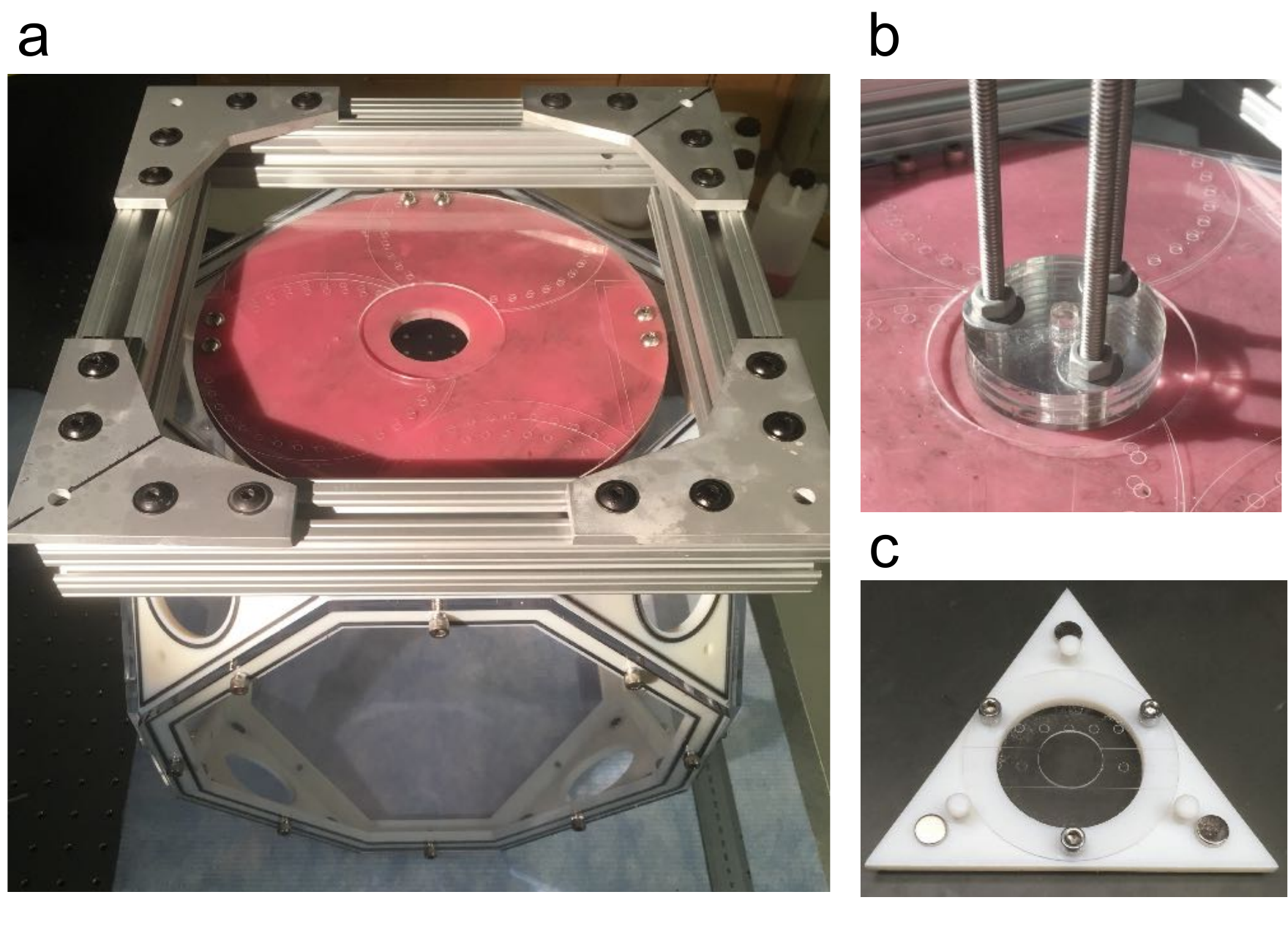}
\caption{\textbf{Components of the experimental chamber.} (a) A silicone rubber (``flap'') with a hole of diameter $D_p$ is attached on the top face to minimize excess flow into/out of the chamber. (b)An acrylic, cylindrical piston is inserted through the hole. (c) A 3D-printed ``holster'' is magnetically attached to each truncated surface of the chamber. An acrylic sheet with an orifice or a 3D-printed helical mask is bolted to the holster.}
\label{fig:chamber_components}
\end{figure}

\subsection{Actuation}
The flow is set up by translating a piston by a linear motor (STA2510S, Copley Controls Corp).. When the piston is actuated impulsively in the vertical direction, it pulls a slug of fluid through each orifice into the flow chamber. This creates a vortex ring at each orifice.

\begin{figure}
\centering
\includegraphics[width=\textwidth]{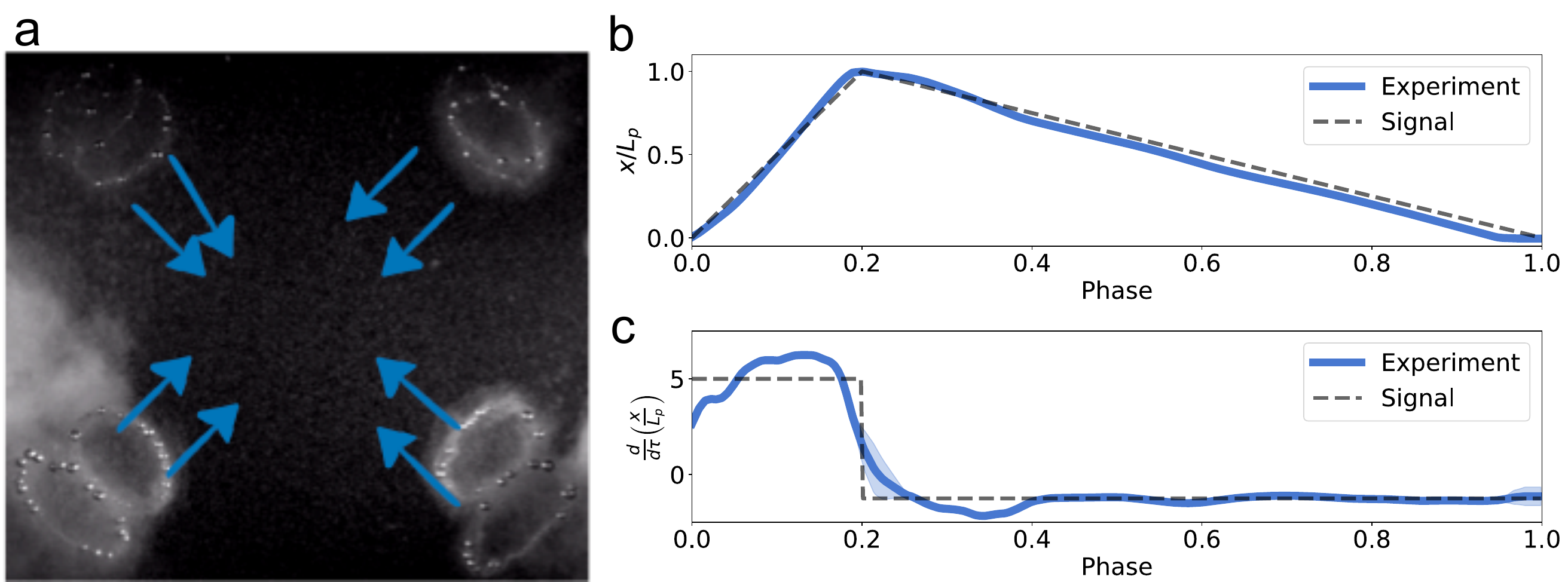}
\caption{\textbf{Controlled generation of vortex rings} (a) Generated vortex rings are visualized by bubbles. (b) An actuating signal (gray) and realized motion of the piston are plotted against the phase. The curve is obtained by averaging the encoder output in a phase-locked manner. $L_p$ is the commanded stroke length. (c) Commanded (gray)  and realized (blue)  velocity profiles are plotted against the phase. The shade represents the standard deviation of the collected velocity profiles ($n=8$).}
\label{fig:bubble_vring_signal}
\end{figure}

The actuation velocity is programmable, and controlled by a National Instrumental instrument DAQ board. A typical signal is shown in Figure \ref{fig:bubble_vring_signal}(b) as a dotted line.  As the piston lifts up every cycle, some volume of fluid is drawn into the chamber through the orifices at the triangular plates, creating a set of vortex rings. The properties of the rings are described in Section \ref{sect:vortex_rings}. For periodic forcing, we program the motor to slowly move the piston back to its starting position. 

\subsection{3D PTV experiments}
\begin{figure}
\centering
\includegraphics[width=0.99\textwidth]{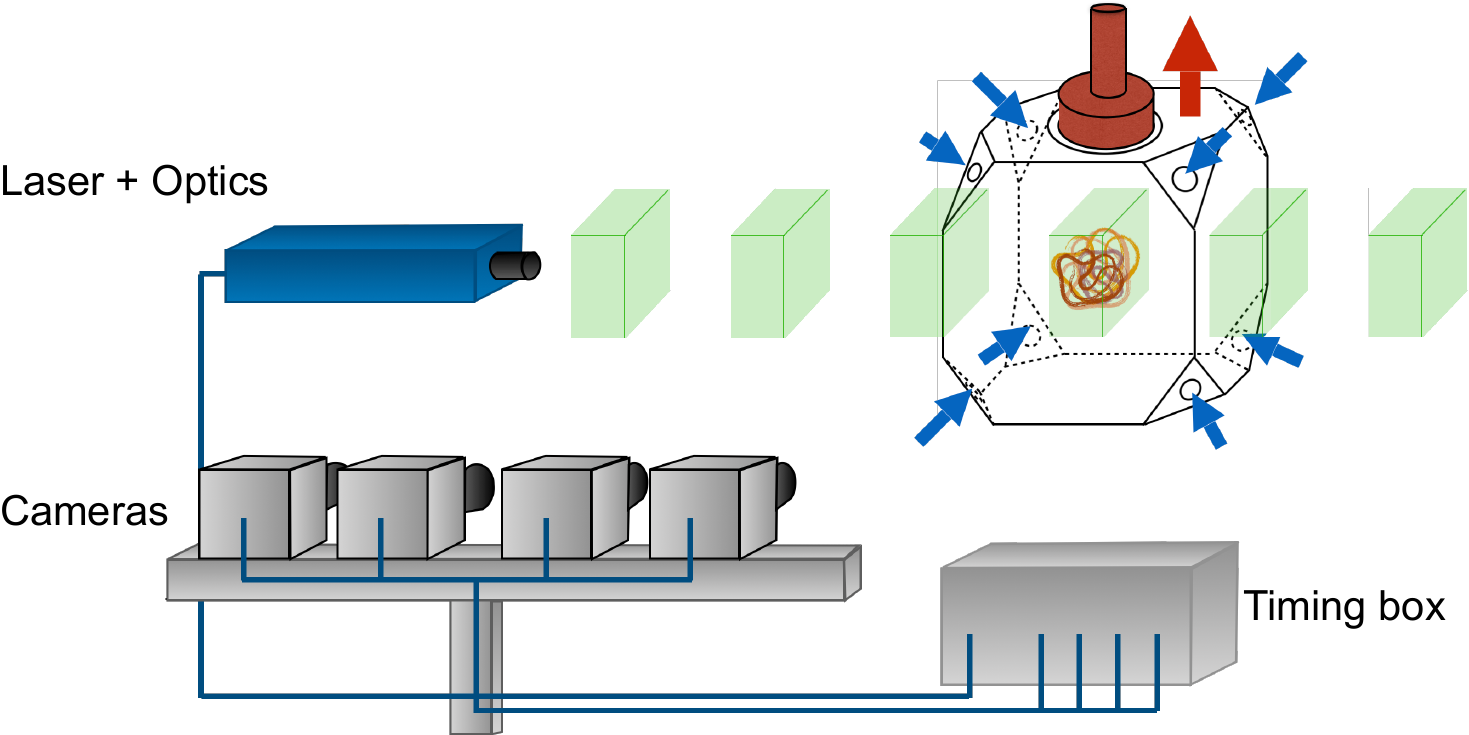}
\caption{{\textbf{A schematic of the 3D PTV experiments.}} A Nd: YLF, pulsed laser and two cylindrical lenses are used to create an illuminated region in the chamber. Four cameras track the $O(10^5)$ particles every frame that are used to reconstruct an Eulerian velocity field.}
\label{fig:stb_setup}
\end{figure}

We perform 3D PTV with three or four high-speed cameras (Phantom VEO640L, Vision Research) in the setup depicted in Figure \ref{fig:stb_setup}). 
Each camera is equipped with a Scheimpflug adaptor to correct the focus as the lens is tilted with respect to the imaging plane. The same band-pass filter as described in the PIV section is used to eliminate the background illumination. 
Two cylindrical lenses are used to expand the beam from an Nd:YLF pulsed laser ($\lambda=526.5$nm) to create an illuminated {\it volume} as opposed to a plane. 
We use the same fluorescent particles as described above.

We extract velocity fields using the LaVision ``Shake-the-box'' package~\cite{schanz_shake--box_2016}.  The 3D PTV experiments include a two-step calibration procedure. The first step uses a static target to create mapping functions between the each camera view and the real 3D space. The second step self-consistently refines the mapping functions using particle images. Details of the method can be found in~\cite{schanz_shake--box_2016}. The maximum volume for the measurement is approximately 200mm x 120mm x 100mm (width x height x depth). In this volume, $O(10^5)$ particles are registered after successful calibrations, and $50-80\%$ are recognized as tracks. 

\begin{table}[]
    \centering
    \begin{tabular}{c c c c c}
         Name & $L_{box}$ (mm) & $D_p$ (mm) & $D_o$ (mm) & Box Type \\ \hline
         Setting 1 & 351.0 & 160.0 & 25.6 & Standard chamber  \\
         Setting 2 & 351.0 & 57.0 & 12.8 & Standard chamber \\
         Setting 3 & 351.0 & 160.0 & 25.6 & Windowed chamber \\
    \end{tabular}
    \caption{Dimensions of the chamber used in this paper}
    \label{tab:exp_dimensions}
\end{table}

\begin{figure}
\centering
\includegraphics[width=0.99\textwidth]{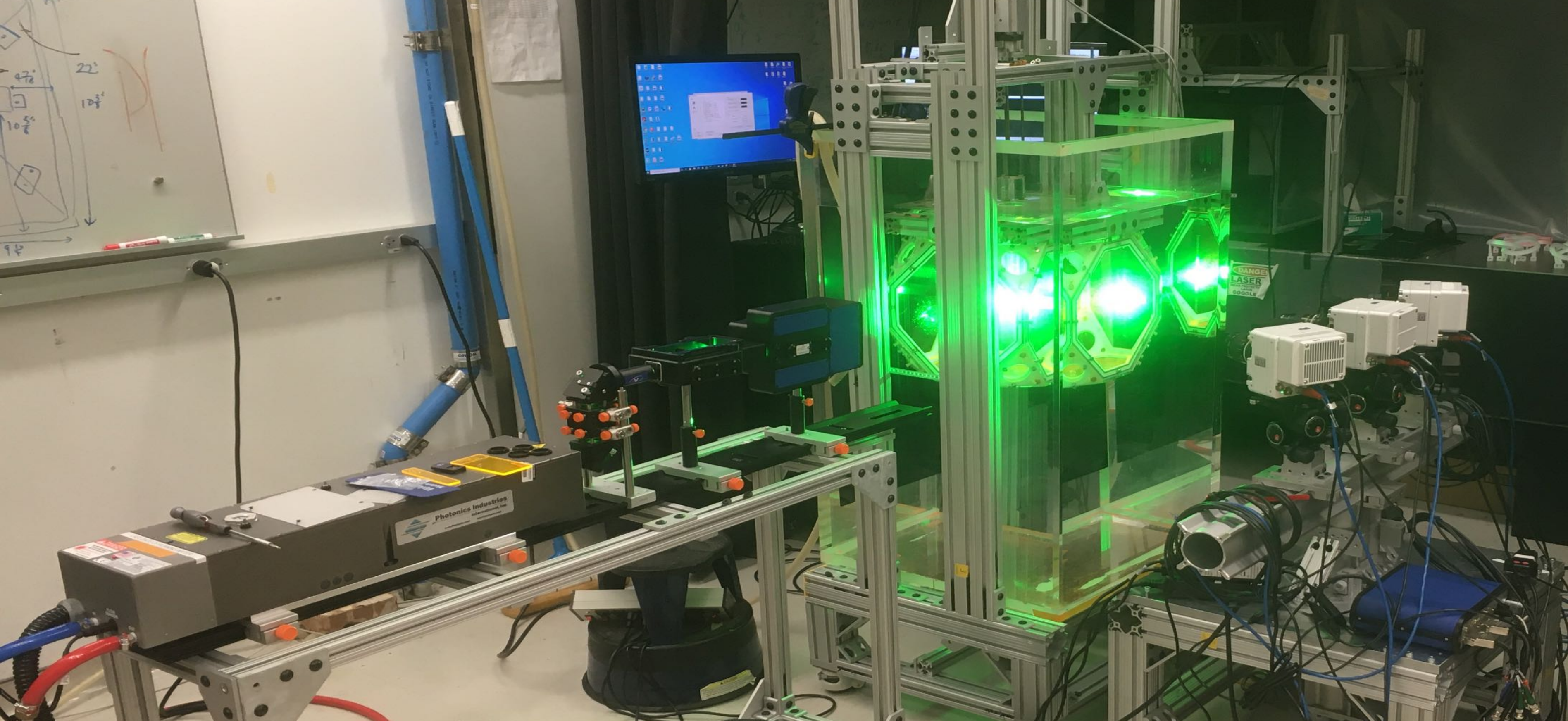}
\caption{{\textbf{The 3D PTV setup}} (left) A Nd: YLF, pulsed laser and two cylindrical lenses are used to create an illuminated region in the chamber. (Middle) The 3D-printed, standard chamber is fixed in a water tank. (Right) Multiple (up to four) high-speed cameras are mounted on the mobile cart.}
\label{fig:stb_setup_photo}
\end{figure}

\clearpage 
\newpage

\section{Vortex rings} \label{sect:vortex_rings}
In this section, we dissect vortex rings generated in our experimental chamber in a following order. First, we review general properties of vortex rings, referring to their canonical theories. Then, we consider the experimental conditions to generate a vortex ring in our chamber that gives a rise to the accessible parameter space. Finally, we present our measurements using 2D PIV and 3D PTV, comparing to the theories and the results obtained by different experimental systems.

\subsection{Anatomy of injected vortex loops}

\begin{figure}[!htbp]
\centering
\includegraphics[width=0.9\textwidth]{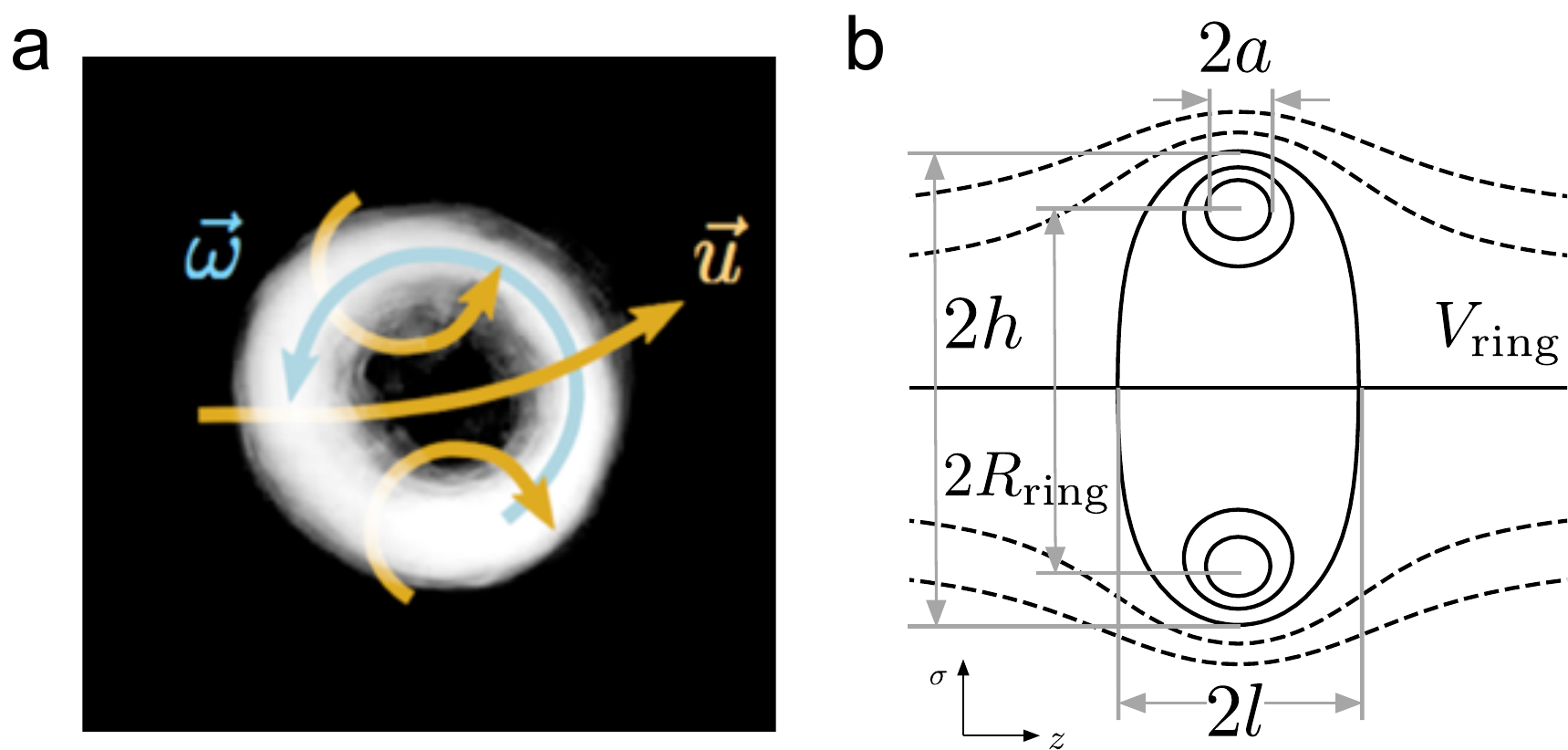}
\caption{(a) A vortex ring (b) Streamline pattern of a vortex ring in the comoving frame}
\label{fig:vring_schematic}
\end{figure}

The vortex loops we generate in our experiment all share a similar anatomy to the ring depicted in Figure~\ref{fig:vring_schematic}.
The vorticity is concentrated in a finite region with the topology of a torus and a geometry somewhere between a thin ring and a sphere. 
The precise distribution of vorticity varies as a function of time under the action of advection and viscous diffusion. At each moment in time the vortex loop propagates under its self-induced velocity field, which can be obtained from the vorticity field by Biot-Savart inversion. 

The resulting flow is naturally sub-divided into an advected `atmosphere' and a surrounding potential flow that is matched to the atmosphere at its surface. 
The latter is most easily identified by plotting the flow streamlines in the co-moving frame of reference (Figure~\ref{fig:vring_schematic}b).

While the vorticity distribution provides a complete specification of a vortex loop, much additional insight can be gained by comparing its instantaneous structure to canonical analytic stationary solutions for vortex loops  in inviscid fluids,  computing their inviscid invariants (Impulse $I$, Angular Impulse $A$, Helicity  $H$, Energy $E$, and circulation $\Gamma$ (see Box 1), and rationalizing their values in terms of the `atmospheric' flow structure.

\begin{tcolorbox}[floatplacement=t,float,title={\bf Box 1: Inviscid invariants},colframe=red!15!black]
The inviscid invariants are important to characterize a vortex ring besides its velocity field~\cite{saffman1995vortex, akhmetov_vortex_2009}. We may interpret that a vortex ring houses these invariants except energy inside its atmosphere. For what fraction of energy is housed inside the atomopshere, see Box 3.
\begin{align}
   {\rm Circulation: }\Gamma &= \int_\Sigma \omega(z, \sigma) d\sigma dz
    \label{eq:circ_func_general_vring}\\
   {\rm Linear~Impulse: }\vec{I} &= \frac{1}{2} \rho \int_\Omega \vec{r} \times \vec{\omega} dV =\rho \int_\Omega \vec{u} dV +  \frac{1}{2} \rho  \int_\Sigma \vec{r} \times (\hat{n} \times \vec{u} ) dA  \\ &= \pi \rho \int_{\Sigma} \omega\sigma^2 d\sigma dz \text{   (axisymmetric)}
    \label{eq:impulse_func_axisym_vring}\\
   {\rm Energy: }K &= \frac{1}{2} \rho \int \vec{u} \cdot \vec{u} dV  \\
                   &= \pi \rho \int_\Sigma \psi \omega d\sigma dz \text{   (axisymmetric)}
    \label{eq:energy_func_axisym_vring}\\
   {\rm Angular~impulse: } \vec{A}&= \frac{1}{3} \rho \int_\Omega \vec{r} \times (\vec{r} \times \vec{\omega}) dV 
    \label{eq:angImpulse_func_general_vring}\\
   {\rm Helicity: } H &= \int_\Omega \vec{u} \cdot \vec{\omega} dV\label{eq:helicity_general_vring}
\end{align}
Here $\Omega$ and $\Sigma$ refer to the volume and cross-sectional area of the vortex atmosphere. $\psi$ is a corresponding streamfunction. 
\end{tcolorbox}

Several analytical models have been developed for vortex loops~\cite{lamb1924hydrodynamics, kambe2007elementary, norbury_family_1973}. 
Canonical examples include vortex loops on the thin toroidal ring end of the geometric spectrum and Hill's spherical vortex on the more spherical end.
A single-parameter family of solution due to Norbury conveniently interpolates between the two~\cite{norbury_family_1973}. See Box 2 for a comparison of the corresponding integral invariants.

While planar (axisymmetric) vortex loops carry non-zero impulse, energy and circulation, non-planar (azimuthally varying) geometries are required to carry, in addition, angular impulse and helicity. 
Canonical models of nonplanar rings include  thin cored helical vortex rings~\cite{scheeler2017complete} and the generalization of Hill's spherical vortex to include swirl~\cite{moffatt1969degree, moffatt1988generalised} each of which carries finite angular impulse and helicity.
Box 2 includes a computation of the inviscid invariants for these canonical flows. 

Physical intuition for the values of the inviscid invariants of vortex loops can be readily obtained by examining their co-moving atmosphere. 
Circulation is given by integrating the velocity along any loop that passes along the axis of symmetry and otherwise is exterior to the vortex atmosphere. 
It represents the topological charge or `strength' of the loop. 
The impulse $I$ (Box 1,2) is given by the product of translational velocity of the atmosphere $V_{\rm ring}$, multiplied by the sum of the mass enclosed in the atmosphere $m = \rho \Omega_{\rm atmosphere}$ and the added mass $m'$ of the potential flow past the atmosphere ($m/2$ for a spherical atmosphere): $I = (m+m')  V_{\rm ring}$. 
The energy $E$ is similarly given by the kinetic energy of the potential flow past the atmosphere $\frac{1}{2} `m' V_{\rm ring}^2$, the kinetic energy associated with the translation of the atmosphere $\frac{1}{2} m V_{\rm ring}^2$ and an additional contribution from the rotational energy within the atmosphere, which is vortex model dependent, but nonetheless proportional to $V_{\rm ring}^2$. 
Similar arguments can be constructed for the angular impulse.

Helicity by contrast, is most intuitively understood in terms of the topology and geometry of the vorticity field. The value of helicity is given by the  average circulation weighted  linking umber between vortex tube pairs~\cite{moffatt1969degree}. In practice, for thin cored helical rings, the effects of viscosity drive the value towards the $\Gamma^2 Wr$, where $Wr$ is the writhe of the vortex loop center line.

\begin{tcolorbox}[title={\bf Box 2: Canonical vortex ring models and their properties},colframe=red!15!black]
\label{box:vring_models}

Below we summarize the properties of two canonical axisymmetric vortex rings with a finite cross-section: the thin-cored model and Hill's spherical vortex. 
In general, incompressibility and the axisymmetry reduce the problem of identifying steadily translating solutions to solving a Poisson equation for the stream function $\psi(z, \sigma)$, $\Delta \psi=-\omega$. The result is
\begin{align}
   \psi(z, \sigma) = \frac{1}{2\pi} \int_\Sigma \sigma^{\frac{1}{2}} 
   \left[ \left( \frac{2}{q} - k \right) K_1(q) - \frac{2}{q} K_2(q)    \right] \omega(z', \sigma') dz' d\sigma',
    \label{eq:strm_func_general_vring}
\end{align}
where $q=\left[ 4r \sigma / \left( (z-z')^2 +  (\sigma-\sigma')^2 \right) \right]$, and $\Sigma$ is the cross-section of the ring. $K_1(q)$ and $K_2(q)$ are the first and the second kind of the elliptic integrals~\cite{lamb1924hydrodynamics, akhmetov_vortex_2009}. The thin-cored model \cite{lamb1924hydrodynamics} assumes $\omega=\text{const.}$ inside the core with radius $a\ll R_{\rm ring}$. The nature of the core such as the medium can be specified by additional coefficients $\alpha$ and $\beta$~\cite{lamb1924hydrodynamics, gharib_rambod_shariff_1998, sullivan_niemela_hershberger_bolster_donnelly_2008, akhmetov_formation_2001}.  As for Hill's vortex ring with its speed $V_{\rm ring}$, it assumes $\omega(z, \sigma)=-15V_{\rm ring}/ (2A_{\rm sph}^2) \sigma$ within its vortex atmosphere with radius $A_{\rm sph}$. To make a meaningful comparison with the thin-cored model, we choose $R_{\rm ring} = \int \omega \sigma dV / \int \omega  dV $ as the radius of the vortex instead of $A_{\rm sph}$\footnotemark[1]. Table \ref{tab:vring_model_summary} summarizes the assumptions of the models, and their integral properties.\\

\captionof{table}{Summary of the canonical vortex models: thin-cored model and Hill's spherical vortex.}
\centering
\begin{tabular}{c|c|c} \label{tab:vring_model_summary}
Quantity & Thin-cored model & Hill's spherical vortex \\ \hline
$\omega(z, \sigma)$ &   $\begin{cases}
                        \text{const.}, &  \|\sigma - R_{\rm ring}\|\leq a\\
                        0 & \text{otherwise}
                        \end{cases}$ 
                    &   $\begin{cases}
                       15V_{\rm ring}/({2A_{\rm sph}^2}) \sigma, & \sigma \leq A_{\rm sph}\\
                        0 & \text{otherwise}
                        \end{cases}$\\
$R_{\rm ring}$ & $R_{\rm ring}$ & $\frac{3}{4}A_{\rm sph}$\\
$\Gamma_{\rm ring}$ & $\pi \omega a^2$ & $5 A_{\rm ring} V_{\rm ring} =  \frac{20}{3} R_{\rm ring} V_{\rm ring}$\\
$I_{\rm ring}$ & $\approx \pi \rho \Gamma_{\rm ring} R_{\rm ring}^2$ & $\frac{2\pi}{5} \rho \Gamma_{\rm ring} A_{\rm sph}^2 = \frac{32\pi}{45} \rho \Gamma_{\rm ring} R_{\rm ring}^2$\\
$K_{\rm ring}$ & $\frac{1}{2}\rho \Gamma_{\rm ring}^2 R_{\rm ring}\left[ \ln{\frac{8R_{\rm ring}}{a}}-\alpha  \right]$ &  $\frac{2\pi}{35} \rho \Gamma_{\rm ring}^2A_{\rm ring} = \frac{8\pi}{105} \rho \Gamma_{\rm ring}^2 R_{\rm ring}$\\
$A_{\rm ring}, H_{\rm ring}$ & 0 & 0\\
$V_{\rm ring}$ & $\frac{\Gamma_{\rm ring}}{4\pi R_{\rm ring}}\left[ \ln{\frac{8R_{\rm ring}}{a}}-\beta  \right]$ & $\frac{\Gamma_{\rm ring}}{5A_{\rm ring}} = \frac{3}{20}\frac{\Gamma_{\rm ring}}{R_{\rm ring}}$\\
\end{tabular}

~\\[12pt]

\includegraphics[width=0.95\textwidth]{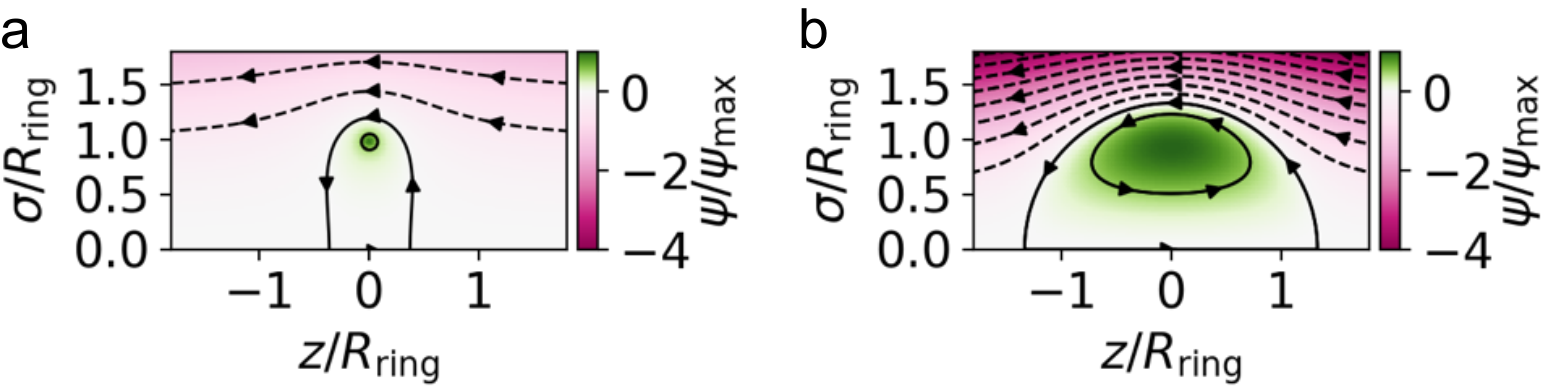}
\label{fig:si_vortex_ring_model_streamlines.pdf}
\captionof{figure}{\textbf{Streamlines of the two canonical models of a vortex ring in the co-moving frame.} (a) Thin-cored model ($R_{\rm ring}/a=60$) (b)Hill's spherical vortex. The color indicates the value of a streamfunction, and the isocontours at every $\Delta \psi / \psi_{\rm max}=0.5$ are shown. (a) and (b) have the same circulation.}

\raggedright

\footnotetext{This choice of $R_{\rm ring}$ is arbitrary but the other candidates are sufficiently close to the first moment of vorticity $0.75A_{\rm sph}$. In the comoving frame, the velocity becomes zero at $1/\sqrt{2}A_{\rm sph}\approx 0.71A_{\rm sph}$ away from the center. The square-root of the second moment is equal to $\sqrt{3/5}A_{\rm sph}\approx0.77A_{\rm sph}$. }
\end{tcolorbox}

\begin{tcolorbox}[title={\bf Box 3: Energy distributions from the atmospheric point of view},colframe=red!15!black]

\label{box:energy distribution}
Energy of a vortex ring may significantly vary depending on the distribution of vorticity within the vortex atmosphere. The atmospheric view provides a simplified view on the inviscid invariants. For example, the impulse of a vortex ring can be readily shown as the sum of the momentum of a fluid within the vortex atmosphere and its added mass in inviscid, incompressible flows~\cite{akhmetov_formation_2001}: $I_{\rm ring}=(m+m')V_{\rm ring}$. How is the energy partitioned from the atmospheric point of view?

As a subject of the study, we consider a family of vortex rings that satisfy $\omega \propto \sigma$. It can be parameterized by $\alpha_{\rm Norbury}= \Sigma / (\pi R_{\rm ring}^2) \in (0, \sqrt{2}]$ where $\Sigma$ is the cross-sectional area of the ring~\cite{norbury_family_1973}. This family smoothly interpolates from a vortex loop with zero cross-sectional area to Hill's spherical vortex. See their vortex atmospheres in Figure \ref{fig:vring_energy_partition_norbury}c.

\centering
\includegraphics[width=0.7\textwidth]{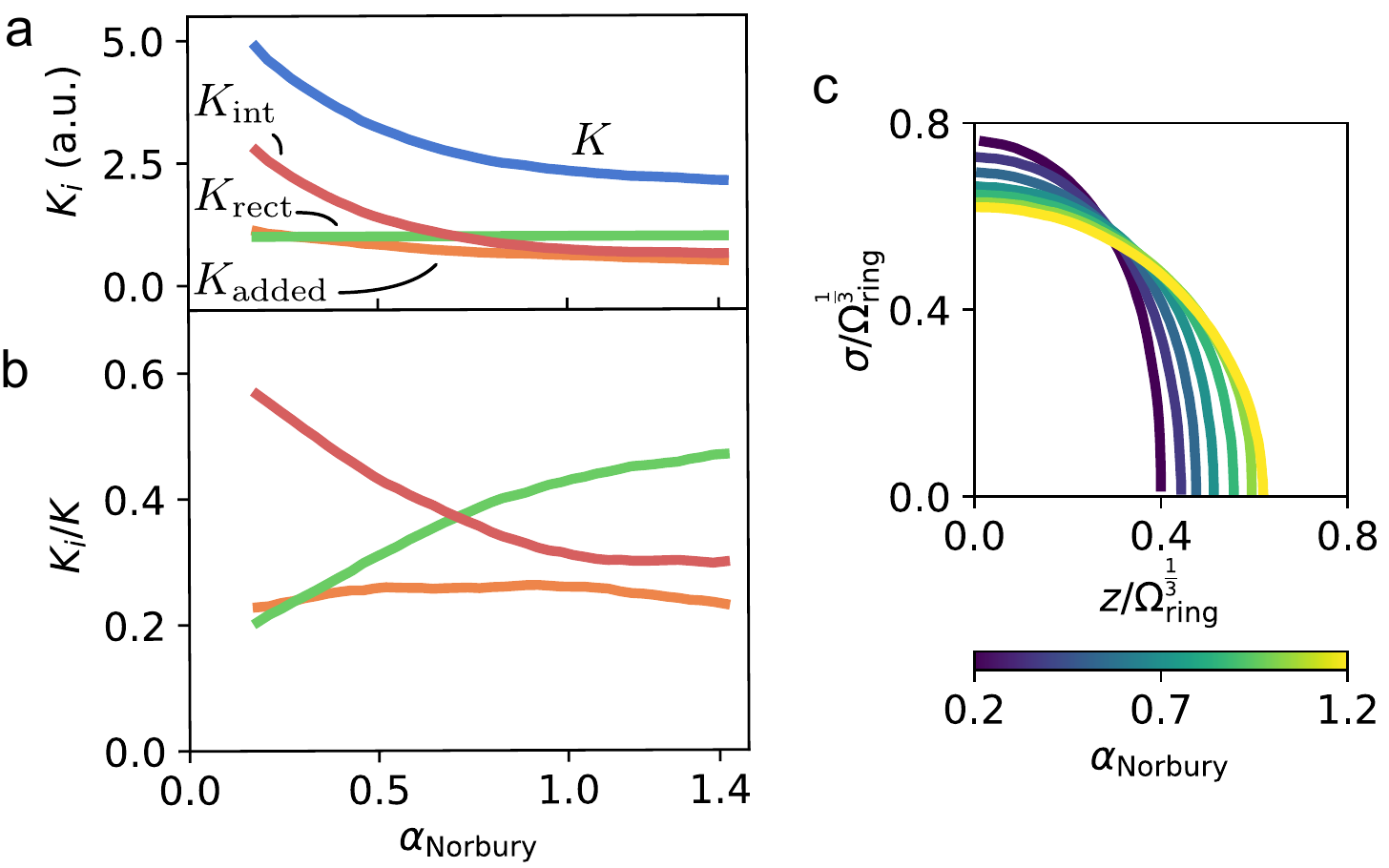}

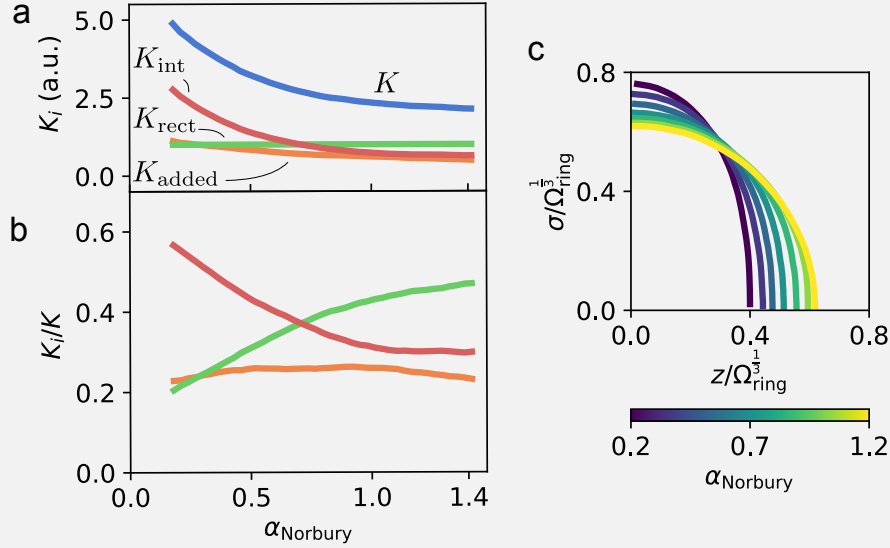
\captionof{figure}{\textbf{Energy partition of Norbury's vortex rings} ($\omega(\sigma) \propto \sigma$)~\cite{norbury_family_1973}. (a) Shape of vortex atmosphere with Norbury's shape parameter $\alpha_{\rm Norbury} \in (0, \sqrt{2}]$. $\Omega_{\rm ring}$ is the volume of each vortex ring.  (b) Relative contribution of energy outside the vortex atmosphere $K_{\rm added}$, energy of the rectilinear motion of the fluids inside the vortex atmosphere $K_{\rm rect}$, and energy of the internal motion $K_{\rm int}$ are plotted against Norbury's shape parameter $\alpha_{\rm Norbury}$.}
\label{fig:vring_energy_partition_norbury}
\raggedright
In general, the energy of a vortex ring can be partitioned as 
\begin{align}
    K &= \underbrace{K_{\rm added}}_\textrm{Motion outside VA} + \underbrace{K_{\rm rect}}_\textrm{Rectilinear motion of VA} + \underbrace{K_{\rm int}}_\textrm{Internal motion inside VA}\\
    &= K_{\rm added} +  K_{\rm ring}.
    \label{eq:ring_energy_partition}
\end{align}

The energy outside the atmosphere is given by $K_{\rm added}= \frac{1}{2} M_{ij} U_i U_j$ where $M_{ij}$ is the added mass tensor of the fluid with the same boundary as the vortex atmosphere. This is because the flow outside the vortex atmosphere in the co-moving frame is identical to the flow of a rigid body in a stream $U_i$ in the absolute frame. Hence, we conclude that the vortex atmosphere encloses $1 - K_{\rm added} / K$ of the energy of the entire flow. Inside the atmosphere, the motion consists of the rectilinear motion and the swirling motion. The exact partition among these modes vary, depending on the vorticity distribution as shown in Figure \ref{fig:vring_energy_partition_norbury}b. Noticeably, the energy inside the atmosphere remains nearly the same for vortex rings with $\alpha_{\rm Norbuary}>0.2$ that resemble the rings generated in our experiments($\alpha_{\rm Norbury}=0.8-1.0$). 

For a spherical vortex, $K_{\rm added}=\frac{1}{4} M_{\rm sph} V_{\rm sph}^2$ and $K_{\rm sph}=\frac{15}{14}M_{\rm sph}V_{\rm sph}^2$ where the mass of the vortex atmosphere is equal to $M_{\rm sph}=\frac{4}{3}\pi \rho A_{\rm sph}^3$. We find $K_{\rm added}: K_{\rm rect}: K_{\rm int}=7:14:9$, meaning that  $76.\overline{6}\%$ of the entire energy is enclosed inside the vortex atmosphere.

\end{tcolorbox}

\subsection{Production of vortex rings}

\begin{figure}[htbp]
\centering
\includegraphics[width=0.8\textwidth]{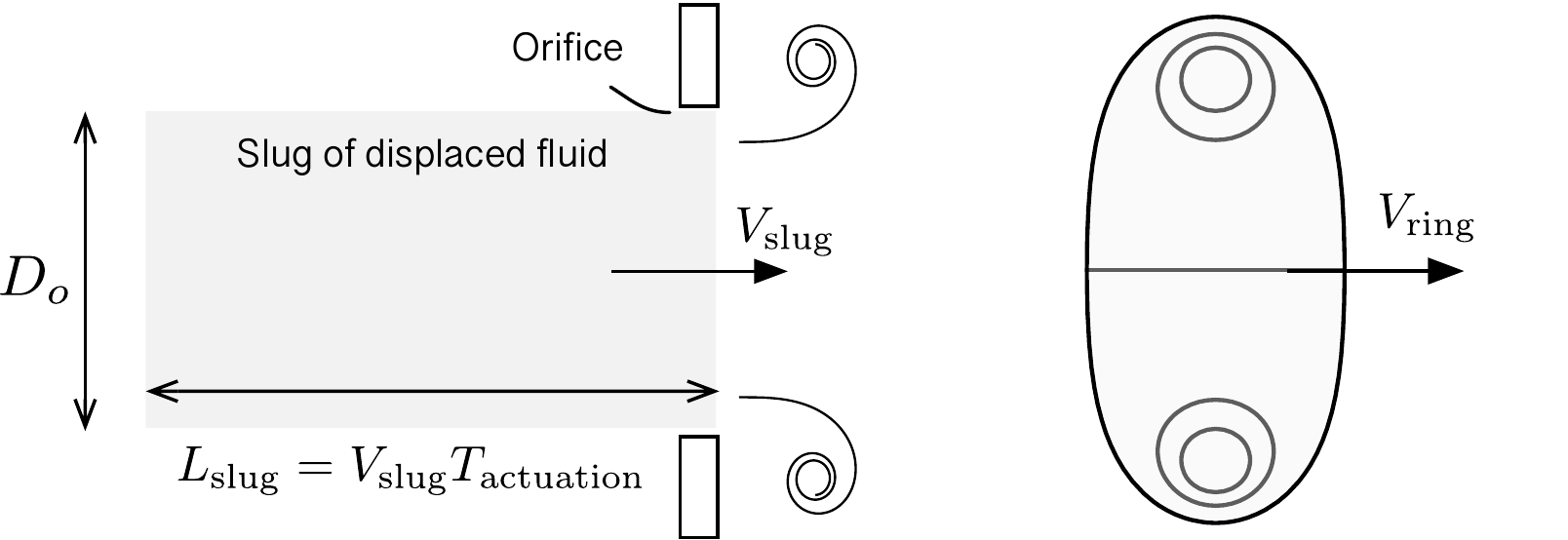}
\caption{\textbf{Production of a vortex ring.} A slug of fluid is extruded through an orifice. The boundary layer at the edge of the orifice separates, and rolls up to form a vortex ring.} 
\label{fig:vring_slug}
\end{figure}

We generate vortex rings by impulsively drawing fluid through an orifice. 
This is a common approach in both natural settings (e.g. dolphin rings, volcanic eruption vortex rings) as well as laboratory settings where it has been studied extensively~\cite{gharib_rambod_shariff_1998, dabiri_gharib_2004, akhmetov_formation_2001}. 
Sketched in Figure~\ref{fig:vring_slug},  the injection of a `slug' of fluid into otherwise quiescent fluid, is followed by the conversion of this slug into a vortex ring that propagates away from the orifice. 
This  conversion  comes about through a subtle combination of inertial and viscous dynamics, including boundary layer separation, vortex sheet roll-up and viscous entrainment~\cite{gharib_rambod_shariff_1998, dabiri_gharib_2004, akhmetov_formation_2001}.

A natural starting point to characterize  the injection process, is to consider a cylinder of fluid with velocity $V_{\rm slug}$,  diameter of an orifice $D_o$, and length $L_{\rm slug} = V_{\rm slug} T_{\rm actuation}$ where $T_{\rm actuation}$ is the time during which the fluid is actuated.
In addition we have the fluid density $\rho$, and kinematic viscosity $\nu$.
From these we end up with dimensionless parameters:
\begin{align}
    L_* = \frac{V_{\rm slug} T_{\rm actuation}}{D_o} \quad , \quad {\rm Re}_{\rm slug}=\frac{V_{\rm slug} D_o}{\nu}, 
    \label{eq: exp_params_nondim}
\end{align}
$L_* = L_{\rm slug}/D_{\rm slug}$ is known as the `formation number' of a vortex ring~\cite{gharib_rambod_shariff_1998, glezer_formation_1988, sullivan_niemela_hershberger_bolster_donnelly_2008} and is known to be a critical parameter that separates regimes in which coherent vortex rings are ejected and a regime in which a ring with a trailing jet is instead produced. 
Additional parameters such as the opening angle of the orifice and a velocity program factor $P=\langle V_p^2 \rangle / \langle V_p \rangle^2$ that captures the temporal dependence of the fluid injection,  can also play a significant role. 

In our setup, in which we draw fluid at the top of the setup which in turn draws fluid through N (typically 8) orifices at the corners of our cubic flow chamber, the length of the cylindrical slugs and their velocity program factor must be derived from the diameter, stroke length and velocity program of the piston at the top of the chamber.

\subsubsection{Relationship between $L_*$ and piston control parameters}
If divided evenly between the slugs emerging from the N (typically 8) orifices, the volume of fluid drawn by the piston $\pi \left(\frac{D_{\rm piston}}{2}\right)^2L_{\rm piston}$ and its flux give:
\[
V_{\rm slug} = \frac{1}{N} \left( \frac{D_p}{D_o} \right) ^2 V_p \quad , \quad
L_* \equiv \frac{L}{D} = \frac{1}{N} \left( \frac{D_p}{D_o} \right)^2 \frac{L_p}{D_o} \label{eq:ld_definition}
\]
There is however an asymmetry between top and bottom orifices in our tank due to the fact that the piston draws in fluid at the top of the tank.
This results in $V_{\rm slug}^{\rm top} = V_{\rm slug} (1 - \delta)$ and $V_{\rm slug}^{\rm bottom} = V_{\rm slug} (1 + \delta)$, with $\delta $, which can be tuned by a combination of $L_*$ and $v$, in the range of $\sim$0-0.3 in our experiments.

\subsubsection{Impulse of a slug with a velocity program}
The injection of a slug of fluid will, in general,  occur at  velocity $V_{\rm slug}(t)$ that varies during the slug injection process. 
This can occur either because of experimental limitations on acceleration, or by design, in an attempt to optimally transfer impulse and energy from the slug to a vortex ring. For example, Glezer~\cite{glezer_formation_1988} showed that different temporal signals with the same $L_*=L/D$ generate vortex rings with different properties. 

\begin{figure}[htbp]
\centering
\includegraphics[width=0.8\textwidth]{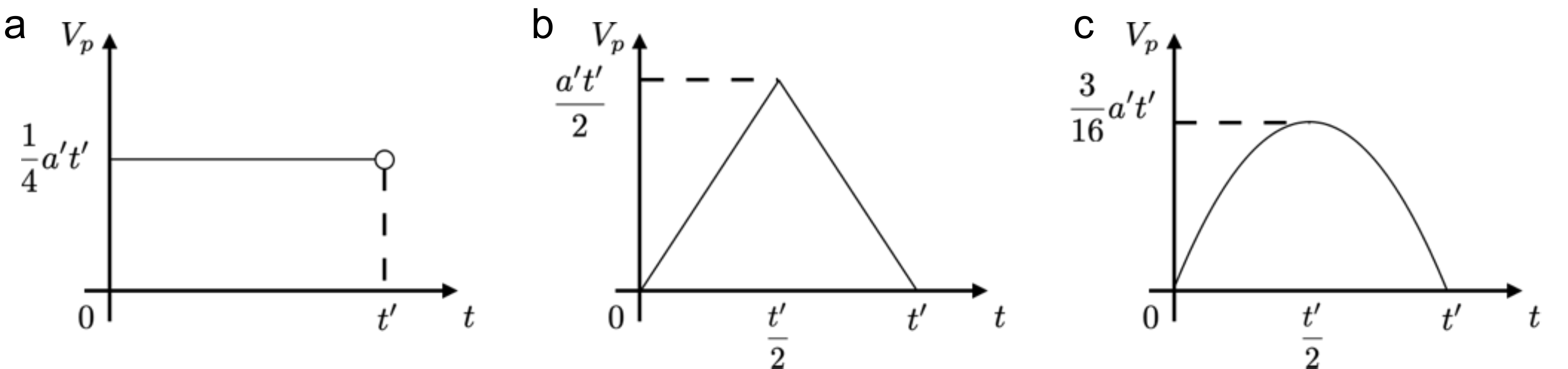}
\caption{\textbf{A velocity program factor $P = \langle V_p^2\rangle / \langle V_p\rangle^2$ characterizes a temporal profile of piston velocity $V_p$.} $a'$ is equal to $4\langle V_p \rangle/t'$. Here are the values of the velocity program factor of simple signals with the same stroke length. (a) A step function: $P=1$ (b) A triangular function: $P=4/3$ (c) A parabolic function: $P=6/5$.}
\label{fig:velocity_program_factor}
\end{figure}

To account for this time variation, it is useful to consider the impulse $I_{\rm slug}$ that represents the momentum or impulse of an isolated slug of fluid, decoupled from the surrounding fluid\footnote{In general the injection of a slug into a surrounding fluid will result in an impulse that includes contributions from the pressure at the walls exterior to the orifice as well as the added mass. Nonetheless $I_{\rm slug}$ represents a useful characterization of the injection process. }. 
The integrated momentum flux gives:
\begin{align}
    I_{\rm slug}  &=  \int_0^{T} V_{\rm slug} \left( \rho  \pi  \left(\frac{D_{\rm slug}}{2}\right)^2  V_{\rm slug}(t)\right) \  dt =
\rho  \pi  \left(\frac{D_{\rm slug}}{2}\right)^2  \langle V_{\rm slug}^2 \rangle_t T \\
 &=  \frac{\pi}{4} \rho  D_{\rm slug}^2 L_{\rm slug} P \langle V_{\rm slug} \rangle_t = \frac{\pi}{4} N L_* \cdot \rho D_{\rm slug}^3 v_{\rm eff}.
    \label{eq:impulse_sys}
\end{align}
where $P= \langle V_p^2 \rangle_t / \langle V_p \rangle_t^2 \geq 1$ and $v_{\rm eff}=P\langle V_p \rangle_t$.
Figure~\ref{fig:velocity_program_factor} shows some examples of velocity programs and their associated program factor $P$.

\subsubsection{Accessible slug parameter space} \label{sect:param_space_limit}
\begin{figure}[htbp]
\centering
\includegraphics[width=0.75\textwidth]{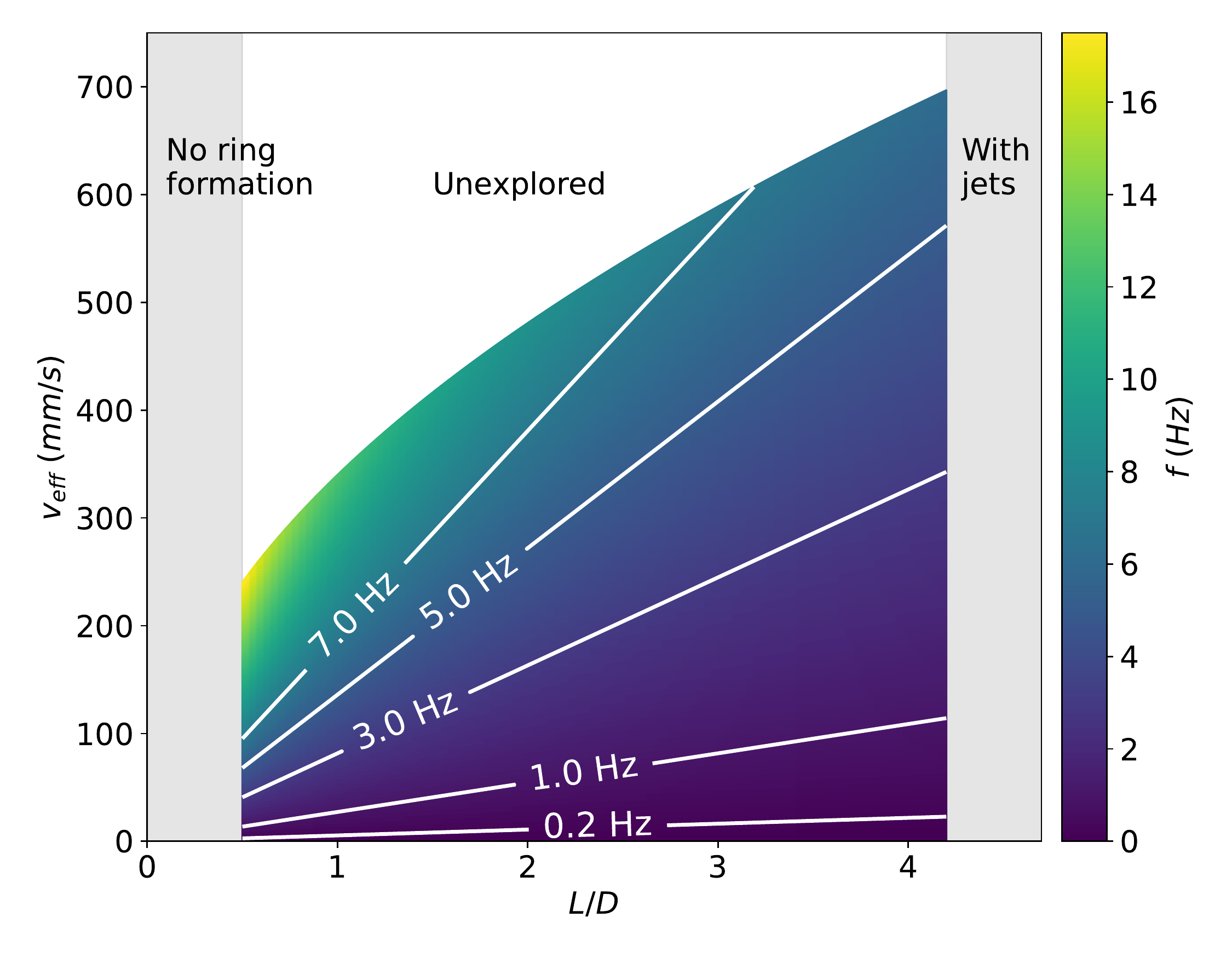}
\caption{\textbf{Accessible slug parameter space is spanned by the non-dimensional stroke length and the effective stroke velocity.} The upper bound of the accessible space is set by maximum acceleration of the piston, and the curve for $(a_{max}, D_o, P)=(1.6\times 10^4 mm/s^2, 25.6mm, 1.0)$ is shown.  The lower bound (white lines) depends on the frequency of the oscillatory motion. The heat map represents the maximum frequency that a vortex ring can be generated in our experiment. }
\label{fig:exp_phase_space}
\end{figure}

Figure~\ref{fig:exp_phase_space} illustrates the accessible/useful slug parameter space.  

An empirical bound on ring formation (gray shaded areas at low and high $\frac{L}{D}$) is given by $0.5\lesssim \frac{L}{D} \lesssim 4.0$. On the lower end, no ring formation is observed, on the upper end a jet trails the vortex ring. These empirical bounds are close to those found in ~\cite{sullivan_niemela_hershberger_bolster_donnelly_2008, gharib_rambod_shariff_1998}.

$v_{\rm eff}$ is limited, on the upper end by the maximum acceleration $a_{max}$ that the piston can attain (upper curved boundary). 
This follows $v_{eff} \leq \frac{P}{2} (a_{max}D)^{\frac{1}{2}} \left( \frac{L}{D} \right)^{\frac{1}{2}}$.

On the lower end, $v_{\rm eff}$ is limited by the requirement  that the stroke time must be less than a forcing period (while, frequency dependent, lines): $v_{eff} \geq (PfD)\frac{L}{D}$. 

\begin{figure}[htbp]
\centering
\includegraphics[width=\textwidth]{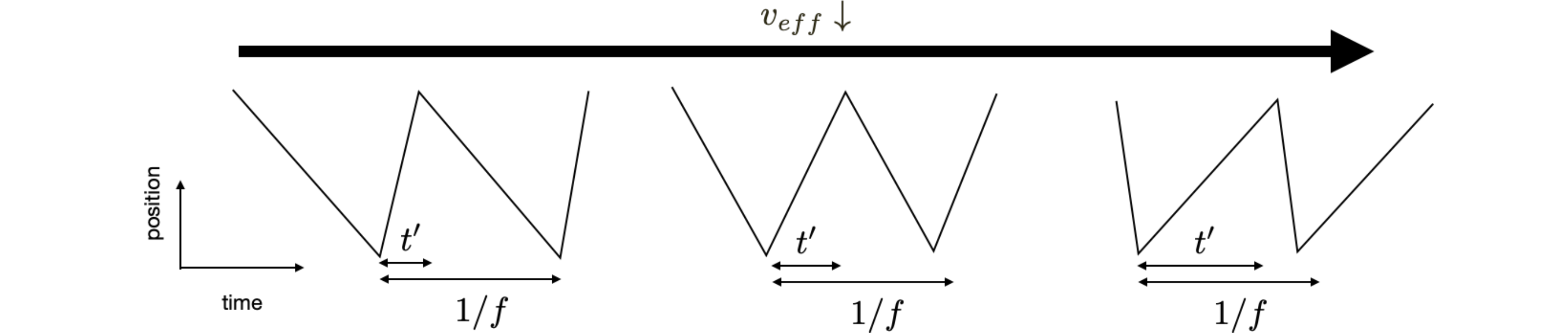}
\caption{\textbf{Experimental constraint on the phase space:} The piston must move by a stroke length $L$ faster than the forcing period $1/f$, limiting the accessible phase space: $v_{eff} \geq (PfD)\frac{L}{D}$. }
\label{fig:exp_const3}
\end{figure}

To characterize the generated vortex rings under different experimental conditions ($L/D, v_{eff}$), we use $f=0.2$Hz or 1Hz. We hereby assume that the same rings are generated at a higher frequency.

\subsection{Characterization of vortex rings}

The vortex rings we produce are the fundamental building block of the turbulence we create. 
We take a three-pronged approach to characterize their properties:
\begin{enumerate}
    \item We perform PIV measurements in a custom tank with side-windows (CW laser sheet, PIV Lab) to efficiently span the accessible phase space
    \item We performed PIV measurements of single vortices created in a custom tank with side-windows by leaving only one orifice open (CW laser sheet, LaVision - DaVis)  
    \item We performed 3D PTV measurements (Pulsed laser, STB) of the 3D dynamics and structure of individual vortex rings created in the 3D-printed, standard chamber leaving only one orifice open. 
\end{enumerate}

\subsubsection{Parameter space}

\begin{figure}[htbp]
\centering
\includegraphics[width=0.99\textwidth]{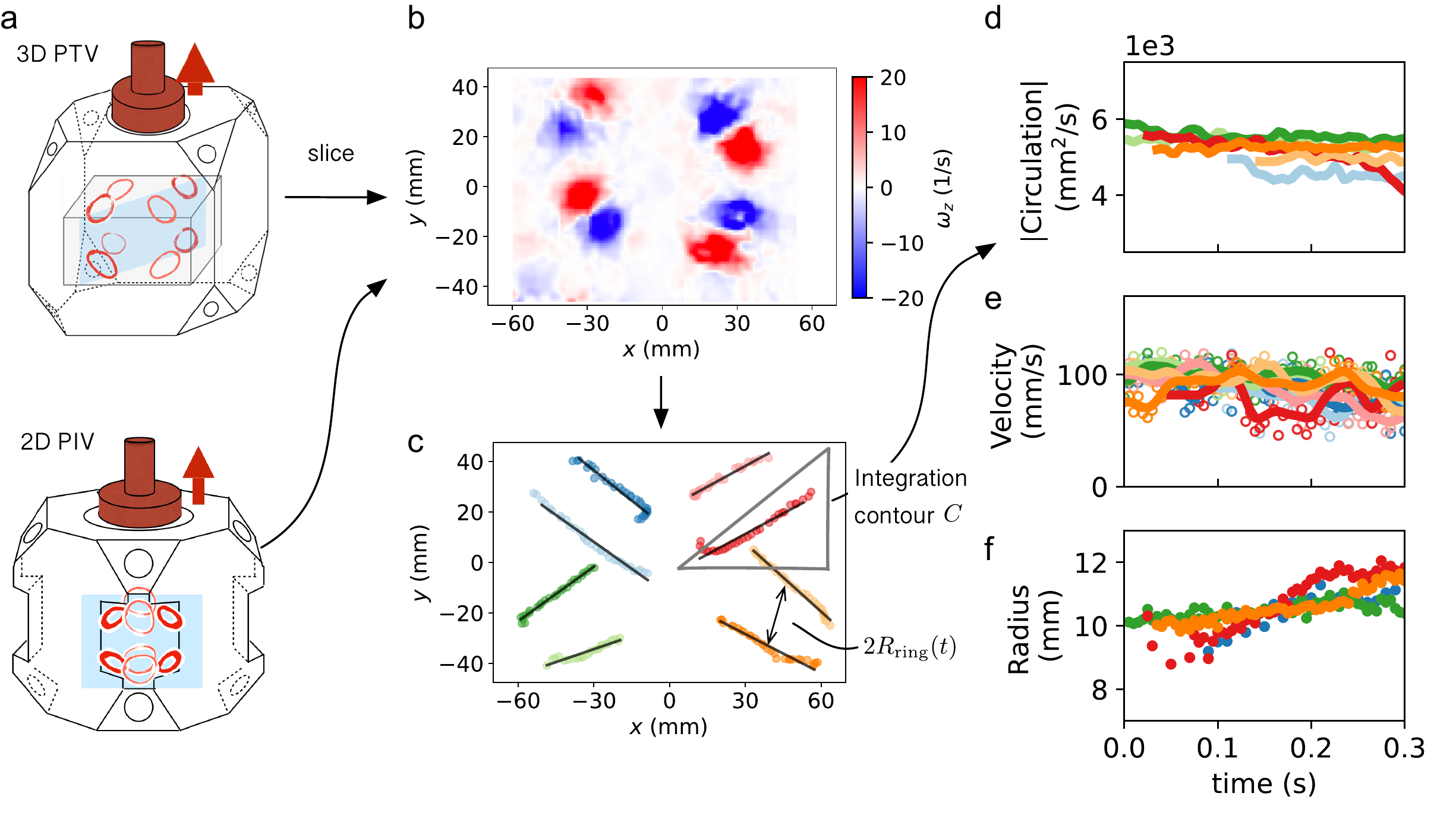}
\caption{\textbf{PIV/PTV measurements on vortex ring properties.} (a) We conducted 3D PTV with the standard chamber and 2D PIV with the windowed chamber. For the former, we obtained a flow field on the plane of interest by interpolation. (b) We then compute the vorticity field by $\Nabla \times~\vec{u}$. (c) Taking a center of voriticity identified each core position at instant of time. (d) By computing $\int_C \vec{u} \cdot d\ell$, we measure circulation. (e-f) We measure the instantaneous velocity and radius based on the extracted trajectories in (c).}
\label{fig:vring_measurements}
\end{figure}

To assemble a basic characterization of the sets of vortex rings produced when the we actuate the piston in our tank, we create sets of 8 vortex rings with varying $L_{\rm piston}/D_{\rm piston}$ in  the box with small windows illustrated in Figure\ref{fig:FlowChmaberGeometry}c-d. 
We illuminate a plane on which half of the vortex rings travel towards the center of the chamber, and track them until they collide.
We performed PIV measurements on the emerging vortices using a CW laser and processed the images with PIV Lab (Figure~\ref{fig:vring_measurements}a).
A typical snapshot is shown in Figure~\ref{fig:vring_measurements}b.
Four vortex dipoles can be clearly seen, corresponding to cross sections of four of the eight vortex rings. 

The core position at an instant of time is then determined by taking a center of the  vorticity field over a region that includes the core. Repeating this process for all the cores until the vortex rings collide, we extracted their trajectories (Figure~\ref{fig:vring_measurements}c).
The vortex ring radius and the translational speed is measured by tracking the core positions.  
The circulation of each ring is then computed from the extracted velocity field via $\Gamma = \int_C \vec{u}\cdot d\vec{l}$ (Figure~\ref{fig:vring_measurements}d).  A sample integration contour is used in Figure~\ref{fig:vring_measurements}c. By dividing the distance between the core positions at distant frames by the time it took, we obtain the instantaneous velocity of each core (Figure~\ref{fig:vring_measurements}e).  We extracted the radius by computing the distance between the instantaneous core locations (Figure~\ref{fig:vring_measurements}f).

\begin{figure}[htbp]
\includegraphics[width=\textwidth]{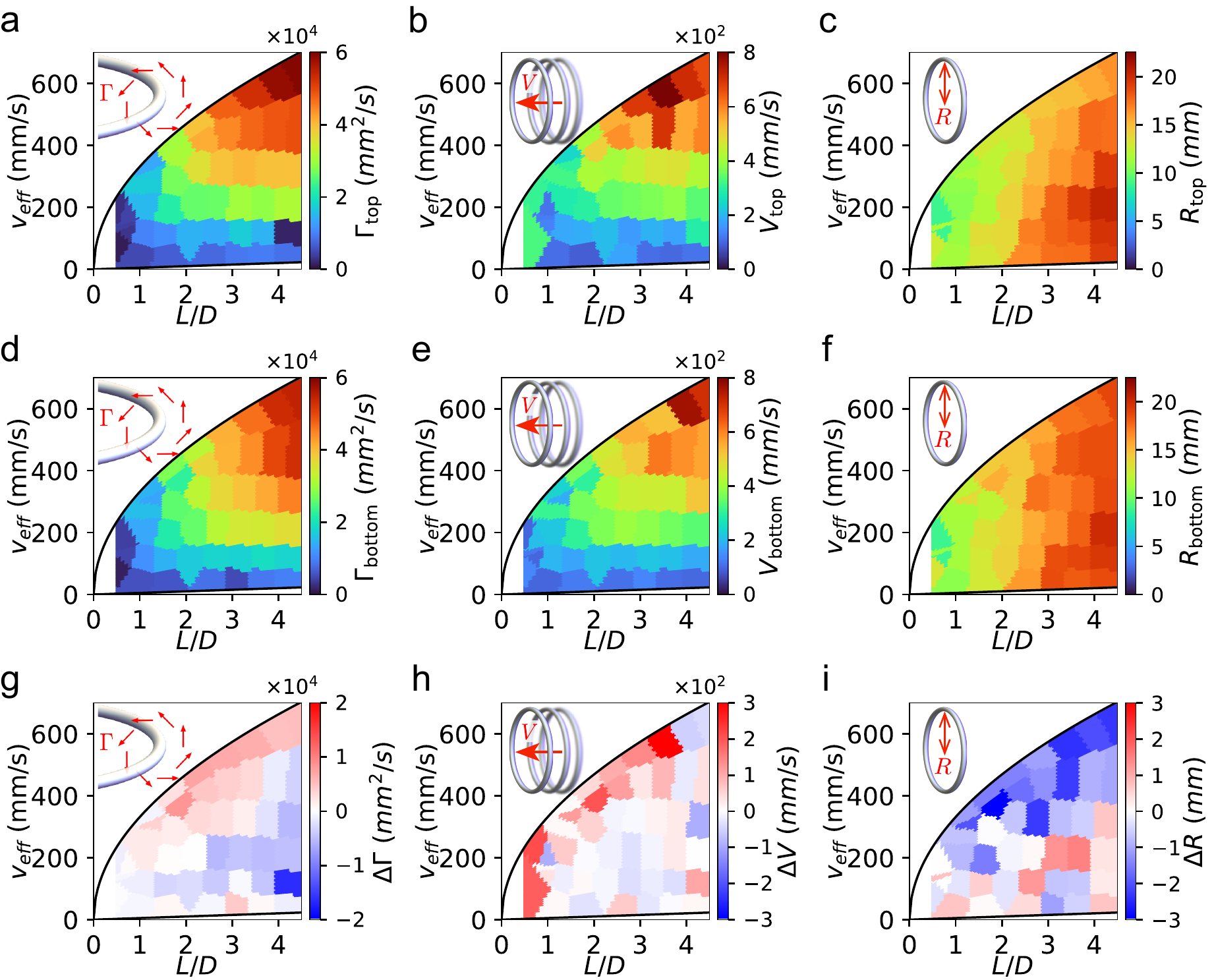}
\caption{\textbf{Measured circulation $\Gamma$, translational velocity $V$, and radius $R$ of the generated vortex rings are plotted in the phase space spanned by formation number $L/D$ and effective piston velocity $v_{eff}$.} (a-c) $\Gamma$, $V$, and $R$ of vortex rings generated at the top orifices are shown. (d-f) $\Gamma$, $V$, and $R$ of vortex rings generated at the bottom orifices are shown. (g-i) Difference of each quantity between the top and the bottom orifices are shown. The solid black lines define the experimentally accessible phase space ($f=0.2$Hz).}
\label{fig:vring_characterization}
\end{figure}

Figure~\ref{fig:vring_characterization} shows circulation, translational velocity, and diameter of vortices produced from the top (a-c) and bottom (d-f) orifices, and their differences (g-i). 
Vortex rings are successfully generated when $L/D > 0.5$, and remain  isolated structures for $L/D \lesssim 4$. 
We observe that rings are accompanied by a trailing jet when $L/D > 4$ as reported by several authors ~\cite{gharib_rambod_shariff_1998}. 

The asymmetry of forcing can cause the rings from the top orifices to have different properties then the ones emerging from the bottom orifices. 
We find  asymmetries of at most $\sim$17\% $\sim$30\%, and $\sim$10\%  in $\Gamma$, $V$, and $D$ respectively. 
The information in Figure~\ref{fig:vring_characterization} enabled choices of actuation parameters that kept asymmetries below $\sim 5\%$ for experimental runs presented in the main text.  
In particular cases, by further detailed characterization,  this can be reduced to $\sim 0\%$ as in the case of the data presented in Figure~2 of the main text ($(L/D, v_{eff}=(2.0, 200$mm/s))).

\subsubsection{Non-dimensionalization and relation to literature}

To compare our vortex production with the literature on vortex production by slug injection, we extracted non-dimensionalized relationships between $R_{\rm ring}$, $\Gamma_{\rm ring}$, $V_{\rm ring}$ and $L/D(=L_*)$ from ~\cite{sullivan_niemela_hershberger_bolster_donnelly_2008, gharib_rambod_shariff_1998, akhmetov_formation_2001} and compared them to our measured values. 

~\cite{akhmetov_formation_2001} adopts the following list of variables to characterize vortices produced by an orifice of a tube:
\begin{align}
    R_{\rm ring}, a, h, l, \Omega_{\rm atmosphere}, V_{\rm ring}, \Gamma_{\rm ring}, \Gamma_{c}, I_{\rm ring} 
    \label{eq: vr_params}
\end{align}
where $R_{\rm ring}$ is the radius of the vortex ring, 
$a$ the radius of the vortex core, $h, l$ the lengths of the semi-axes of the vortex atmosphere, $V_{\rm ring}$  the self-induced velocity of the vortex ring, $\Gamma_{\rm ring}$  the total circulation, $\Gamma_c$ the circulation around the vortex core, $\Omega_{\rm atmosphere}$ the volume of the vortex atmosphere  and $I_{\rm ring}$ the impulse of the vortex ring. 
They are non-dimensionalized as:
\begin{align}
    R_* = \frac{R_{\rm ring}}{D_o}, a_* = \frac{a}{D_o},  h_* = \frac{h}{D_o}, l_*=\frac{l}{D_o}, \Omega_*=\frac{\Omega_{\rm ring}}{D_o^3}, \label{eq: vr_params_nondim_i}\\
    \Gamma_* = \frac{\Gamma_{\rm ring}}{D_o V_{\rm slug}},  \Gamma_{c*} = \frac{\Gamma_c}{D_o V_{\rm slug}},  V_* = \frac{V_{\rm ring}}{V_{\rm slug}}, I_* = \frac{I_{\rm ring}}{\rho D_o^3 V_{\rm slug}}.   \label{eq: vr_params_nondim_ii}
\end{align}

\begin{figure}[htbp]
\centering
\includegraphics[width=0.99\textwidth]{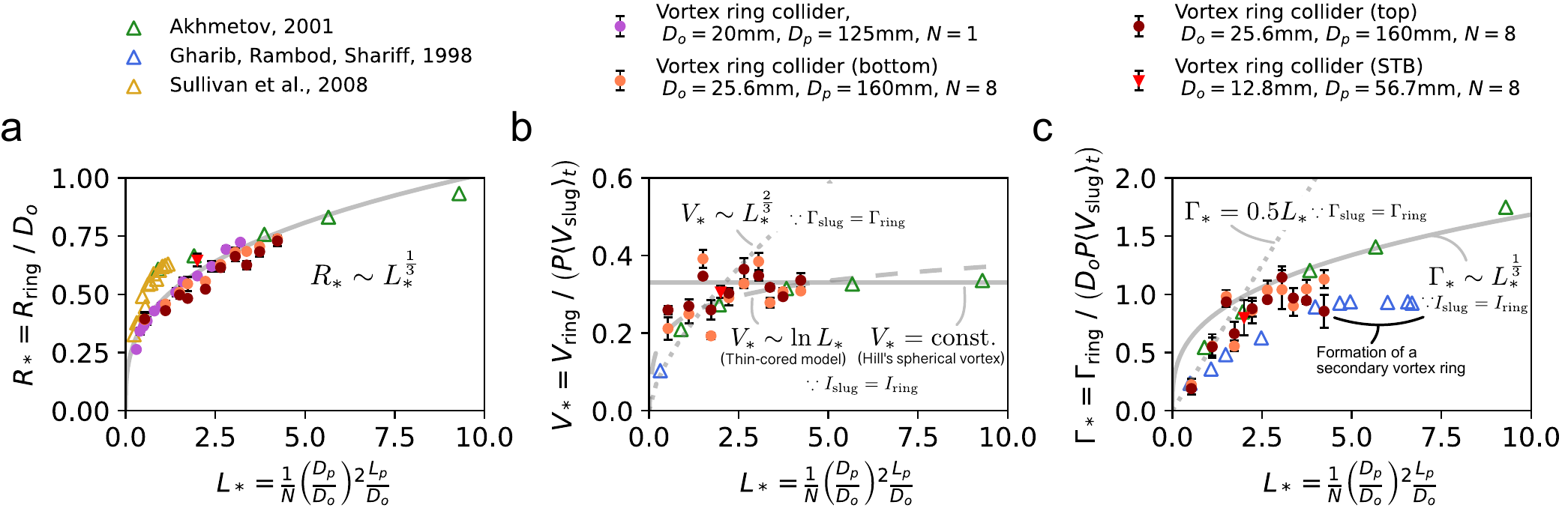}
\caption{\textbf{Scaling laws of vortex ring radius $R_{\rm ring}$, velocity $V_{\rm ring}$, and circulation $\Gamma_{\rm ring}$ on the formation number $L_*$} The values of the vortex rings generated by submerging jets through a nozzle are shown as a reference, and are consistent with the measurements of the vortex ring collider. (a) Non-dimensionalized radius shows $R_* \sim L_*^{1/3}$, reflecting the conservation of mass. (b) Non-dimensionalized velocity of a vortex ring monotoniously increases with $L_*$, and plateaus out above $L_*=3$. (c)Non-dimensionalized circulation agrees with the slug model (circulation matching) $\Gamma = 0.5L_*$ below $L_*=2$ but agrees better with $\Gamma_*\propto L_*^{1/3}$ at high $L_*$. }
\label{fig:vring_master}
\end{figure}

Figure~\ref{fig:vring_master}a-c displays the non-dimensionalized radius $R_*$, velocity $V_*$, and circulation $\Gamma_*$ against the formation number $L_*$ along with  literature values from ~\cite{gharib_rambod_shariff_1998, sullivan_niemela_hershberger_bolster_donnelly_2008, akhmetov_formation_2001}.
The non-dimensionalization of our  data-points assumes zero top-bottom asymmetry and we included our program factor for the slug velocity. 
Our re-scaled radii, velocities and circulations for both top and bottom sets of vortex rings, are  consistent with literature values obtained ejecting individual slugs from tubular vortex ring generators.  
This supports our experimental strategy of using a single piston to actuate slugs through $N$ orifices. 

We note that our agreement with literature values is within the range of agreement between different published experiments. 
These relatively small variations emerge naturally from differences in the location of the measurement (distance from the orifice), the orifice geometry and actuation program which result in vortices with slightly different atmosphere shape as well as varying levels of entrainment. For example,  the vortices studied in  Sullivan et al.~\cite{sullivan_niemela_hershberger_bolster_donnelly_2008} are more ellipsoidal than  ~\cite{akhmetov_formation_2001} and our measurements (an eccentricity of 0.60 compared to our 0.15-0.40) and travel somewhat faster.

In addition to the comparison with literature values, also shown in  Figure~\ref{fig:vring_master}a-c are gray lines that represent the scalings predicted by different matching criteria between slug and ring properties.
In the case of the vortex radius (Figure~\ref{fig:vring_master}a), matching the mass of the slug and the mass of the vortex atmosphere predicts $R_* \propto L^{1/3}$
In the case of $V_*(L_*)$ and $\Gamma_*(L_*)$ we show three different possible scalings, each of which may be dominant in different ranges of $L_*$: (i) matching the circulation of a vortex with the circulation of an isolated slug  gives ($\Gamma_* \propto L_*$, $V_*\propto L_*^{2/3}$). (ii) matching the impulse of the slug with the impulse of a thin-cored vortex ring gives ($\Gamma_* \propto L_*^{1/3} V_* \propto \ln{L_*})$), $V_*\propto L_*^{2/3}$) (iii) matching the impulse of the slug with the impulse of Hill's spherical vortex model gives ($\Gamma_* \propto L_*^{1/3}, V_* \propto \text{const.}$)

\subsubsection{Individual vortex ring structure and properties - 2D}
\label{sect:vortex_ring_characterization}
As discussed in Box 3 above, an accurate measurement of the vortex energy requires precise knowledge of the vorticity distribution within the vortex atmosphere.
To characterize the detailed structure of our vortices, we covered seven out of the eight orifices and performed PIV measurements on single vortex rings as they traversed the experimental chamber.

An instantaneous vorticity field can be obtained by the direct differentiation of the velocity field. Noise in a velocity field becomes amplified by differentiation. 
Thus we averaged the vorticity field over neighboring frames assuming that the field was steady during the recording. 
For incompressible, axisymemtric flows, the streamline function can be computed by $\psi(\sigma, z) = \int_A^B \sigma (u_z d\sigma - u_\sigma dz)$ in cylindrical coordinates. Choosing the reference point $A$ at a different location only shifts the stream function by some constant.

\begin{figure}[htbp]
\centering
\includegraphics[width=0.99\textwidth]{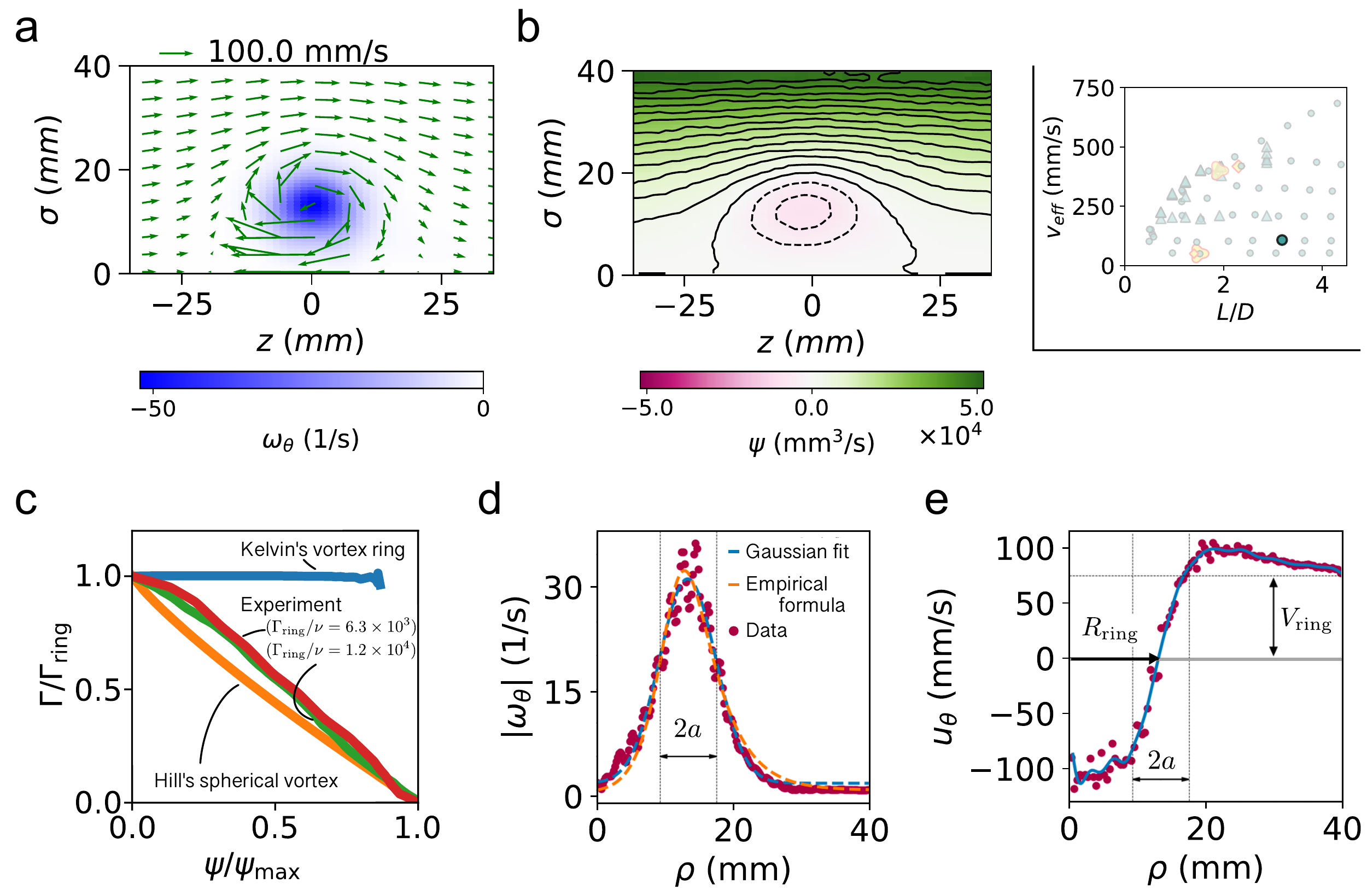}
\caption{\textbf{Characterization of a vortex ring in the experiments reveals its similarity to Hill's spherical vortex.}(a) A velocity and vorticity field of a typical vortex ring in the comoving frame, obtained by PIV (b) Streamfunction and isocontours ($\Delta \psi = 5 \times 10^3$mm$^3$/s) (c) Normalized circulation along the closed streamlines in the vortex atmosphere is plotted against the normalized streamfunction. (d) Voticity distribution at z=0 (e) Axial velocity profile in the plane z=0. The top right plot indicates the location of the measurement in the phase space.}
\label{fig:vring_std_characteristics}
\end{figure}

The vorticity distribution of a typical vortex ring generated in the windowed chamber is shown in Figure\ref{fig:vring_std_characteristics} (d). 
It is in good agreement with a  a Gaussian distrubution, as well as with the empirical formula found by Akhmetov~\cite{akhmetov_vortex_2009} for the traditional vortex ring generator (vortex ring gun):
\begin{align}
    \omega(z_*, \sigma_*) = \frac{ \omega_{\rm max} \sigma_* } { \cosh{ \left[ 
    c_1  \left( z_*^2 + (\sigma_* -1)^2  \right) ^{\frac{1}{2}} 
    \right]} } \text{; $z_* = \frac{z}{R_{\rm ring}}$, $\sigma_* = \frac{\sigma}{R_{\rm ring}}$}
    \label{eq: vort_dist}
\end{align}
where $c_1=11.7$ for a vortex ring gun~\cite{akhmetov_vortex_2009}. 

When compared to the canonical Lamb-Oseen and Hill vortices, the distribution of vorticity can be clearly be seen to be much closer to a diffuse spherical vortex than a concentrated Lamb-Oseen vortex. 

\begin{figure}
\centering
\includegraphics{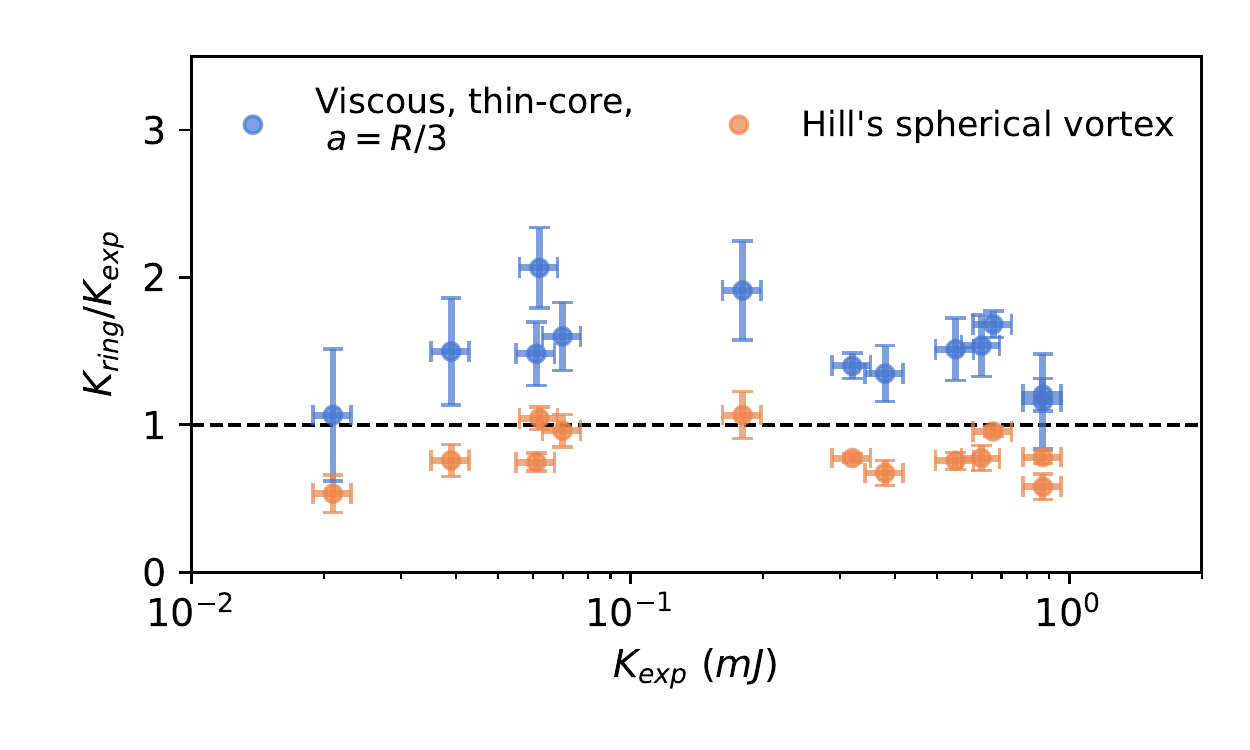}
\caption{\textbf{Kinetic energy enclosed in the vortex atmosphere is consistent with the formula of Hill's spherical vortex.} The kinetic energies inside vortex atmosphere, predicted by the thin-cored and Hill's spherical vortex models(Box \ref{box:vring_models}) with measured radius and circulation, quantitatively agree with the measurements by 2D PIV. We compute $K_{\rm ring}=cK(\Gamma_{\rm ring}, R_{\rm ring})$ with $R_{\rm ring}/a=3, \alpha=2.04$ for the thin-cored model and $A_{\rm sph} = 4/3R_{\rm ring}$ for Hill's spherical vortex model (see Box 2). For both models, $c = K_{\rm ring}/K=0.77$ is used.}
\label{fig:vring_energy_comparison}
\end{figure}

For a planar vortex ring, such 2D PIV measurements are sufficient to further compute its inviscid invariants listed in Eq. \ref{eq: vr_params}. The important quantities are  radius $R_{\rm ring}$, velocity $V_{\rm ring}$, circulation $\Gamma_{\rm ring}$, vorticity distribution $\omega(z, \sigma)$, impulse $\vec{I}_{\rm ring}$, and energy inside the vortex atmosphere $K_{\rm ring}$. 
For example, in the case of a vortex ring in Figure \ref{fig:vring_std_characteristics}, we find $|I_{\rm ring}| = 3.7\times 10^{-3}$kg$\cdot$m/s compared to $|I_{\rm Kelvin}| = 5.0\times 10^{-3}$kg$\cdot$m/s=1.3$|I|$, and  $|I_{\rm sph}| = 3.5\times 10^{-3}$kg$\cdot$m/s=0.95$|I|$. For energy, we find $K_{\rm ring} = 0.42$mJ compared to $K_{\rm ring, Kelvin} = 1.0$mJ=2.4$K_{\rm ring}$ and $K_{\rm ring, sph} = 0.34$mJ=$0.83K_{\rm ring}$. Figure \ref{fig:vring_energy_comparison} compares $K_{\rm ring}$ between our measurements and the predictions ($K_{\rm ring, Kelvin}$ and $K_{\rm ring, sph}$) for various vortex rings generated by different experimental protocols. For the predicted values, we used the measured radius $R_{\rm ring}$, circulation $\Gamma_{\rm ring}$, and the ratio $K_{\rm ring}/K=0.77$ (see Box 3) for the expressions in Box 2. The result shows that Hill's spherical vortex model underestimates the energy by $<30\%$ and the thin-cored model frequently overestimates the value over $50\%$ in the range of vortex rings we used to create a turbulent state. This intermediate behavior is consistent with our measurement on the vorticity distribution in Figure \ref{fig:vring_std_characteristics}c.

\subsubsection{Individual vortex ring structure and properties - 3D}
\label{sect:vring_3d_measurements}

While axisymmetric vortex rings carry energy, circulation, and impulse, for a vortex loop to carry angular impulse and helicity, a departure from axisymmetry is required. Helical vortex loops, in particular, carry all of the inviscid invariants. 

\begin{figure}[htbp]
\centering
\includegraphics[width=0.9\textwidth]{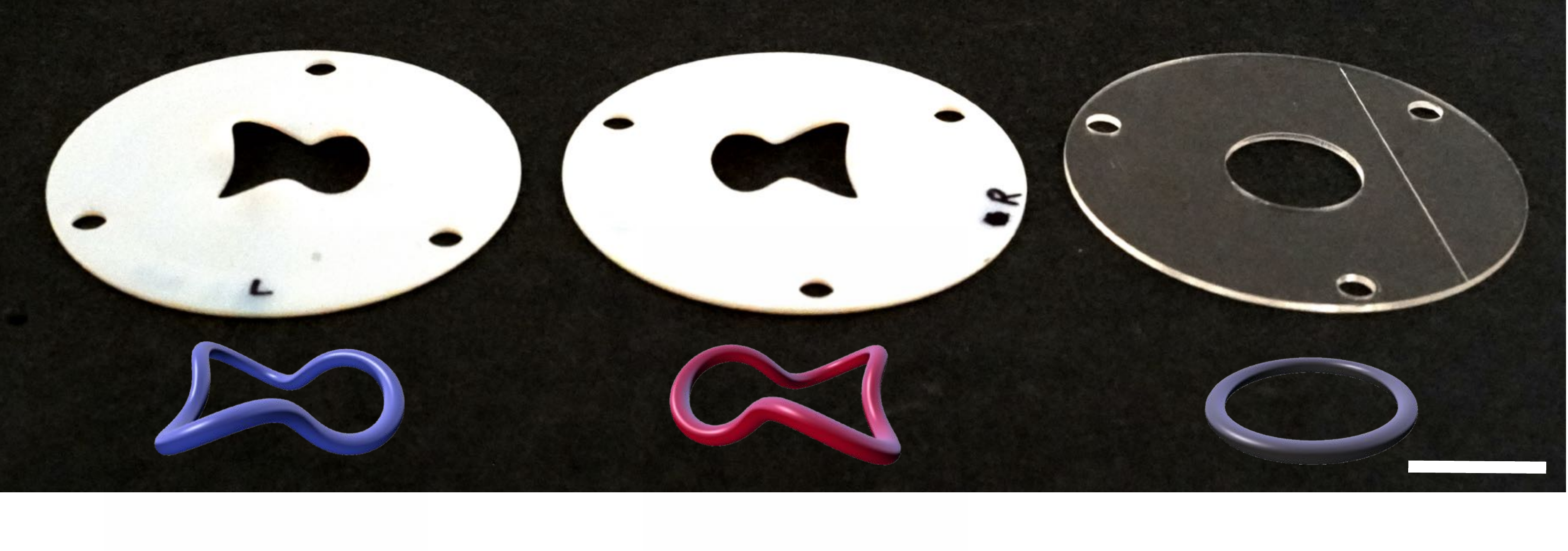}
\caption{\textbf{Helical masks generate vortex rings with nonzero helicity.}(Left) A left-handed mask (Middle) A right-handed mask (Right) A planar mask as a reference. The cartoons illustrate the height profiles of the orifices.  Scale bar: 25.6mm.}
\label{fig:helical_mask}
\end{figure}

To generalize the set of inviscid invariants that we can inject into our turbulent blob, we created nozzles whose boundary is helical, as shown in Figure \ref{fig:helical_mask}.
To study the properties of vortex rings produced through these nozzles, we generated individual helical vortices in our chamber and  then measured the resulting flow using 3D PTV (LaVision, Shake the box).

\begin{figure}[htbp]
\centering
\includegraphics[width=0.90\textwidth]{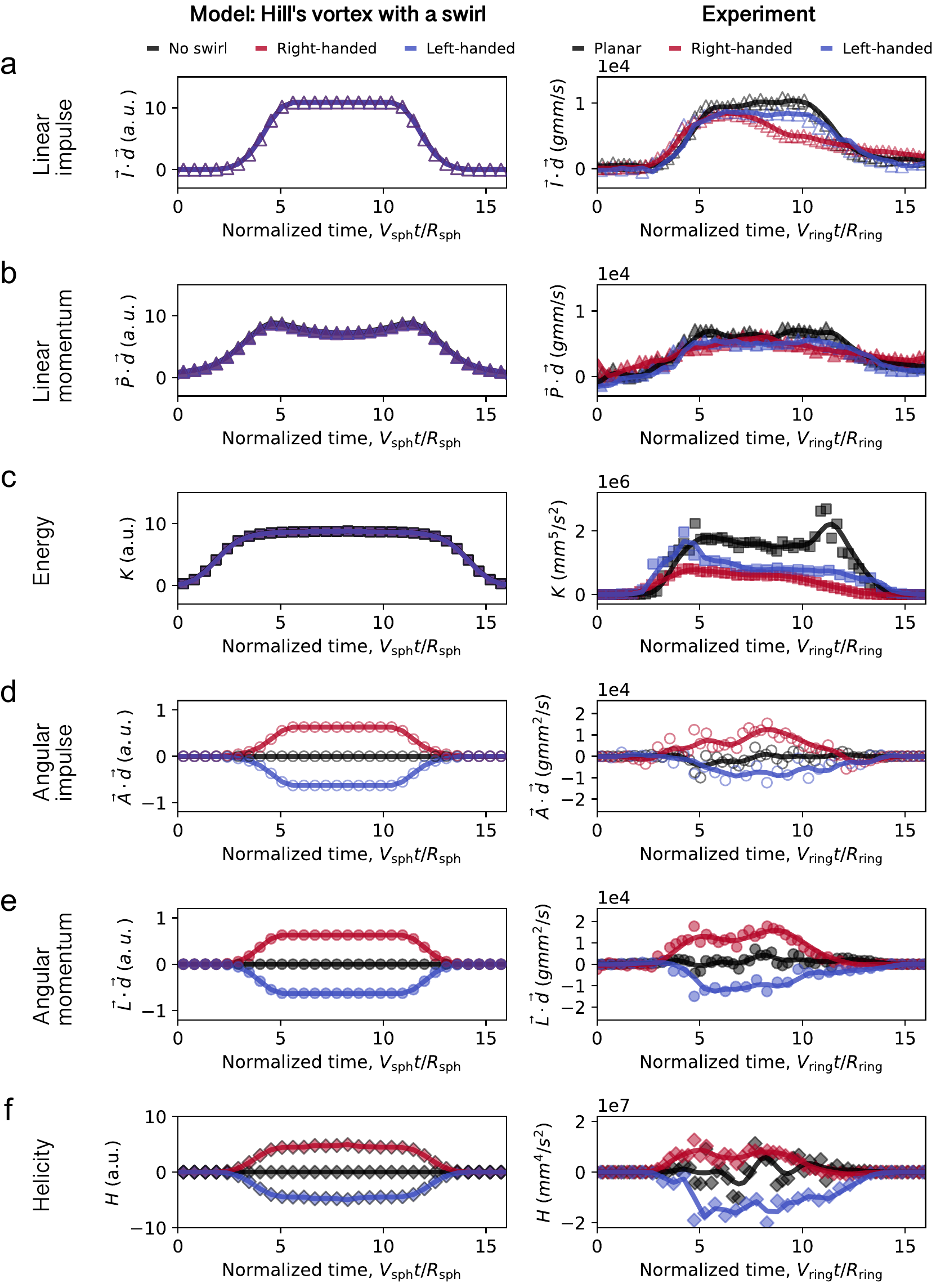}
\caption{\textbf{Measured inviscid invariants of a planar, and helical vortex rings qualitatively agree with simulations. (Left column: Hill's spherical vortex ring with a swirl. Right column: Experiment).} $\vec{d}$ is a unit vector that denotes the traveling direction of a vortex. The domain of the integration is a moving sphere with radius $2R_{\rm ring/sph}$ that encloses vortex atmosphere throughout the computation. (a) Linear impulse (b) Linear momentum (c) Energy (d) Angular impulse (e) Angular momentum (f) Helicity.}
\label{fig:inviscid_inv_measurement}
\end{figure}

For comparison we measured the 3D flow field of individual rings produced through our conventional circular orifices.  The helical apertures are designed to have the same mean radius as in the planar case, and the same stroke length and velocity profile ($L/D=1.5, v_{eff}=50$mm/s) was used to generate a vortex ring. Time-resolved, 3D velocity fields over the illuminated volume were obtained for a planar, a right-handed, and a left-handed vortex ring. The handedness is determined by the respective orientation of the angular momentum to the linear momentum: it is right-handed if they point to the same direction, and is left-handed if they point to the opposite directions. 
Angular impulse $\vec{A} = \int_{\mathcal{V}} \frac{1}{3} \rho \vec{x} \times ( \vec{x} \times  \vec{\omega})  dV$ (Figure~\ref{fig:inviscid_inv_measurement}d) and angular momentum $\vec{L} = \rho  \int_{\mathcal{V}} \vec{x} \times \vec{u}  dV$ (Figure\ref{fig:inviscid_inv_measurement}e) are then computed. The difference between these quantities are due to the contribution of the surface integral in $\vec{L}$. When all vorticity is included inside the domain of integration, and $\vec{\omega} \cdot \hat{n}=0$ is satisfied, these quantities agree to each other. 

To compare with a theory, we numerically compute the same quantities for a steady spherical vortex with a swirl~\cite{moffatt1969degree}. This flow is a solution of the incompressible Navier-Stokes equation with nonzero angular momentum and helicity due to the swirling motion inside the vortex atmosphere. The velocity field is given by
\[ 
\vec{u} = (u_r, u_\theta, u_\phi) = \left( \frac{1}{r^2\sin{\theta}}  \partial_\theta \psi, -\frac{1}{r\sin{\theta}}  \partial_r \psi, \frac{k \Xi}{r \sin{\theta}}  \right)\\
\]
where the (Stokes) stream function $\psi$ is identical to Hill's spherical vortex
\[
\psi = 
\begin{cases}
    -\frac{3}{4} V_{\rm ring} r^2 \left(1-\frac{r^2}{A_{\rm sph}^2} \right) \sin^2{\theta},& \text{ } r\leq A_{\rm sph}\\
    \frac{1}{2} V_{\rm ring} r^2 \left(1-\frac{A_{\rm sph}^3}{r^3} \right) \sin^2{\theta},& \text{ } r> A_{\rm sph}
\end{cases}
\]
and
\[
\Xi = 
\begin{cases}
    -\frac{3}{2} V_{\rm ring} \frac{J_{3/2}(kA_{\rm sph})}{J_{5/2}(kA_{\rm sph})} \left[ \left( \frac{A_{\rm sph}}{r}\right)^{\frac{3}{2}} \frac{J_{5/2}(kA_{\rm sph})}{J_{3/2}(kA_{\rm sph})}  -1 \right] r^2\sin^2\theta
    ,& \text{ } r\leq A_{\rm sph}\\
    \frac{V_{\rm ring}}{2} \left( 1 - \frac{A_{\rm sph}^3}{r^3}\right) r^2 \sin^2 \theta,& \text{ } r> A_{\rm sph}
\end{cases}
\]
is responsible for the azimuthal motion with a real parameter $k$. Here, $J_{3/2}$ and $J_{5/2}$ are the Bessel function of the first kind.

We perform the simulation as following.
\begin{enumerate}
    \item Define a simulation box with the same aspect ratio of the measured domain in the experiments.
    \item Place Hill's vortex with( or without) a swirl ($k=\pm 1$, or 0).
    \item Let the vortex propagate at $V_{\rm sph}$. 
\end{enumerate}
We compute the inviscid invariants that are listed in Box 2, linear momentum $\vec{P}=\int_\mathcal{V} \vec{u} dV$ and angular momentum $\vec{L}=\int_\mathcal{V} \vec{r} \times \vec{u} dV$. We integrated over a moving control volume (a sphere of radius $2R_{\rm ring}$) that encloses nearly all vorticity in the flow. Any contribution from outside the simulation box or the measured domain to the integration is considered zero. 
As summarized in Figure \ref{fig:inviscid_inv_measurement}, the experimental results show the qualitatively same behavior as the simulation with the swirly spherical vortex for all inviscid invariants.
While we are most likely not resolving the vorticity field fully, our measurements indicate the genration of helical vortex ring using a helical mask.

\clearpage

\subsubsection{Effect of coarse-graining on the measurement of helicity}\label{sect:coarse_graining_helicity}

In main Figure 5 and \ref{fig:inviscid_inv_measurement}, we presented the direct measurement of helicity $\mathcal{H} = \int_\mathcal{V} \vec{u} \cdot \vec{\omega} dV$ over a domain that encloses a turbulent blob and a helical vortex ring. In either case, a velocity and a vorticity field were not completely resolved, underestimating  the helicity. In this section, we study how coarse-graining a velocity field affects the resulting helicity. To do so, we generate analytical expressions for flow fields of a helical vortex tube whose vorticity field lines are organized around its center line. (See ~\cite{scheeler2017complete} SI Eq.43-44 for details). This construction enables us to tune the writhe of the center-line, the twisting of the field-lines, and the core size independently. The vorticity distribution inside the core is approximately Gaussian. Their construction is based on a complex scalar potential $\psi=z_1^m/ Q(z_1, z_1^*, z_2, z_2^*)$ where $(z_1, z_2)$ are coordinates on $S^3$. $m$ controls the number of windings of the field-lines, and  $Q(z_1, z_1^*, z_2, z_2^*)$ encodes the center-line. The velocity field is expressed as 
\begin{align}
    \vec{u} = \frac{a'^2}{4\pi} \left( \exp{ \left( - \frac{1-\chi'^2}{a'^2 \chi'} \right)} \Nabla \eta' + \frac{1}{2 i} \Nabla \log{(\frac{Q(z_1, z_1^*, z_2, z_2^*)}{Q^*(z_1, z_1^*, z_2, z_2^*)})} \right)
\end{align}
where $\eta'=\log{(\psi/\psi^*)}/(2\pi) \in [0, 2\pi)$, and $a'$ is a clipping length scale. To mimic the three-fold helical vortex generated in the experiment, we used
\begin{align*}
Q(z_1, z_1^*, z_2, z_2^*)&=  \frac{R_{\rm minor}}{R_{\rm major}}z_1^3 + z_2;\\
z_1 &= \frac{2R_{\rm major}(x+iy)}{(R_{\rm major}^2 + x^2 + y^2 + z^2)}\\
z_2 &= \frac{2R_{\rm major} z+ i(x^2 + y^2 + z^2  - R_{\rm major}^2 )} { (R_{\rm major}^2 + x^2 + y^2 + z^2)}
\end{align*}
where $R_{\rm major}$, and $R_{\rm minor}$ represent the major and minor radius of the helical ring respectively. $R_{\rm major}/R_{\rm minor}=4$ was chosen to be consistent with the center line of the experimental helical ring. 

 $\chi' \in [0.9, 0.99], m \in [1, 5]$

\begin{figure}[htbp]
\centering
\includegraphics[width=\textwidth]{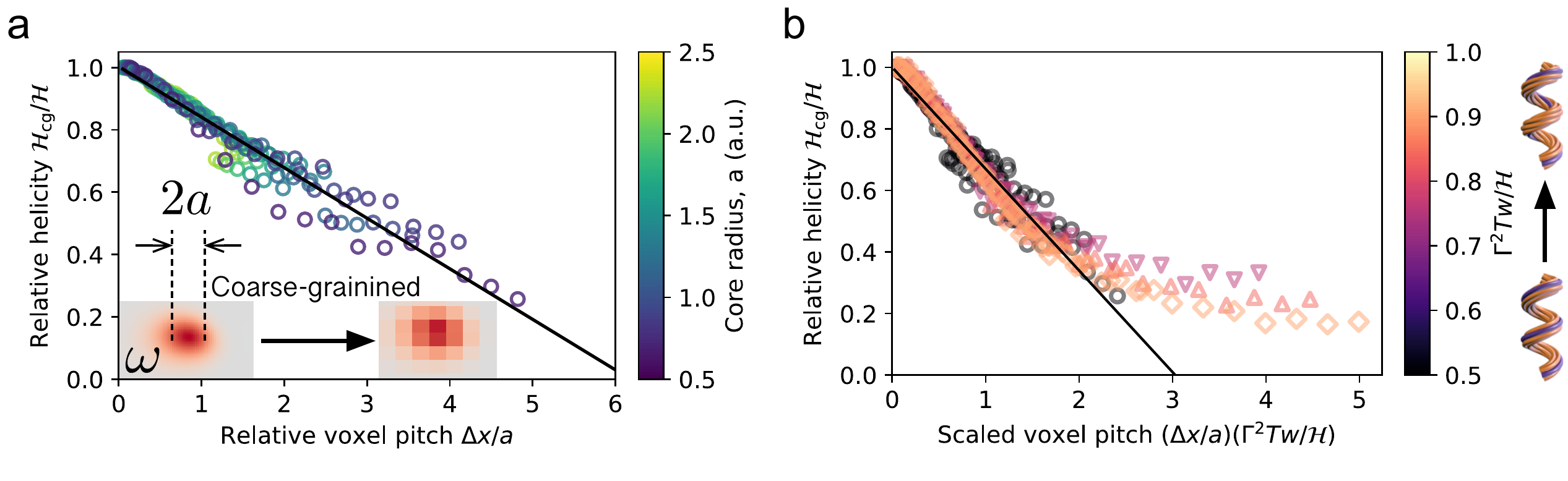}
\caption{\textbf{Coarse-graining a velocity field attenuates helicity.} (a) Scaled helicity with a coarse-grained field collapses to a line. (d) Data collapse of the relative helicity when plotted against the voxel pitch scaled with a twist-to-helicity ratio and a core radius.}
\label{fig:helicity_coarse_graining}
\end{figure}

To mimic a flow field that is obtained by 3D PTV, we coarse-grained a spatially resolved flow field, then computed the helicity. A coarse-graining procedure is a combination of convolution with a box kernel, followed by undersampling (See insets in Figure \ref{fig:helicity_coarse_graining}a\footnote{We coarse-grained a \textit{velocity} field even though the inset figures show an example using a vorticity field. To compute the helicity in Figure \ref{fig:helicity_coarse_graining}, a vorticity was derived from the coarse-grained velocity field.}). Coarse-graining a velocity field indeed attenuates the measured helicity as shown in Figure\ref{fig:helicity_coarse_graining}a for helical rings with various core size set by  $\chi' \in [0.9, 0.99]$. These curves collapse onto a line when plotted against the voxel size scaled by each core size (Figure\ref{fig:helicity_coarse_graining}a). 

Coarse-graining attenuates more helicity for a helical vortex ring with a higher twist\footnote{A more twisted helical vortex ring was generated by increasing $m \in [1, 5]$.} $Tw = (\mathcal{H}/\Gamma^2 - Wr)$ where $Wr$ is the writhe of the center-line.  Unlike the writhe, twisting is local winding of the vorticity field-lines; hence, it is more susceptible to the coarse-graining. We find a data collapse as we scaled the voxel pitch 
$\Delta x$ to  $(\Delta x / a)(\Gamma^2 Tw / \mathcal{H})$ (Figure \ref{fig:helicity_coarse_graining}b). In reality, helicity tends to writhe ~\cite{scheeler2017complete}, and a case with a low twist-to-helicity ratio is more common in nature. Under moderate coarse-graining when a segment of a vortex core is represented by more than 2$^3$=8 voxels, the attenuation of helicity is linear to $\Delta x / a$. This means for both main Figure 5 and Figure~\ref{fig:inviscid_inv_measurement}, the expected attenuation is approximately 10\% as $(\Delta x / a)(\Gamma^2 Tw / \mathcal{H})$ is estimated to be less than $(\Delta x / a) \approx 0.3$.

\newpage

\subsubsection{Summary of experimental runs}
\begin{figure}[htbp]
\centering
\includegraphics[width=0.5\textwidth]{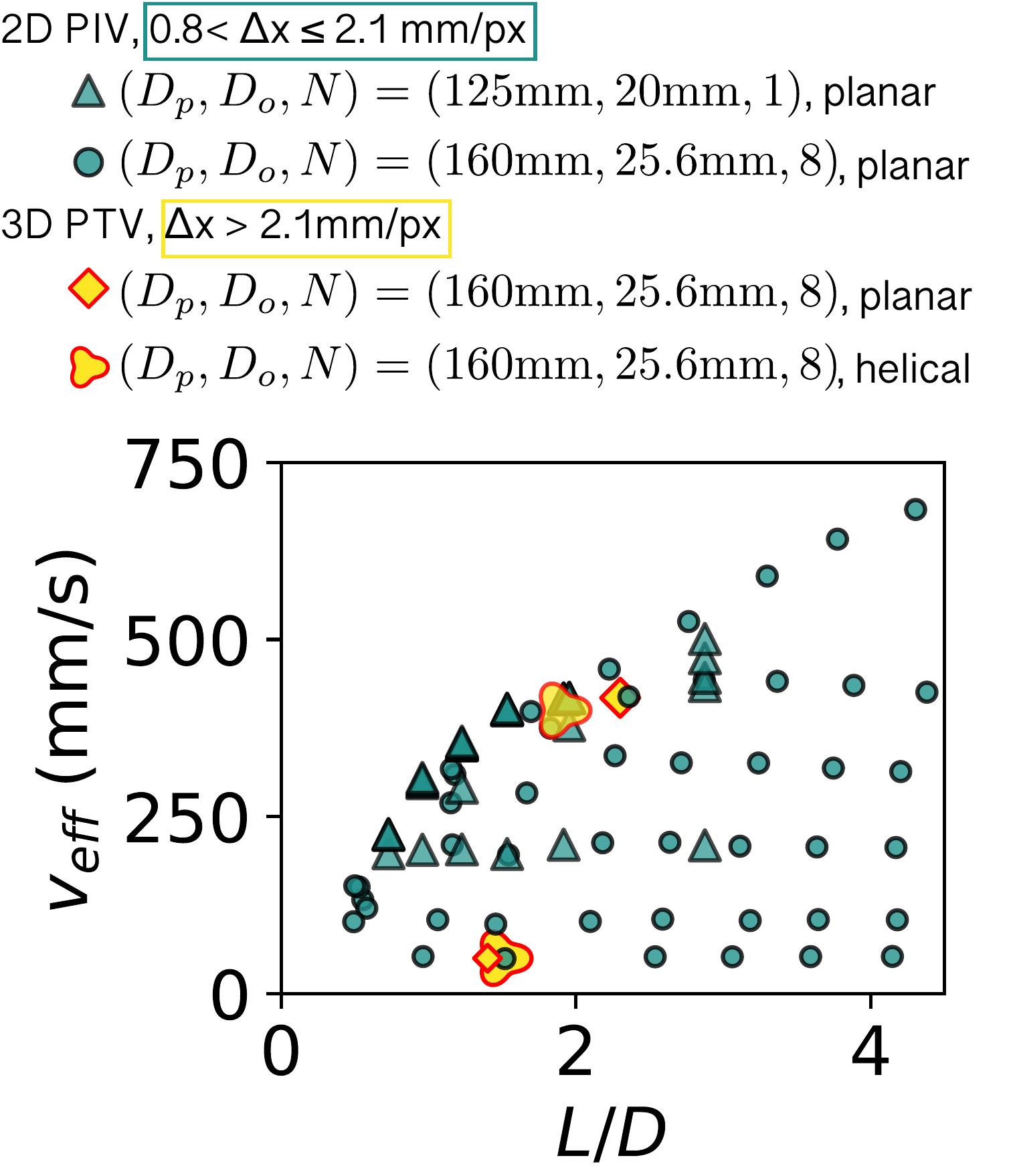}
\caption{\textbf{Summary of measurements performed to characterized vortex rings in the parameter space spanned by  effective piston velocity $v_{eff}$ and formation number $L/D$.} }
\label{fig:data_points_vring}
\end{figure}

Figure \ref{fig:data_points_vring} shows the measurements performed to characterize a single vortex ring, and a helical ring.

\clearpage
\newpage

\section{Vortex ring collisions - Summary of experimental runs}
\label{sect:Vortex ring collisions}
As described in the main text, firing together sets of eight vortices in our setup at repeated intervals, results, at sufficiently high frequencies, in the formation of an isolated blob of turbulence, surrounded by quiescent fluid.

\begin{figure}[htbp]
\centering
\includegraphics[width=0.95\textwidth]{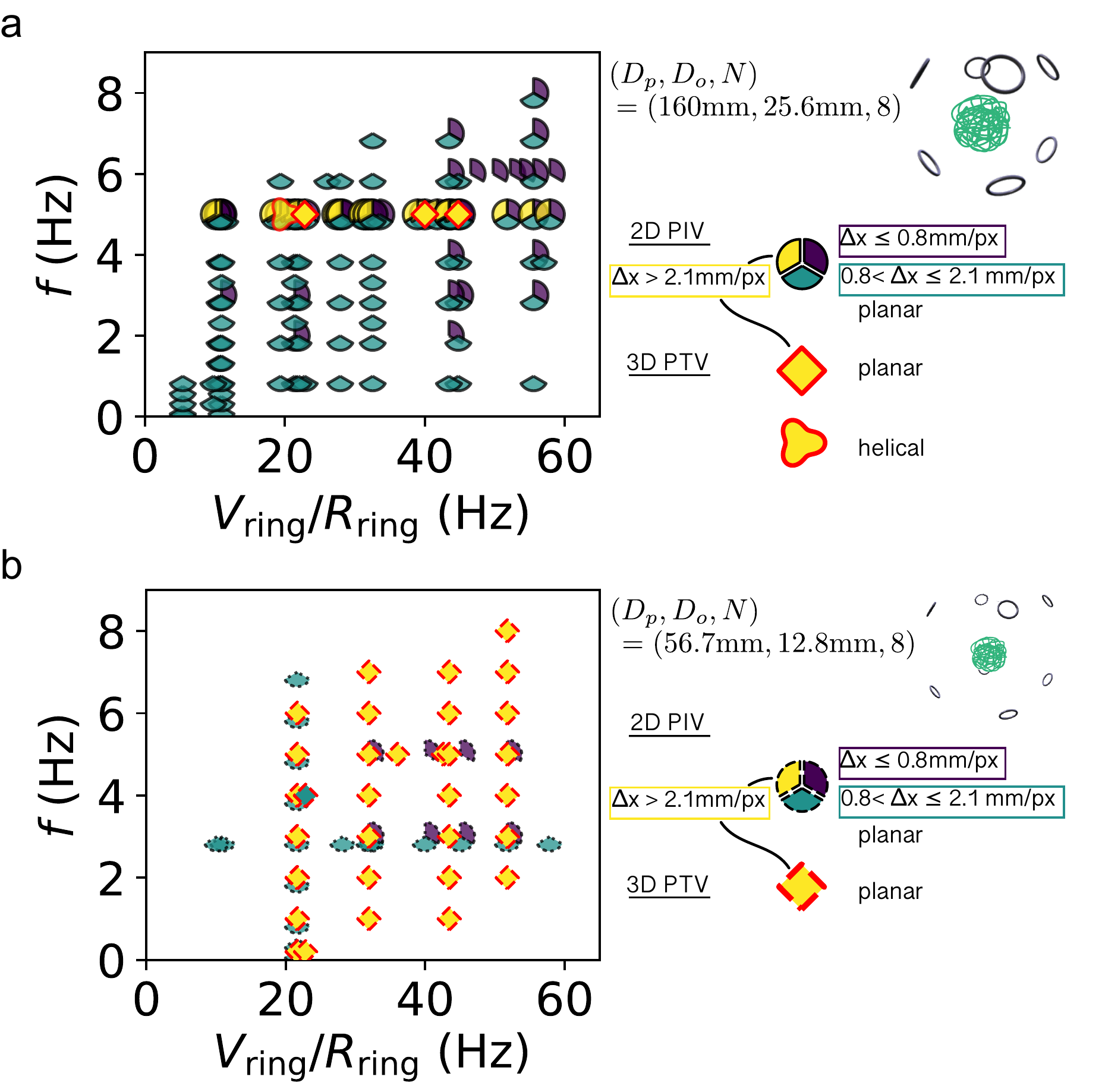}
\caption{\textbf{Summary of measurements to study confined turbulence involving different techniques (2D PIV and 3D PTV) on various experimental configurations.} Altering the size of the piston diameter $D_p$ and the orifice diameter $D_o$ enables to vary the size of a turbulent region formed by the vortex ring collisions. Numerous PIV measurements with different resolutions allow us to investigate turbulent statistics thoroughly. (a/b) This setting enables to create a turbulent region with $R_{\rm blob}\approx 60/30$mm. We investigate the transfer of mass, energy, and enstrophy using a small blob because it can be fully contained in the measurement volume. }
\label{fig:data_points_blob}
\end{figure}

To characterize the physics of the turbulent blob state, we performed a number of experiments in which we fired sets of vortex rings together while varying ring speeds, radii and the interval between sets. We then measured the ensuing dynamics using 2D PIV at different magnifications, as well as 3D PTV.
Figure~\ref{fig:data_points_blob} summarizes the experimental runs. 
A version of this figure, with specific points highlighted is reproduced as an inset in SI figures to make it clear which datasets went into making each derived plot.

\clearpage
\newpage

\section{Anatomy of a turbulent blob}
In this section we examine the structure of a turbulent blob via Reynolds decomposition $U_i(\vec{x}, t)= \langle U_i \rangle_t(\vec{x}) + u_i(\vec{x}, t)$.
\subsection{Structure of the mean flow}
\begin{figure}[htbp]
\centering
\includegraphics[width=0.99\textwidth]{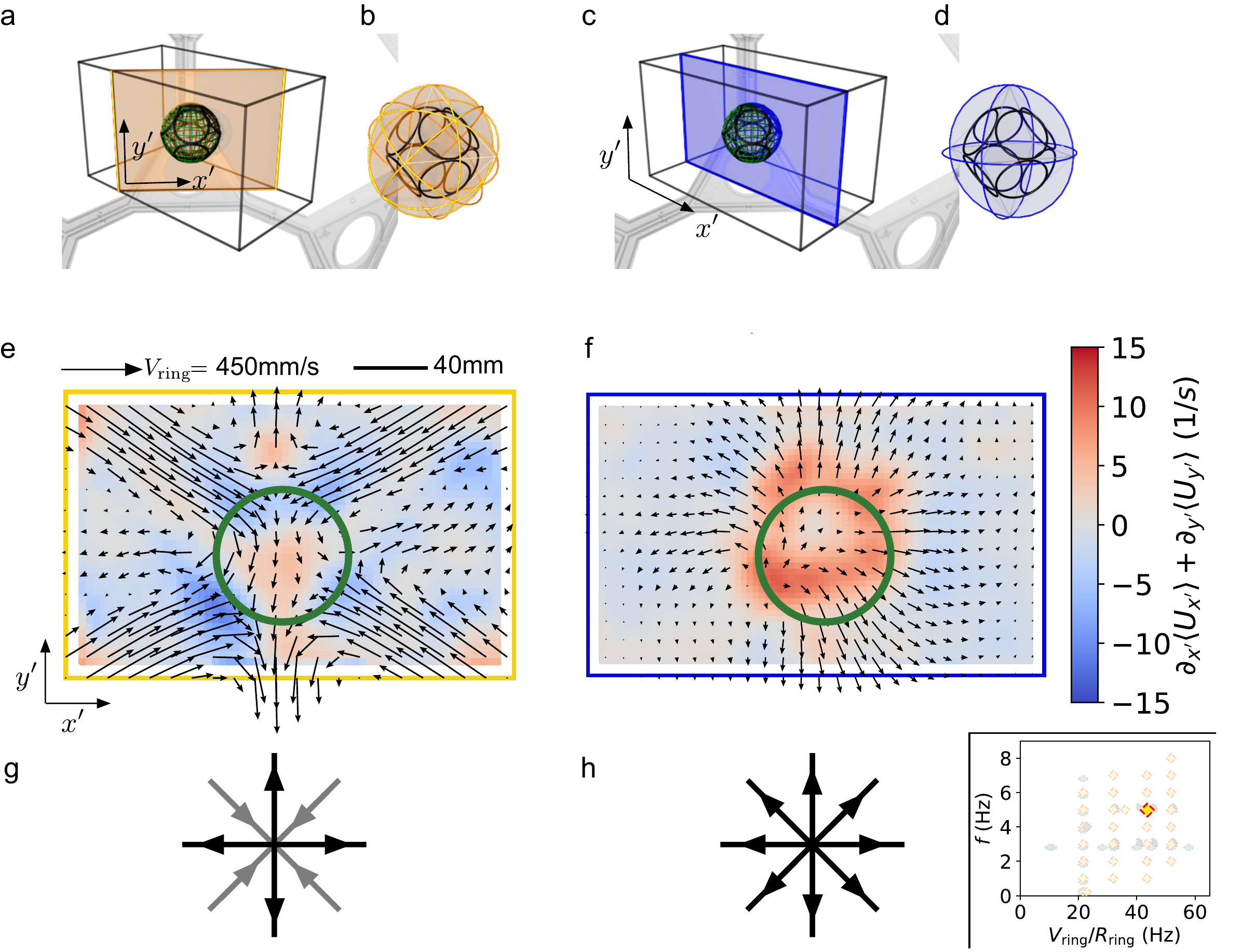}
\caption{\textbf{Mean flow structure and the octahedral symmetry.} There are nine planes of reflection. (a) A diagonal plane cuts through four rings oriented on the faces of an octahedron. (b) Six diagonal planes of reflection are shown. (c) A horizontal plane is oriented between the rings. (d)Three horizontal planes of reflection are shown. (e)The mean flow on the diagonal plane shows the influx from the diagonal directions due to the vortex rings, accompanied by weak outflux in the horizontal and vertical directions. (f) The mean flow on the horizontal plane exhibits a divergent nature. (g-h) A schematic of the flow directions is shown on the diagonal/horizontal plane.}
\label{fig:mean_flow}
\end{figure}

The (temporally averaged) mean flow reveals the strong inflow at the spots where the vortex rings arrive, and nearly uniform outflow elsewhere (Figure \ref{fig:mean_flow}e-f). Such a mean flow acts to cage the injected energy and enstrophy inside the turbulent core with small ejection. This corroborates with a simple picture that energy and enstrophy are supplied by vortex rings, getting dissipated mostly inside the blob, and a small fraction of them leaves. 

Inspecting two planes would be sufficient to explain the mean flow of our experiments because of the octahedral symmetry in the vorticity field. The corresponding point group is $O_h$ in the Schönflies notation, and has two kinds of reflection operations  $\sigma_v$, and $\sigma_h$. $\sigma_v$ corresponds to a reflection about the plane that cuts four out of eight rings into halves (Figure \ref{fig:mean_flow}a). $\sigma_h$ corresponds to a reflection about the plane on which no rings are present (Figure \ref{fig:mean_flow}b). There are six $\sigma_h$ and three $\sigma_v$ in $O_h$ as illustrated in Figure \ref{fig:mean_flow}a-b. Figure \ref{fig:mean_flow}c-d show the measured, mean flow field on each plane. The $\sigma_v$ plane exhibits the strong
inflow in the diagonal directions due to the vortex rings, accompanied by outflow in the $\pm x'$ and $\pm y'$ directions. Meanwhile, the $\sigma_h$ plane contains a source at the center.

In a simplified case, one can infer the strength of the outflow solely by the symmetry. The 6+4=10 reflection planes uniformly split the space. On the $\sigma_v$ plane, there are four incoming directions (diagonal) and four outgoing directions (horizontal and vertical) (Figure \ref{fig:mean_flow}g)). The $\sigma_h$ plane is divergent (Figure \ref{fig:mean_flow}h)). Because of incompresibility, the sum of the incoming and outgoing flows must be zero. There are $4 \times 6=24$ directions of the inflow and $4 \times 6 + 8 \times 3 = 48$ directions of the outflow. If one assumes the strength of the outflow on both planes to be the same, we expect that the outflow speed is a half of the inflow speed.

\newpage

\subsection{Fluctuating flow}
\begin{figure}[htbp]
\centering
\includegraphics[width=0.99\textwidth]{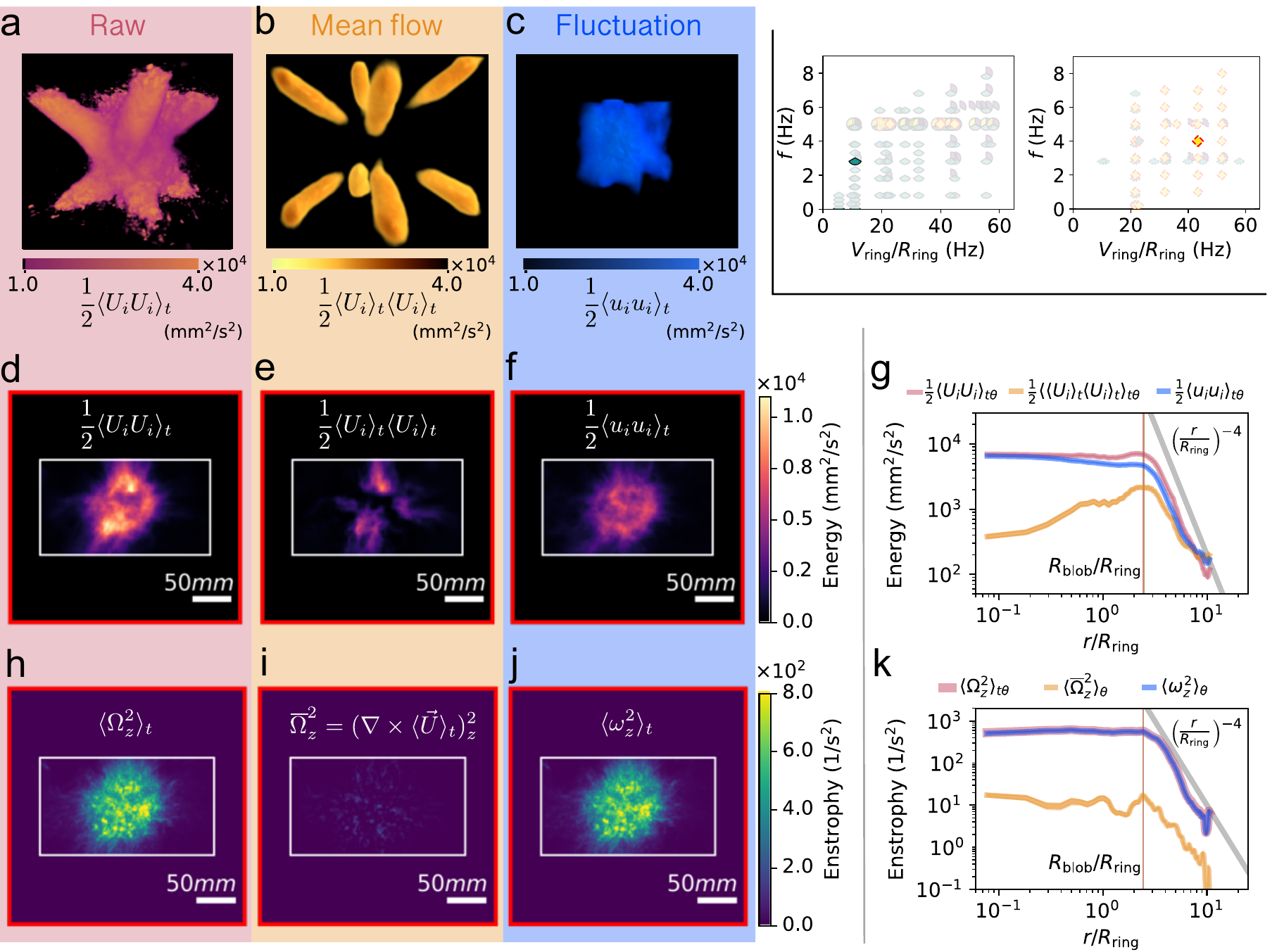}
\caption{\textbf{The fluctuating velocity field contains the majority of energy and enstrophy inside a turbulent blob.}(a-c) Energy (density) fields of a 3D PTV measurement (a)Temporally-averaged energy (b)Mean flow energy (c)Temporally-averaged fluctuating energy (d-k) Energy and enstrophy fields of a 2D PIV measurement on the central slice (d)Temporally-averaged energy (e)Mean flow energy (f)Temporally-averaged fluctuating energy (g) Average energy (density) at distance $r$ away from the center of the energy
(h)Temporally-averaged enstrophy (i)Mean flow enstrophy (j)Temporally-averaged fluctuating enstrophy (k) Average enstrophy (density) at distance $r$ away from the center of the enstrophy.}
\label{fig:fluctuation}
\end{figure}

We shall see that the fluctuation dominates inside the blob compared to the mean flow. The Reynolds decomposition yields 
\[
\frac{1}{2} \langle U_i U_i\rangle_t = \frac{1}{2} \langle U_i \rangle_t \langle U_i\rangle_t + \frac{1}{2} \langle u_i u_i\rangle_t
\]
if the fluctuation has zero mean  $\langle u_i \rangle_t =0$. We will examine this assumption in a later section. As for enstrophy, if the voriticity of the fluctuating field $\omega=\Nabla \times \vec{u}$ has zero mean, the temporally-averaged enstrophy is given by
\[
\langle \Omega_i \Omega_i \rangle_t = \overline{\Omega}_i \overline{\Omega}_i + \langle \omega_i \omega_i\rangle_t; ~~~\overline{\Omega}_i = \epsilon_{ijk} \partial_j \langle U_k \rangle_t.
\]
Figure \ref{fig:fluctuation}a-f, and h-j show the energy and enstrophy in the raw, (temporally-averaged) mean, and fluctuating velocity fields of a turbulent blob obtained by 3D PTV and 2D PIV. The fluctuating field contains the majority of both energy and enstrophy inside a turbulent blob. Averaging the energy and enstrophy density over angle yields Figure \ref{fig:fluctuation}g and k respectively, revealing that the fluctuating energy is approximately 20 times greater than the mean flow energy at the center of the blob. The exact ratio may largely vary from 10 to 1000, depending on the experimental conditions ($L_*$, $v_{eff}, f, D_p,$ and $D_o$). Outside the blob ($r>R_{\rm blob}$), both energy and enstrophy decay as $r^{-4}$. The same dependency is observed for the local dissipation rate $\epsilon(r)=2\nu \langle s_{ij}s_{ij} \rangle_{t\theta}$ (main Figure 3b).

\clearpage
\newpage

\section{Turbulent statistics}
In this section we discuss the velocity fluctuations in our turbulent blob.   
We first examine the raw fluctuating velocity spectra, addressing inhomogeneity and anisotropy of the measured fluctuating velocity field in sections \ref{sect:def_vel_fluctuations} and \ref{sect:inhomo_anisotropy}.
 Section \ref{sect:Energy spectra and structure function}-\ref{sect:windowing} discusses how we compute the spectra and the structure functions from PIV and PTV measurements. Section \ref{sect:Foundation of the Local Energy Spectrum Estimate} discusses how we computed these functions using only the regions inside the blob. 

The authors find it pedagogical to summarize the methods to compute 3D and 1D turbulent energy spectra and structure functions from experimental data. As PIV and PTV became an important method to study fluid mechanics, the authors hope that this section serves as a guide. The methods are standard but we explicitly discuss issues of computing the spectra from experimentally obtained 2D/3D fields without using Taylor's hypothesis. For the turbulent blob, the mean field inside the blob is substantially smaller than the fluctuating field, making Taylor's hypothesis inapplicable. 

\subsection{Definition of velocity fluctuations}\label{sect:def_vel_fluctuations}
There is a subtlety in defining the fluctuation of the flow because the mean flow may in principle depend on the phase with respect to the oscillatory motion. This may favor the use of a phase-averaged flow $\langle U_i \rangle_n$ as a mean flow over a time-averaged flow $\langle U_i \rangle_t(\vec{r})$. For the definition of the phase-locked (or phase-averaged) velocity $\langle U_i \rangle_n(\vec{r}, \Theta)$, see Section \ref{sect:nomenclature}.

\begin{figure}[htbp]
\centering
\includegraphics[width=0.99\textwidth]{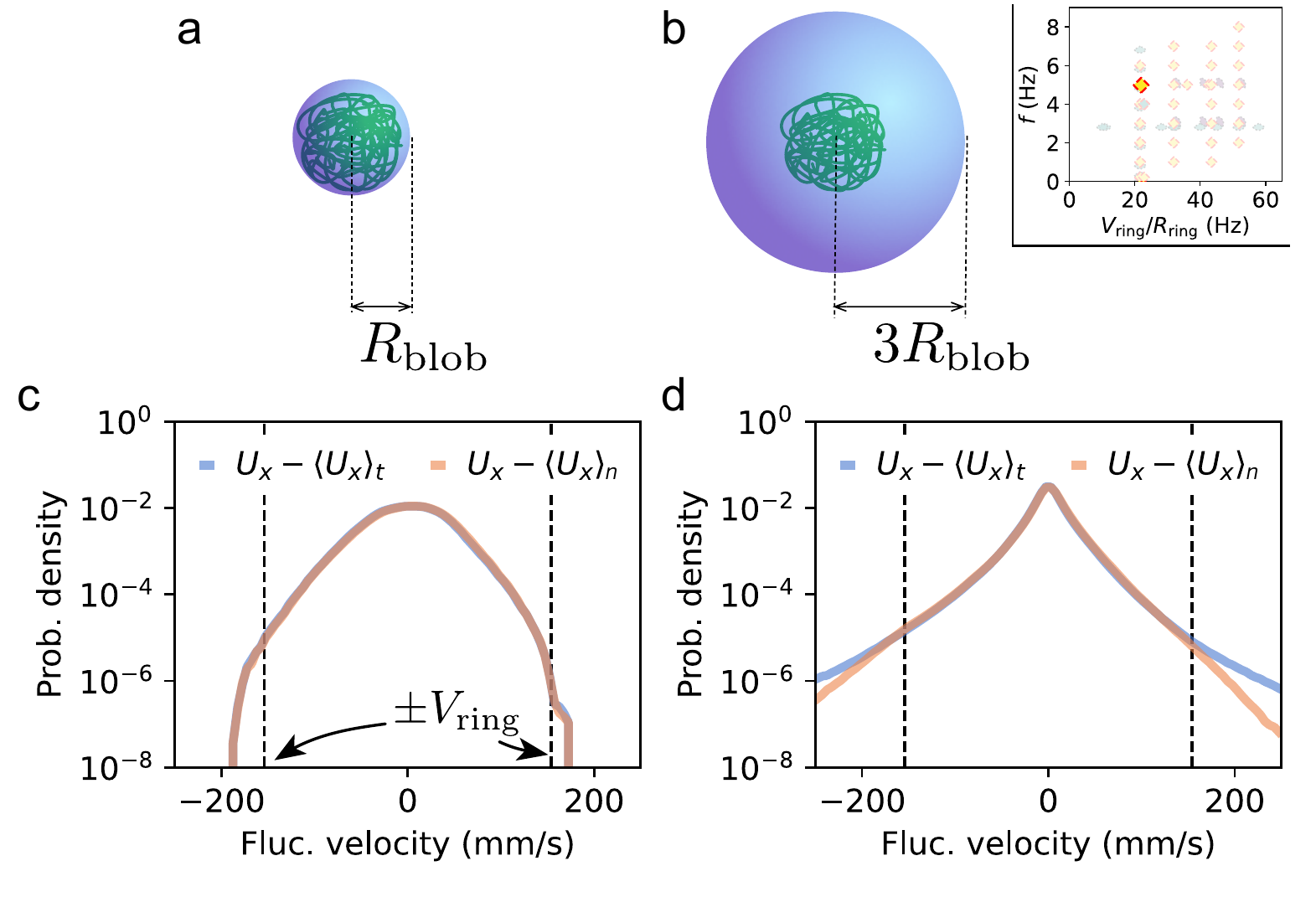}
\caption{\textbf{Fluctuations inside the turbulent blob are phase-independent regardless of the choice of the mean flow between time-averaged flow and the phase-locked flow.} (a-b) Two domains are considered to examine statistics of a fluctuating velocity field: a region within a turbulent core $r<R_{\rm blob}$ and a spherical region containing the turbulent core and its ambient surroundings $r<3R_{\rm blob}$. (c) Probability distribution functions (PDF) of the two fluctuating velocities for ($L/D, v_{eff}, f$)=(2., 200mm/s, 4Hz) with $R_{\rm blob} = $32mm show no difference between how the fluctuations are defined. (d) The fluctuations outside the turbulent blob depend on the phase because the vortex rings appear at specific phase in this domain $r \in (R_{\rm blob}, 3R_{\rm blob}]$, giving rise to the difference between the PDFs of the fluctuations.  }
\label{fig:def_fluc_vel}
\end{figure}

Consider two spherical domains with a radius $r=R_{\rm blob}$ and $r=3R_{\rm blob}$ (Figure \ref{fig:def_fluc_vel}a-b). We find that both definitions of the fluctuating velocity have the same distribution inside the core, reflecting zero mean velocity throughout a cycle(Figure \ref{fig:def_fluc_vel}c). The difference of these two conventions are evident as we inspect the distribution inside the second domain that includes the ambient surrounding in addition to the turbulent core (Figure \ref{fig:def_fluc_vel}d). The overpopulation at high velocity for $ U_i - \langle U_i \rangle_t$ is caused by ignoring the phase-dependence of the vortex ring locations. In summary, the (fluctuating) velocity statistics inside the turbulent core does not depend on the phase. All turbulent analyses presented in this paper use the time-averaged flow as a mean flow. 

\subsection{Inhomogeneity and isotropy of the flow}\label{sect:inhomo_anisotropy}
\begin{figure}[htbp]
\centering
\includegraphics[width=0.98\textwidth]{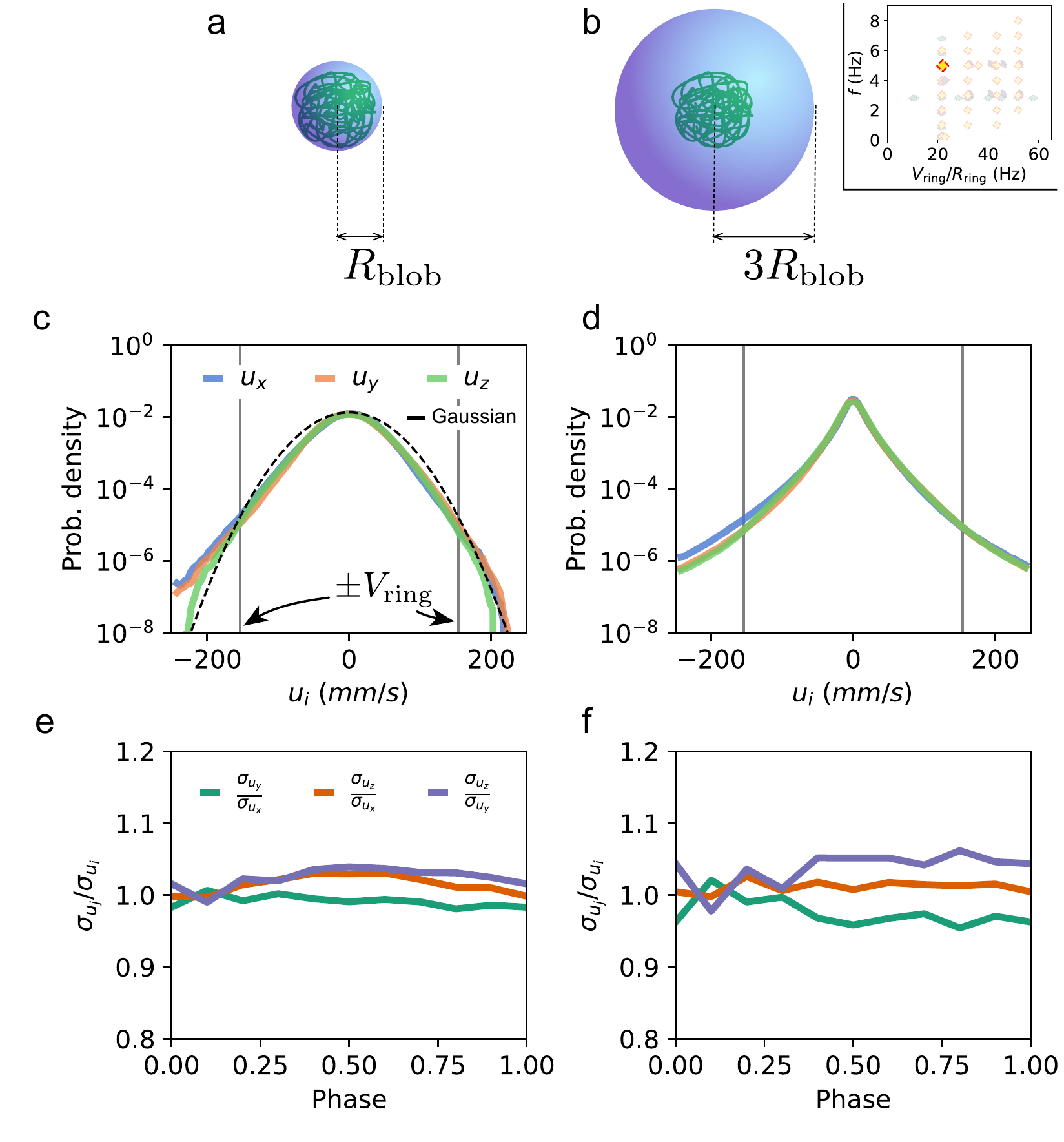}
\caption{\textbf{Inhomogeneity and anisotropy of velocity fluctuations $u_i = U_i - \langle U_i \rangle_t$} (a-b) Two domains are considered to examine statistics of a fluctuating velocity field: a region within a turbulent core $r<R_{\rm blob}$ and a spherical region containing the turbulent core and its ambient surroundings. (c-d) Probability distribution functions (PDF) of fluctuating velocities for ($L/D, v_{eff}, f$)=(2.0, 200mm/s, 4Hz) with $R_{\rm blob}$ = 32mm  are shown for the two domains, indicating isotropy inside the core and increase of anisotropy as the region includes its surroundings. (e-f) Ratios between the two second moments of the velocity PDFs.}
\label{fig:inhomogeneity_anisotropy}
\end{figure}

The flow induced by vortex ring collision at $f>f_c$ is clearly inhomogeneous and anisotropic as it consists of a turbulent core and a mean flow structure originated from the vortex rings (Figure 3a). However, the turbulent flows inside the core $r\leq R_{\rm blob}$ is isotropic and homogeneous. The anisotropy and inhomogeneity grows as the domain of consideration includes its surroundings. Consider two spherical domains with a radius $r=R_{\rm blob}$ and $r=3R_{\rm blob}$ each (Figure \ref{fig:inhomogeneity_anisotropy}a and b). Now we examine the statistics of all three components of a fluctuating velocity field of a turbulent blob, obtained by a 3D PTV experiment($L/D, v_{eff}, f$)=(2., 200mm/s, 4Hz), $R_{\rm blob}=32mm$, recorded volume: 200mm $\times$ 108mm $\times$ 123mm, duration: 4s = 40 cycles). The probability distribution functions (PDFs) of the fluctuating velocity field inside the turbulent core (Figure \ref{fig:inhomogeneity_anisotropy}c) shows almost no anisotropy. Quantitatively, the isotropy can be assessed by comparing the standard deviations for each component $\sigma_{u_i}/\sigma_{u_j}~(i\neq j)$. Figure \ref{fig:inhomogeneity_anisotropy}e indeed shows high isotropy inside the turbulent core as well as inside the larger domain (Figure \ref{fig:inhomogeneity_anisotropy}f); however, the flucutations are inhomogeneous. 

\begin{sidewaysfigure}
\centering
\includegraphics[width=0.9\textwidth]{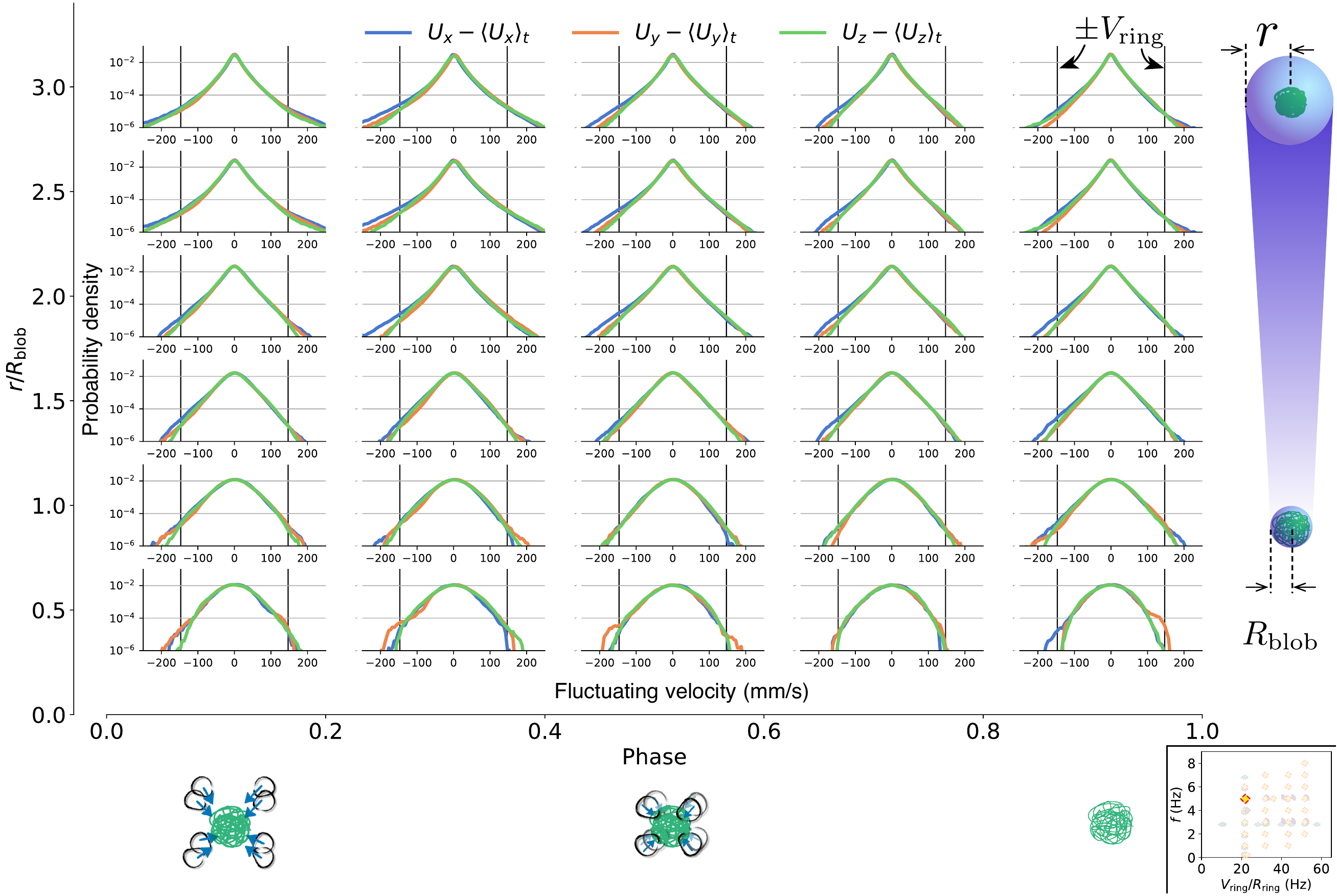}
\caption{\textbf{Isotropy, homogeneity, and phase-dependence of fluctuating velocity fields $u_i = U_i - \langle U_i \rangle_t$.} 
Probability distribution functions (PDFs) of $u_i$ over spherical domains of various radii $r$ show high isotropy, indicated by overlap of three colors $u_x, u_y, u_z$. The changes in the shape of the PDFs over various $r$ reflect the inhomogeneity of fluctuations. Small changes over the phase across different $r$ indicates that the fluctuation is almost independent of the phase. The data is obtained by a 3D PTV experiment for ($L/D, v_{eff}, f$)=(2., 200mm/s, 4Hz) with a recording time of 10s = 40 cycles. The range of the velocity vectors is roughly the propagation speed of the injected vortex ring $V_{\rm ring}$.}
\label{fig:pdf_fluc_vel}
\end{sidewaysfigure}

The velocity PDFs over multiple spherical domains with radius $r$ at different phases are shown in Figure \ref{fig:pdf_fluc_vel}. One first finds that the statistics does not depend on the phase, and the flow is nearly isotropic in all of the considered domains. The fluctuations are nearly Gaussian inside the core $r\leq R_{\rm blob}$. The skewness is -0.1 and kurtosis is 4.0 for all velocity components within $r\leq R_{\rm blob}$, compared to 0 and 3 if the fluctuations were Gaussian. The PDF is altered as the considered domain grows in size, including the ambient, less turbulent regions. This is reflected by a growing peak at $u_i = U_i - \langle U_i \rangle_t = 0$ in Figure \ref{fig:pdf_fluc_vel} as $r/R_{\rm blob}$ increases.

\newpage 

\clearpage

\subsection{Turbulent statistics and vortex ring metrics}
Table \ref{tab:turbulence_statistics} summarizes the turbulent statistics and the vortex ring metrics that were measured in the two settings (See Table \ref{tab:exp_dimensions}).

\begin{table}[]
    \centering
    \begin{tabular}{c c c}
         \\
         Quantity &Setting 1 & Setting 2 \\ \hline 
         $R_{\rm blob}$ & $50-60mm$ & $30-40mm$  \\
         $R_{\rm ring}$ & $13-18mm$ & $7-9mm$  \\
         $V_{\rm ring}$ & $100-800mm/s$ & $50-400mm/s$  \\
         $\Gamma_{\rm ring}$ & $0.1-1\times 10^5 mm^2/s$ & $0.3-4\times 10^4 mm^2/s$  \\
         $\Gamma_{\rm ring}/\nu$ & $1-6\times 10^4$ & $0.3-4\times 10^4$ \\
         $\epsilon_0$ & $4\times10^5 - 4\times 10^6 mm^2/s^3$ & $1\times10^4 - 1\times 10^5 mm^2/s^3$  \\
         $\eta=\left( \frac{\nu^3}{\epsilon_0}\right)^{\frac{1}{4}}$ & $0.02-0.04mm$ & $0.06-0.1mm$  \\
         $\lambda$ & $ 1-5mm$ &  $1-5mm$ \\
         $\mathcal{L}=\frac{u'^3}{\epsilon_0}$ & $100-120mm$ & $60-80mm$  \\
         $Re_{\lambda}=\frac{u' \lambda}{\nu}$ & $50-300$ & $50-200$  \\
         $u'=\sqrt{\frac{1}{3} \langle u_i u_i \rangle}$ & $50-250mm/s$ & $25-150mm/s$  \\
    \end{tabular}
    \caption{\textbf{Vortex ring metrics and turbulence statistics inside the blob for the two experimental settings- Setting 1, 2: $(D_p, D_o)=(160.0mm, 25.6mm), (57.0mm, 12.8mm)$}. The listed dissipation rates $\epsilon_0$ are values based on the rate-of-strain tensor.}
    \label{tab:turbulence_statistics}
\end{table}

\newpage 
\section{Computation of energy spectra and structure function from experimental real space velocimetry data} \label{sect:Energy spectra and structure function}
\begin{figure}[htbp]
\centering
\includegraphics[width=0.9\textwidth]{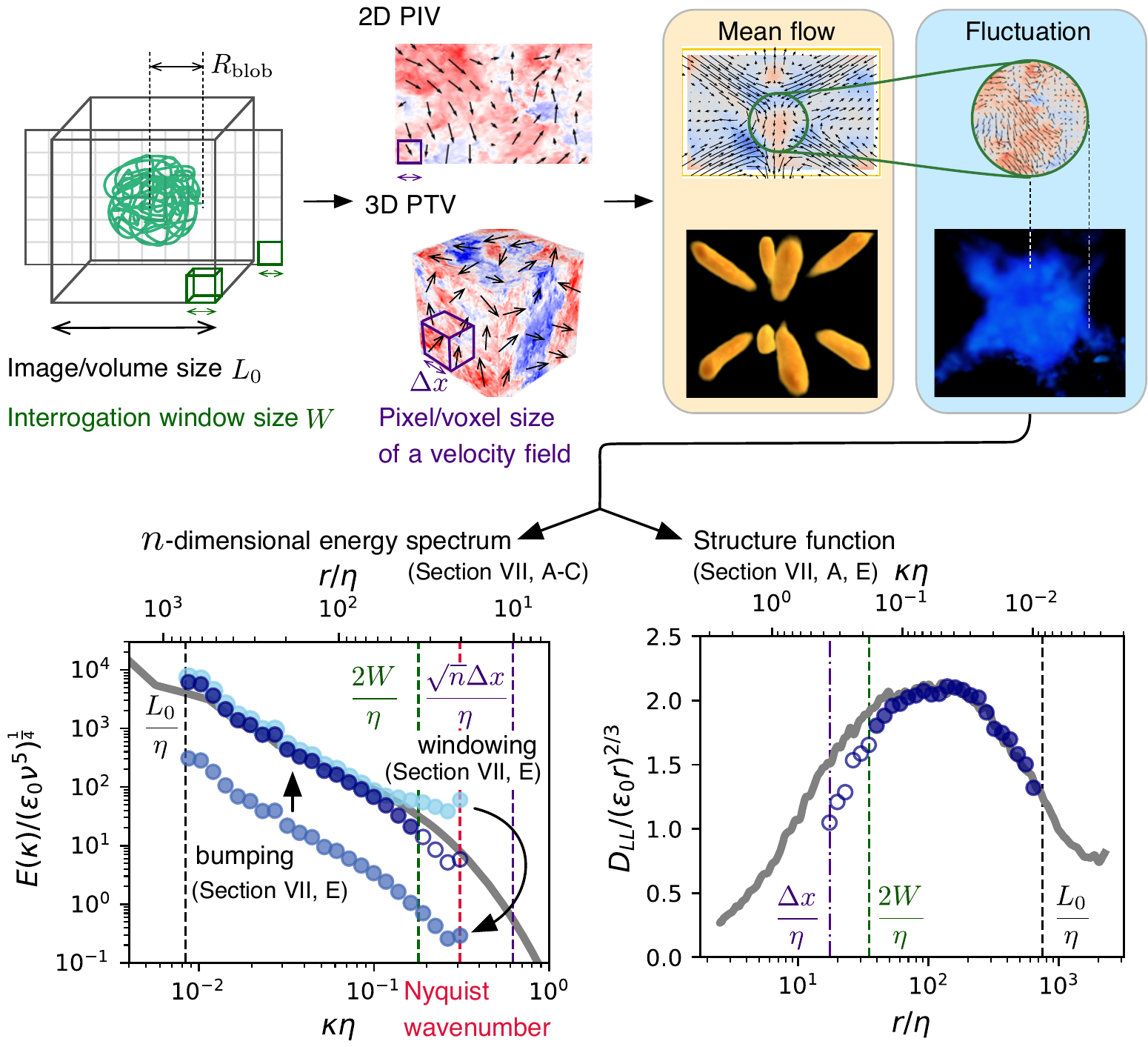}
\caption{\textbf{Guide to the computation of the energy spectrum and the structure function without applying Taylor's frozen turbulence hypothesis.} 1. Extract a velocity field via cross-correlation of the adjacent images, or directly tracking the particles in 3D. 2. Decompose the velocity field into the mean flow and the fluctuations (Reynolds decomposition) 3. Compute a local turbulent energy spectrum or a structure function inside the region of interest. For the former, windowing combats against undesired effects (aliasing and spectral leakage) of DFT for the Fourier analysis, and bumping accounts for the attenuation of the signal by windowing. The attenuation of the magnitude by windowing is exaggerated for this illustration.}
\label{fig:energy_spectrum_guide}
\end{figure}

The fluctuation energy spectra and structure functions are an essential component of canonical characterizations of turbulent flows.
For a homogeneous isotropic flow, the three-dimensional energy spectrum  $E(\kappa)$, one-dimensional energy spectrum $E_{11}(\kappa_1)$ and the structure functions $D_{ij}=\langle \delta u_i \delta u_j \rangle$  contain equivalent information about the fluctuating velocity field. 
However, the computation of each is affected differently by experimental considerations such as the dimensionality of the velocity data (e.g. point measurements with anemometers or laser doppler velocimetry vs spatially resolved PIV/PTV measurements) and experimental limitation (e.g. the extent of the measurement area, resolution, noise level). 

In the case of spatially resolved velocimetry, a natural approach to measuring the fluctuation energy spectrum is Fourier analysis; however, aperiodicity, discreteness, and finiteness of experimental data can give rise to artifacts.
In this context, the real-space velocity correlation function $D_{ij}$ is sometimes preferred as a measure of turbulent fluctuations because it is less susceptible to measurement artifacts  as illustrated later in this section. 

We begin this section by reviewing the definitions of each fluctuation-characterizing function, before studying the effects of experimental noise, finite window size and finite resolution on each measurement.

\subsection{Definitions and relations of energy spectra}

Consider the fluctuating component of a turbulent velocity field $u_i({\bm{x}})$ in a domain $V$. 
The velocity two-point correlation tensor $R_{ij}$ is defined as: 
\begin{align}
    R_{ij}({\bm{r}}) = \langle u_i({\bm{x} + \bm{r}}) u_j({\bm{x}})\rangle.
\end{align}
Its Fourier transform is the velocity spectrum tensor $\Phi_{ij}$:
\begin{align}
    \Phi_{ij}({\bm{\kappa}}) \equiv  \frac{1}{ (2\pi ) ^3} \iiint_{-\infty}^{\infty} R_{ij}({\bm{r}}) e^{-i {\bm{\kappa \cdot r} }} d{\bm{r}}.
    \label{eq:vel_spectrum_tensor}
\end{align}
The one-dimensional energy spectra $E_{ij}$ are considered when three-dimensional Fourier transformation is not available, and are defined as twice the one-dimensional Fourier transform of the two-point correlation tensor:
\begin{align}
    E_{ij}(\kappa_1) &\equiv  2\cdot \frac{1}{ 2\pi} \int_{-\infty}^{\infty} R_{ij}(r_1\hat{e}_1) e^{-i \kappa_1 r_1} dr_1.
    \label{eq:one_dimensional_spectra}
\end{align}
Eq. \ref{eq:vel_spectrum_tensor} and \ref{eq:one_dimensional_spectra} are expressed interms of the two-point correlation function; however, they can be alternatively expressed using the Fourier transform of a turbulent velocity field, $\tilde{u}_i({\bm{\kappa}}) = (1/2\pi) \int u_i({\bm{x}}) \exp{(-i {\bm{\kappa}} \cdot {\bm{x}})} d{\bm x}$, via the Wiener-Khinchin theorem.  Eq. \ref{eq:vel_spectrum_tensor} becomes:
\begin{align}
    \Phi_{ij}({\bm{\kappa}}) = \langle \tilde{u}_i({\bm \kappa}) \tilde{u}_j^*({\bm \kappa}) \rangle.
    \label{eq:vel_spectrum_tensor_using_uk}
\end{align}
As for  Eq. \ref{eq:one_dimensional_spectra}, let us denote the one-dimensional Fourier transform operator along the direction $\hat{x}_i$ as $\mathcal{F}_{x_i}$. The one-dimensional spectra $E_{ij}$ are then:
\begin{align}
    E_{ij}(\kappa_1) &= 
    2 \left\langle
        \mathcal{F}_{x_1}[u_i({\bf x})] 
        \mathcal{F}_{x_1}[u_j({\bf x})]^* 
    \right\rangle_{x_2, x_3} \\
    &=  2 \left\langle 
        \left[ \frac{1}{2\pi} \int_{-\infty}^{\infty} u_i({\bf x}) e^{-i\kappa_1 x_1} dx_1 
        \right]  
        \left[ \frac{1}{2\pi}  \int_{-\infty}^{\infty} u_j({\bf x}) e^{i \kappa_1 x_1} dx_1
        \right] 
    \right\rangle_{x_2, x_3}.
    \label{eq:1d_energy_spectrum_function_using_uk_cft}
\end{align}

The definitions of the velocity spectrum and the one-dimensional energy spectrum functions in terms of the two-point correlation functions (Eq. \ref{eq:vel_spectrum_tensor} and \ref{eq:one_dimensional_spectra}) 
are well suited to  theoretical considerations of random processes with known two-point correlation functions, and have provided a means to compute the spectra from the point-wise measurements historically. A further discussion on the two equivalent representations of the energy spectrum can be found in Chapter 11 of \cite{Bracewell1978-uj}(p.285-288, 3rd ed.). 

With the assumption of isotropy, the energy spectrum function $E(\kappa)$ can be obtained by stripping off the angular information from $\Phi_{ii}$.
\begin{align}
    E(\kappa) = \frac{1}{2} \iiint_{-\infty}^{\infty} 
    \Phi_{ii}( {\bm{\kappa}} ) \delta(|{\bm{\kappa}}| - \kappa) d{\bm {\kappa} }
    \label{eq:energy_spectrum_function}
\end{align}

Similarly for an isotropic fluctuating field, $R_{ij}$ can be expressed in terms of two scalar functions:
\begin{align}
    R_{ij}(r) = u'^2 \left( g(r) \delta_{ij} + [f(r) - g(r)]\frac{x_i x_j}{r^2}    \right).
    \label{eq:rannk2_isotropic_tensor}
\end{align}
Here, $f(r)$ and $g(r)$ are longitudinal and transverse two-point correlation functions respectively:
\begin{align}
    f(r) &= \frac{\langle u_\parallel({\bm{x}}) u_\parallel({\bm{x} + {r}}) \rangle}{\langle u_\parallel^2 \rangle}
    \label{eq:two_point_correlation_func_longitudinal} \\
    g(r) &= \frac{\langle u_\perp({\bm{x}}) u_\perp({\bm{x} + {r}}) \rangle}{\langle u_\parallel^2 \rangle}
    \label{eq:two_point_correlation_func_transverse}
\end{align}
where, $u_\parallel$ is the velocity component parallel to the displacement vector ${\bm{r}}$. 
With homogeneity and isotropy, the diagonal elements of $E_{ij}$ can be expressed using $f(r)$ and $g(r)$:
\begin{align}
    E_{11}(\kappa_1) &= \frac{2}{\pi} \langle u_1^2 \rangle \int_0^\infty f(r_1) \cos{(\kappa_1 r_1)} dr_1
    \label{eq:e11_def}\\
    E_{22}(\kappa_1)=E_{33}(\kappa_1) &= \frac{2}{\pi} \langle u_2^2 \rangle \int_0^\infty g(r_1) \cos{(\kappa_1 r_1)} dr_1.
\end{align}
The (three-dimensional) energy spectrum $E(\kappa)$ can be expressed by using the one-dimensional energy spectra as $E(\kappa)=-\frac{\kappa}{2} \frac{d}{d\kappa} E_{ii} (\kappa)$.

\subsection{Energy spectrum computation by Fourier analysis}

 Nearly all the energy spectra have been measured by hot-wire anemometry. Such measurements involved measuring the autocorrelation functions, and converting them into the two-point correlation functions with Taylor's frozen turbulence hypothesis. The one-dimensional energy spectrum is readily obtained by Eq. \ref{eq:e11_def}. This method is only valid when the turbulent velocity is sufficiently smaller than the mean flow velocity, and is known to fail for free shear flows~\cite{moin_2009}. 

If presented with spatially resolved velocity field snapshots that are sampled on a uniform grid, the Fourier transform of the velocity is replaced by the discrete Fourier transform:
\[
\breve{u}_i ({\bm \kappa}) = \sum_{n=0}^{N-1} u_i({\bm x}_n) e^{ - i{\bm \kappa } \cdot  {\bm x_n } } 
\]
To derive an expression for the energy spectrum $E(k)$ in terms of $\breve{u}_i ({\bm \kappa})$, we consider the average kinetic energy:
\[
k = \langle \frac{1}{2} u_i({\bm x}) u_i({\bm x})\rangle = \int_0^\infty E(\kappa) d\kappa.
\]
On a discretized velocity field this becomes:
\begin{align}
    k & =\frac{1}{2N}  \sum_{n=0}^{N-1} u_i({\bm x_n}) u_i({\bm x_n})\\
      & = \frac{1}{2N^2} \sum_{n=0}^{N-1} \breve{u}_i({\bm \kappa_n}) \breve{u}^*_i({\bm \kappa_n})  \because \text{Parseval's theorem}\\
      & \approx \frac{1}{2N^2 \Delta \kappa_1 ... \Delta \kappa_d} \sum_{n=0}^{N_r-1} \langle \breve{u}_i({\bm \kappa_n}) \breve{u}^*_i({\bm \kappa_n}) \rangle_{\kappa_\theta, \kappa_\phi} |J({\bm \kappa_n}))| \Delta \kappa  \because \text{isotropy}
    \label{eq:average_tke_intermed}
\end{align}
where $\kappa_n = 2 \pi n / N$ and $|J({\bm \kappa}_n)| = 4 \pi \kappa^2$ ($d=3$) or $2 \pi \kappa$ ($d=2)$ where $d$ is the dimension of the data set being analyzed. For example, $d=2$ for a  2D slice of a velocity field embedded in 3D flow (2D PIV) or for a purely 2D flow. The approximation in Eq. \ref{eq:average_tke_intermed} originates from the transformation of ${\bm \kappa}_n$ from the Cartesian basis to the spherical basis. The discreteness of ${\bm \kappa}_n$ and the rectangular spatial domain also contributes to this approximation. $\Delta \kappa=\sqrt{\Delta \kappa_1^2 + ... + \kappa_d^2}$ is the wavenumber spacing in the spherical coordinate system, and $N_r=\lfloor \max{(|{\bm \kappa}_n|})/\Delta \kappa \rfloor$. 

If one has access to only two components of the 3D velocity field, which is  typical  for  PIV experiments based on a single camera and a laser-sheet, one must further multiply Eq.~\ref{eq:average_tke_intermed} by $3/2$ (appealing to isotropy).

From this we can read off:
\begin{align}
    E(\kappa) \approx \frac{1}{ N^2 \Delta \kappa_1 ... \Delta \kappa_d } \frac{|J|}{2}\langle \breve{u}_i( {\bm \kappa} ) \breve{u}^*_i( {\bm \kappa}) \rangle_{\kappa_\theta, \kappa_\phi}
    \label{eq:energy_spectrum_function_using_uk_dft}
\end{align}

To derive the one-dimensional energy spectrum $E_{ij}$, let us denote the one-dimensional discrete Fourier transform operator along the direction $\hat{x}_i$ as $\breve{\mathcal{F}}_{x_i}$:
\[
\breve{\mathcal{F}}_{x_1} [u_j ({\bm x}_n)] = \sum_{n=0}^{N_1-1} u_j({\bm x}_n) e^{ - i \kappa_1 x_{1, n} }
\]
where $N_1$ is the number of samples along the $\hat{x}_1$. Note that the exponent is $- i \kappa_1 x_{1, n}$ when $i=1$ and not the sum $\kappa_i x_{i, n}$. 
The one-dimensional energy spectra are then
\begin{align}
    E_{ij}(\kappa_1) &= 
    2 \left\langle
        \mathcal{F}_{x_1}[u_i({\bf x})] 
        \mathcal{F}_{x_1}[u_j({\bf x})]^* 
    \right\rangle_{x_2, x_3} \\
    &\approx 2 \left\langle \frac{1}{N_1 \Delta\kappa_1} \sum_{n=0}^{N_1-1} u_i({\bm x}_n) e^{-i\kappa_1 x_1} \cdot \frac{1}{N_1 \Delta\kappa_1} \sum_{n=0}^{N_1-1} u_j({\bm x}_n) e^{i\kappa_1 x_1} \right\rangle_{x_2, x_3}\\
   \therefore~ E_{ij}(\kappa_1) &= 2\cdot \frac{1}{N_1^2 \Delta \kappa_1^2} \left\langle \breve{F}_{x_1}[ u_i({\bm x})] \breve{F}_{x_1}[ u_j({\bm x})]^* \right\rangle_{x_2, x_3}.
    \label{eq:1d_energy_spectrum_function_using_uk_dft}
\end{align}

\begin{figure}[htbp]
\centering
\includegraphics[width=\textwidth]{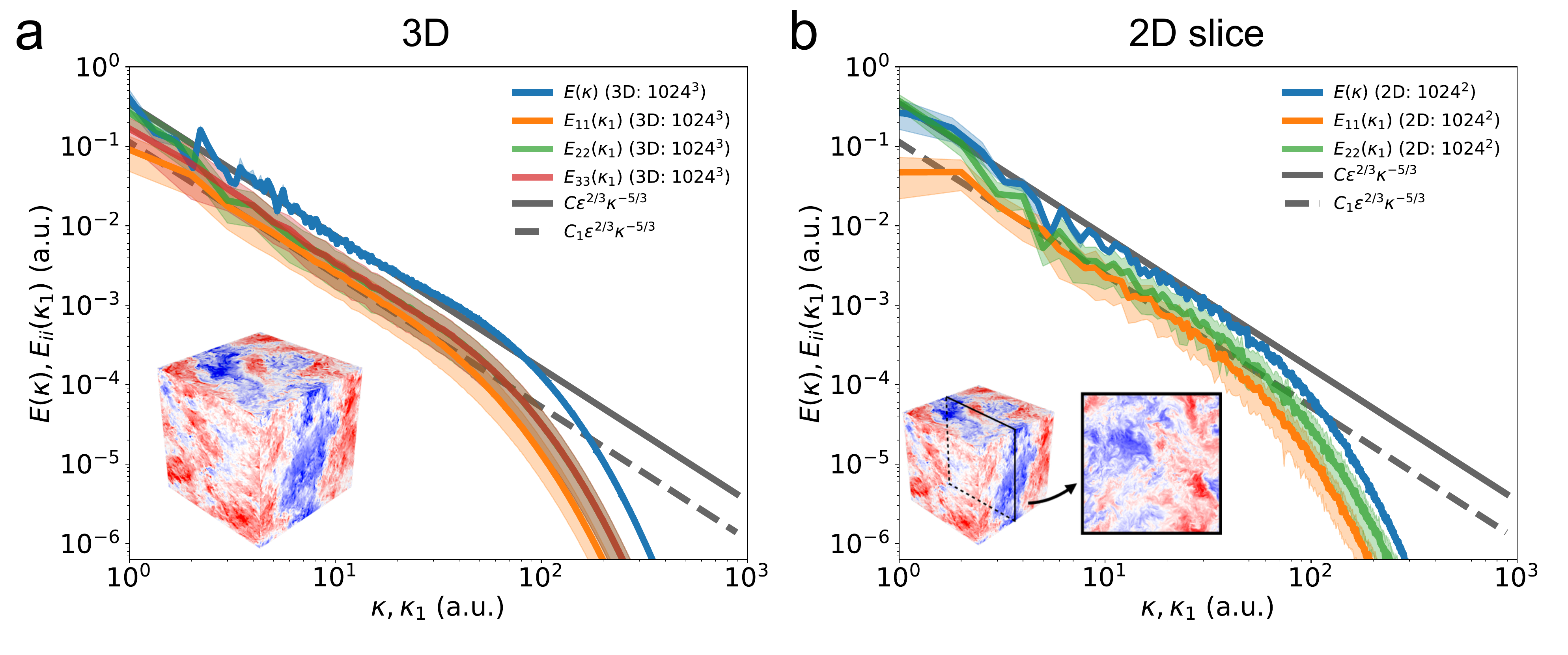}
\caption{\textbf{Direct computation of three- and one-dimensional energy spectra: $E(\kappa)$ and $E_{ii}(\kappa_1)$.} (a) Three- and one-dimensional energy spectra of DNS isotropic, homogeneous turbulence\footnote{The DNS data were obtained from Johns Hopkins Turbulence Database (JHTD).} with a periodic boundary condition were computed by Eq.~\ref{eq:energy_spectrum_function_using_uk_dft} and \ref{eq:1d_energy_spectrum_function_using_uk_dft}. 
(b) Energy spectra of a 2D velocity field embedded in the 3D volume are consistent with the underlying truth if the velocity statistics on the 2D slice is a representative sample of the population.}
\label{fig:energy_spectrum_demo}
\end{figure}

Figure~\ref{fig:energy_spectrum_demo}a displays the three- and one-dimensional energy spectra computed by Eq. \ref{eq:energy_spectrum_function_using_uk_dft} and \ref{eq:1d_energy_spectrum_function_using_uk_dft} for DNS data of isotropic, homogeneous turbulence obtained from Johns Hopkins Turbulence Database\cite{jhtd_li2008public, jhtd_perlman2007data}. In the inertial subrange, they agree with the Kolmogorov spectra ($E(\kappa) = C\epsilon^{2/3} \kappa^{-5/3}$, and $E_{11}(\kappa_1) = C_1\epsilon^{2/3} \kappa_1^{-5/3}$) using the reported dissipation rate $\epsilon$=0.103 a.u.\footnote{The value can be found in the documentation of the forced isotropic turbulence data set.} and the Kolmogorov constants $C=1.6$\cite{mccomb1990physics} and $C_1=0.52$\cite{sreenivasan1995universality}). For the transverse, one-dimensional energy spectra $E_{22}(\kappa_1)$ and $E_{33}(\kappa_1)$, the coefficient becomes $4C_1/3$, assuming isotropy\cite{pope_turbulent_2000}.

Figure~\ref{fig:energy_spectrum_demo}b displays the three- and one-dimensional energy spectra from a 2D slice of the same 3D HIT field  as Figure~\ref{fig:energy_spectrum_demo}a. 
It can be seen that  the spectrum computed from the 2D slice approximates well the spectrum computed from 3D data. The slight jaggedness in Figure~\ref{fig:energy_spectrum_demo}b can be reduced by averaging successive velocity snapshots.

In the case of the DNS data from a periodic box, in which all turbulent scales are well-resolved, the spectra computed using equation~\ref{eq:energy_spectrum_function_using_uk_dft} shown in Figure~\ref{fig:energy_spectrum_demo} are in excellent agreement with what we expect from continuum theory for HIT. 

In the case of experimental data in which the resolution is limited on both ends (the voxel/pixel spacing at the low end and the imaging region size at the high end), a number of artifacts can affect the estimation of $E(\kappa)$. 
We discuss these, and simple methods to correct their effects in the following sections using $E(\kappa)$ as an example (all considerations apply equally to $E_{ij}(\kappa_1)$).

Despite these potential artifacts, obtaining spatially resolved data using PIV/PTV and computing the spectrum via the Fourier method has several advantages over  canonical methods used in many classical turbulent fluctuation measurements that are based on the point-wise  hot-wire anemometry and Laser Doppler velocimetry measurements: 
\begin{enumerate}
    \item One can directly compute the energy spectrum from a single velocity field snapshot.
    
    \item This method does not require Taylor's frozen turbulence hypothesis, and can be applied to flows without a strong mean flow. 
    
    \item An instantaneous planar velocity field is sufficient to acquire the energy spectrum if  the field is isotropic. (Figure \ref{fig:energy_spectrum_demo}b) 
    
    \item This method is more efficient than computing the two-point velocity correlation function $R_{ij}$ thanks to Fast Fourier Transform (FFT). The temporal complexity to compute $R_{ij}$ is $O(n^2)$, whereas FFT takes $O(n\log{n})$, offering a faster turnaround.
\end{enumerate}

\subsection{Effect of domain truncation and noise}
Experimentally obtained velocity fields are in general truncated, and contain noise. 
In this subsection we elucidate the effect of the domain truncation and noise on the energy spectrum by  truncating and adding noise to the same DNS data as used in Figure~\ref{fig:energy_spectrum_demo},  computing spectra of the modified data and comparing the resulting spectrum to the original.

\begin{figure}[htbp]
\centering
\includegraphics[width=\textwidth]{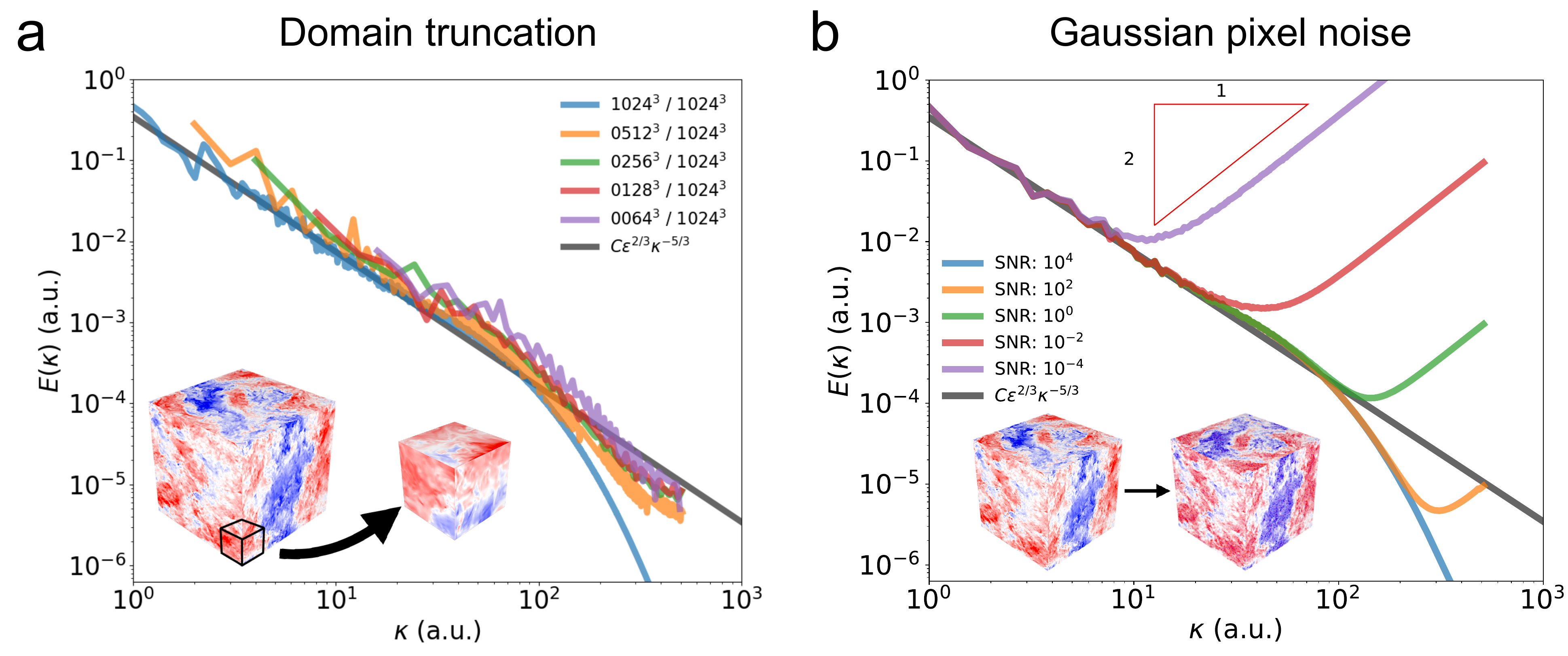}
\caption{\textbf{Effects of domain truncation and gaussian noise on the energy spectrum} (a)Domain truncation- the loss of periodicity leads to overestimate the spectral contribution at high wavenumber due to
aliasing and spectral leakage. (b) Gaussian noise- The noise becomes dominant at high wavenumber, depending on the signal-to-noise ratio. The power at high wavenumber depends on the Jacobian and the wavenumber dependence of noise. When the latter can be ignored, the dependence becomes $\kappa^2$ for 3D DFT and  $\kappa$ for 2D DFT.}
\label{fig:energy_spectrum_truncation_gauss_noise}
\end{figure}

Figure~\ref{fig:energy_spectrum_truncation_gauss_noise} shows the 3D energy spectrum obtained by truncating the domain and computing $E(\kappa)$ using Equation~\ref{eq:energy_spectrum_function_using_uk_dft}. 
The spectrum agrees with the one computed in the full domain at low wave-numbers but differs at high wavenumbers. 
This can be rationalized by considering the effect of a discrete Fourier transform on a finite domain. For an instantaneous velocity field  $u_i(\vec{x}_n)$, the truncated velocity field is given by $u^{\rm truncated}_i(\vec{x}_n) = u_i(\vec{x}_n)G(\vec{x}_n)$ where $G(\vec{x}_n)=1$ inside the (measured) domain, and 0 otherwise. 
Its discrete Fourier transform is then given by the convolution of $\breve{u}_i(\vec{\kappa}_n)$ with $\breve{G}(\vec{\kappa}_n)$.
For example, consider $u_i(x_n)=\sin(\kappa_0 x)$ and a rectangular window with length $\ell \neq 2m\pi/\kappa_0$ ($m\in \mathbb{Z}$). The power spectrum of the DFT $\breve{u}_i(\vec{\kappa}_n)$ has a broad peak centered at $\kappa=\kappa_0$ and non-zero values in its neighborhood (`leakage'). This is a consequence of the convolution with $\breve{G}(x_n)$ that is a sinc function in the Fourier space.

The degree of the leakage in the Fourier space depends on both the signal and the window function. Hence it is important to study how the expected spectrum is affected by a rectangular window. In the case of a turbulent energy spectrum, the `leakage' leads to overestimation at high wavenumber (Figure \ref{fig:energy_spectrum_truncation_gauss_noise}a).
Note that in Figure \ref{fig:energy_spectrum_demo}a-b, the spectral leakage does not occur because the considered velocity field comes from a periodic DNS simulation and the entire domain was included.

To investigate the effect of the noise on the energy spectrum, we add Gaussian noise with its magnitude scaled by the speed at each position $u^{\rm noise}_i(\vec{x}) = u_i(\vec{x}) ( 1 +  f(\mu=0, \sigma))$ where $f(\mu, \sigma)$ is drawn from a Gaussin distribution. Figure \ref{fig:energy_spectrum_truncation_gauss_noise}b shows that the noise dominates above a certain wavenumber that depends on the signal-to-noise ratio (SNR$\equiv$Var[$u_i$]/Var[$u_if(0, \sigma)$]). The exponent depends on the dimension of performed DFT, and the wavenumber-dependence of noise. When the latter can be ignored, the dependence becomes $\kappa^2$ for 3D DFT and $\kappa$ for 2D DFT.

In summary, the spectral leakage and noise lead to the overestimation of the spectral density at high wavenumber as shown by the light blue curve in Figure \ref{fig:energy_spectrum_guide}. For example, \cite{foucaut_piv_2004, atkinson_appropriate_2013} report an energy spectrum with this trait. Improving the image qualities and using the appropriate PIV parameters (`quarter rule'\cite{raffel_particle_2018}) may reduce the noise\cite{westerweel_particle_2013}. In the next section, we shall see that a procedure called `windowing' mitigates the spectral leakage.

\subsection{Windowing and bumping}\label{sect:windowing}
Removing the spectral leakage is not possible because the inverse of a convolution operator does not exist; however, a process called `windowing' can mitigate the leakage by enforcing periodicity to the signal.

\begin{figure}[htbp]
\centering
\includegraphics[width=\textwidth]{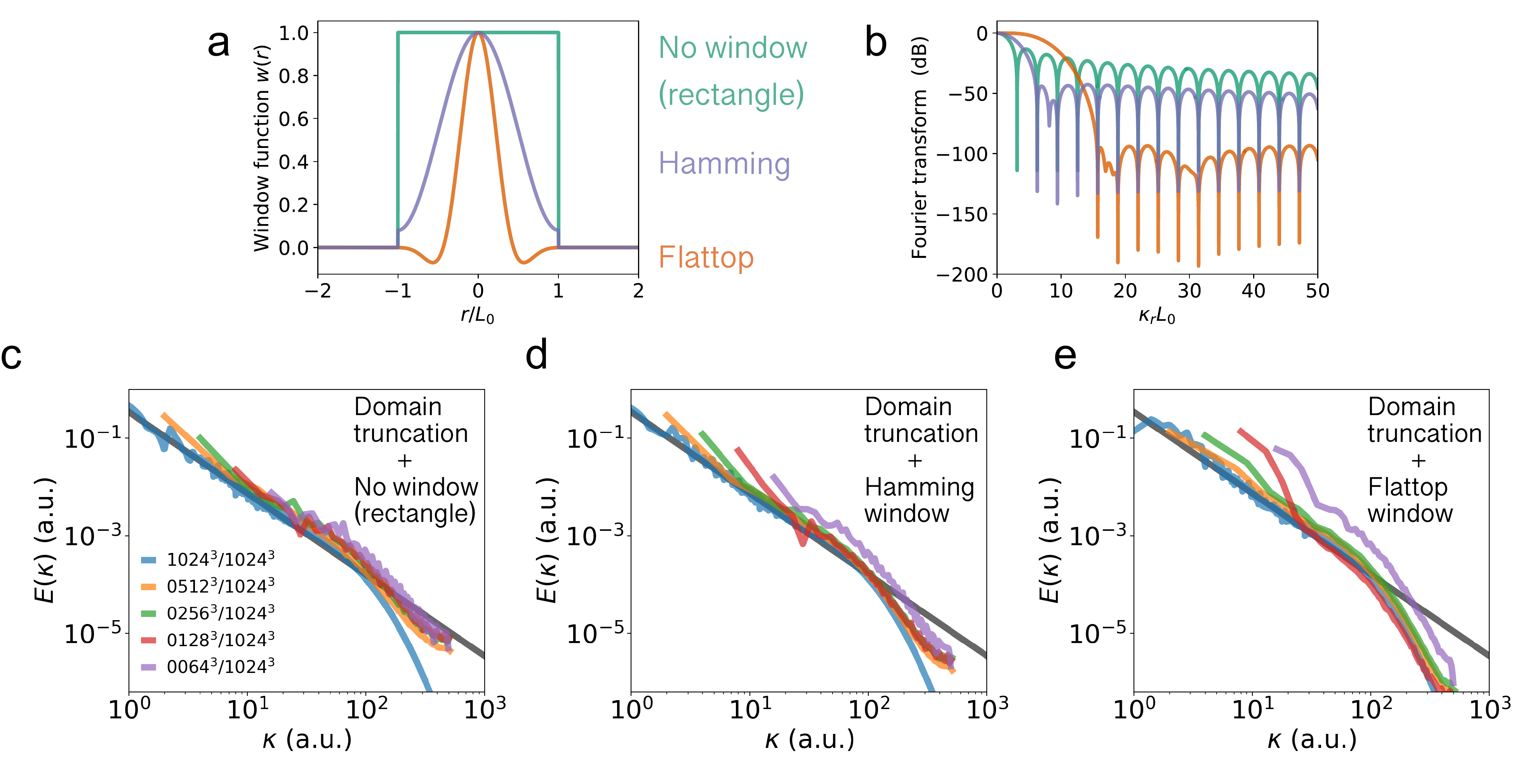}
\caption{\textbf{Windowing mitigates spectral leakage originated from the domain truncation.}(a) Amplitude of the rectangle, hamming, and flattop windows are shown. Multiplying the rectangle window to the true velocity field results in a truncated velocity field with width $L_0$. (b) The Fourier transforms of the considered window functions reveal the degree of spectral leakage. The rectangular window has high spectral resolution but a low dynamic range (strong leakage), whereas the flattop window has the opposite character. The hamming window has an intermediate characteristics.
 (c) Effect of truncating a velocity field with no windowing (d) Applying the hamming window to the truncated velocity field mitigates the leakage due to the domain truncation. The shown spectral density is bumped by a correction factor $\zeta$. (e) Same as (d) but with the flattop window.}
\label{fig:energy_spectrum_windowing}
\end{figure}

The truncated velocity fields are first multiplied by the window function before taking the Fourier transform. The optimal choice of the window function depends on the underlying spectrum. In general,  wideband windows are suited for signals with a high dynamic range like the turbulent energy spectrum but offer less sensitivity, whereas the narrowband windows such as a rectangular window have a low dynamic range but high sensitivity. The dynamic range and sensitivity cannot both be maximized simultaneously. Therefore, it is critical to test the effect of different windows on the expected spectrum to make an informed decision. Figure~\ref{fig:energy_spectrum_windowing}a-b shows rectangular (narrowband), hamming (intermediate), flattop (wideband) windows in the real and Fourier space. The wider the main lobe becomes, the broader a peak becomes in the Fourier space. The decay rate of the side lobe peaks determines the strength of the leakage. 
When we compute the energy spectrum of the windowed velocity field, we multiply the spectrum computed using  Eq.~\ref{eq:energy_spectrum_function_using_uk_dft} by the correction factor $\zeta$  (bumping) to compensate for the loss of the signal (velocity) by windowing.
\begin{align}
    \zeta = \frac{\int_\mathcal{V} u_i({\vec{x}}) u_i({\vec{x}}) d{\vec x} }{\int_\mathcal{V} w^2({\vec{x}}) u_i({\vec{x}}) u_i({\vec{x}}) d{\vec x}}
    \label{eq:windowing_factor}
\end{align}
Here, $w({\vec{x}})$ is a window function. Figure\ref{fig:energy_spectrum_windowing}(c-e) shows the energy spectra with no window, a flattop, and a hamming window respectively. The flattop window enables to resolve the dissipation range thanks to its high dynamic range. 

\subsection{PIV algorithm and its nature as a low-pass filter}
In this section, we demonstrate that it is possible to obtain an energy spectrum that is consistent with the ground truth via a `synthetic PIV' measurement that mimics  ideal PIV conditions. 
The observed deviations can be explained by considering  spectral leakage, and noise (Figure \ref{fig:energy_spectrum_truncation_gauss_noise}).  
We further compare the effects of the PIV interrogation window size to  a low pass filter\cite{willert_digital_1991}, and find that the effects of PIV interrogation window size are similar to those of box filtering. 

\begin{figure}[htbp]
\centering
\includegraphics[width=\textwidth]{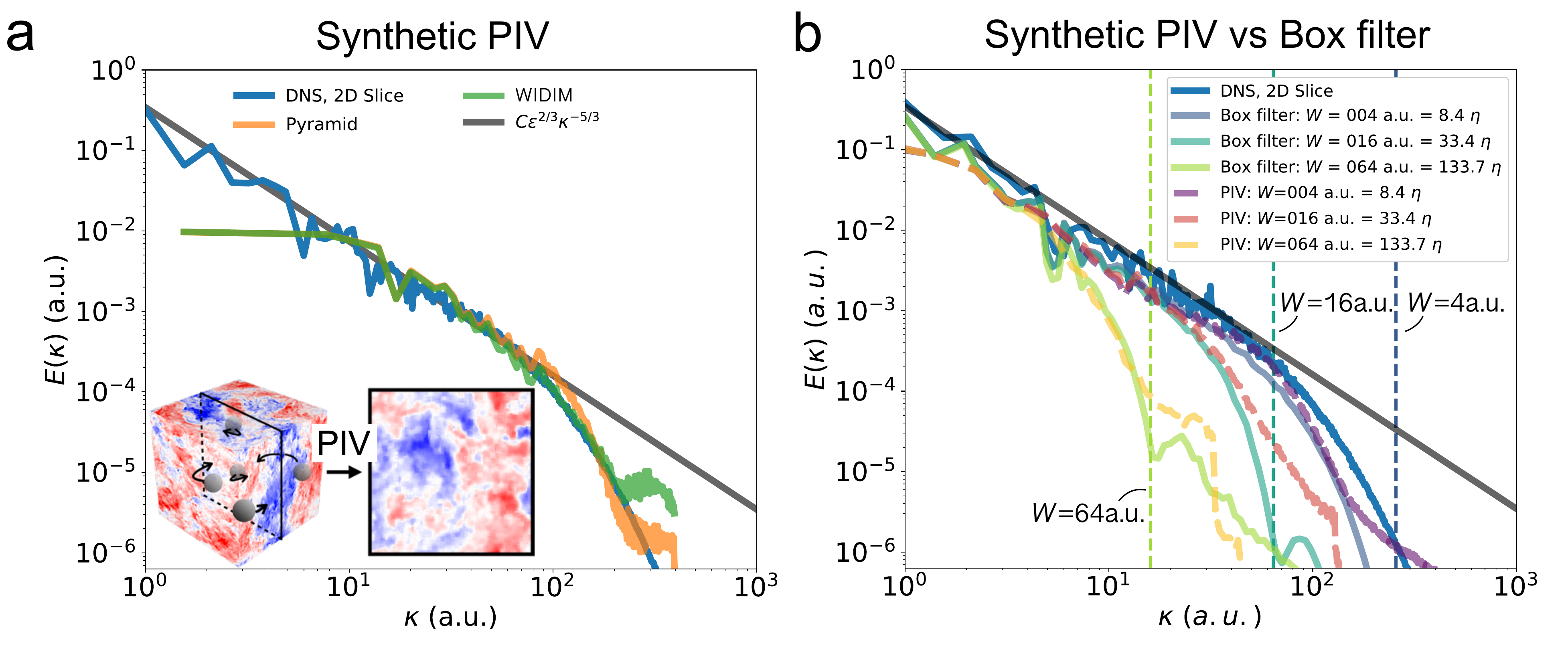}
\caption{\textbf{Energy spectrum of a velocity field extracted by running PIV on synthetic data} (a) Given images with highly seeded particles and zero noise, the pyramid algorithm (orange)~\cite{sciacchitano_multi-frame_2012} yields a more resolved energy spectrum than the standard cross-correlational algorithm (WIDIM, green). (b)PIV operation is comparable to box filtering with its kernel size as same as the interrogation window $W$. The energy spectrum of the PIV-generated field (algorithm: pyramid) deviates from the ground truth above $\kappa=\pi / W$.}
\label{fig:energy_spectrum_piv_effect}
\end{figure}

To assess the spectrum of the PIV-generated velocity fields, we created synthetic images of particles advected by a turbulent field from JHTD data. The field is resolved spatially and temporarily with a periodic boundary condition. With the optimized images and PIV settings (particle-per-pixel: 0.012,  $L$=2048px, total number of particles per image: $5\times 10^4$,  initial interrogation window size: 32px, final interrogation window size $W=4\eta$=4px), it is indeed possible to extract the accurate three-dimensional energy spectrum from the PIV-generated field (Figure \ref{fig:energy_spectrum_piv_effect}a). 
In terms of the image quality, we find that images with high seeding density (ppp$\approx$ 0.01) is critical. The temporal spacing between the adjacent frames should be less than 0.5 Kolmogorov time scales. As for the PIV settings, the smallest interrogation window in the multi-pass algorithm should be comparable to the Kolmogorov length scale. Further discussion on the required criteria to resolve a turbulent field can be found in \cite{westerweel_particle_2013, raffel_particle_2018}. 
We tested two PIV algorithms (1. the standard cross-correlation algorithm called WIDIM~\cite{scarano_iterative_2002} 2. Pyramid algorithm~\cite{sciacchitano_multi-frame_2012}), and observed that the pyramid algorithm extracts the small-scale motion better than the WIDIM; the pyramid algorithm extends the band of accurate measurement by 50\% in the wavenumber space, compared to WIDIM (Figure \ref{fig:energy_spectrum_piv_effect}a). The obtained spectra are in agreement with the reference up to the detection limit. 

Any interrogation-based PIV acts as a low-pass filter on the true velocity field~\cite{willert_digital_1991}. This is because the fundamental operation of PIV is peak finding of a cross-correlation map inside the interrogation windows. 
It assigns essentially an average displacement vector inside the interrogation window. Hence, the dimension of the interrogation window ($W_x \times W_y \text{px}^2$) sets the smallest motion PIV could detect. This effect could be modeled by a $W_x \times W_y$ box filter (Figure \ref{fig:energy_spectrum_piv_effect}b). The actual PIV operation is more complex than box filtering; however, we find that the spectra of the PIV-generated fields are similar to that of the box-filtered field. If not compensated for, this effect could lead turbulence studies using PIV to report a narrower inertial subrange than the truth. Understanding this effect is therefore critical to correctly estimate the $\kappa$ dependence on the energy spectrum. The same low-pass filtering is expected for  3D PTV if the Eulerian field is reconstructed by binning the Lagrangian velocities into voxels. 

Figure \ref{fig:energy_spectrum_piv_effect}b suggests that the any motion smaller than the interrogation is not resolved by PIV. Based on this result, we conclude that the energy spectrum above $\kappa \approx \pi / W$ is significantly underestimated. In main Figure 3f and Figure~\ref{fig:energy_spectrum_guide}, unfilled markers indicate the underestimation  due to this effect of PIV. 

\subsection{Computation of structure function}
The structure function $D_{ij}$ possesses equivalent information as the one-dimensional, energy spectra. With homogeneity and isotropy, it is defined as
\begin{align}
    D_{ij}(r) = \langle \delta u_i \delta u_j \rangle
\end{align}
with $\delta u_i(\vec{x}, \vec{r}) = u_i(\vec{x} + \vec{r}) - u_i(x)$. The longitudinal, structure function is related to the one-dimensional, longitudinal energy spectrum. For example, $D_{11}(r\hat{e_1}) = 2\int_0^\infty E_{11}(\kappa_1) [1- \cos{(\kappa_1 r)}] d\kappa_1$. 

We compute $\delta u_i$ at every possible point on the 2D velocity field of a set of snapshots, then take an average over time and space. The number of available samples $\delta u_i$ with a separation distance $r$ depends on its position $\vec{x}$. Take a square velocity field as an example. The number of available samples with $r=$(diagonal distance of the square) is much more limited than small $r$. To ensure the same statistical weight at every $r$, we randomly sample $m$ pairs for each $r$ to compute $D_{ij}(r)$.
We also make sure that each snapshot is statistically independent by only considering the snapshots that are temporally separated by integral time scale $\tau_\mathcal{L}= \mathcal{L}/u'$. 

By definition, the structure function does not face the undesired effects in the spectral space as explained in the previous sections. It is, however, susceptible to the low-pass character of the PIV operation. Hence, we apply the same criteria as the energy spectrum, and consider that the structure function values at a scale below $r \approx 2W$ are significantly underestimated. In main Figure 3g and Figure \ref{fig:energy_spectrum_guide}, unfilled markers indicate the underestimation due to PIV.

\subsection{Energy spectrum over a region of interest}\label{sect:Foundation of the Local Energy Spectrum Estimate}
Eq. \ref{eq:energy_spectrum_function_using_uk_dft} and \ref{eq:1d_energy_spectrum_function_using_uk_dft} provide means to compute the energy spectrum over a region of interest (ROI). The observable band $[\kappa_{min}, \kappa_{max}]$ is limited by the sampling frequency of the data and the size of the ROI. These \textit{local} energy spectra do not necessarily match a part of the energy spectra of the entire flow. To match these two spectra, it is necessary that the turbulent field is homogeneous. 

In the case of inhomogeneous turbulence, the global energy spectra represent a mixture of different statistical states, and deviate from the results of HIT. In our experiments, the flow induced by the repeated collisions of vortex rings consists of a homogeneous, isotropic turbulent core and comparatively quiescent surroundings. Therefore, we present the \textit{local} energy spectrum over a rectangular (or cuboidal for 3D PTV) region of the homogeneous, turbulent core in Figure 3f.

\subsection{3D energy spectrum derived from  3D PTV data}\label{sect:3d_energy_spectrum}
\begin{figure}[!htb]
\centering
\includegraphics[width=0.99\textwidth]{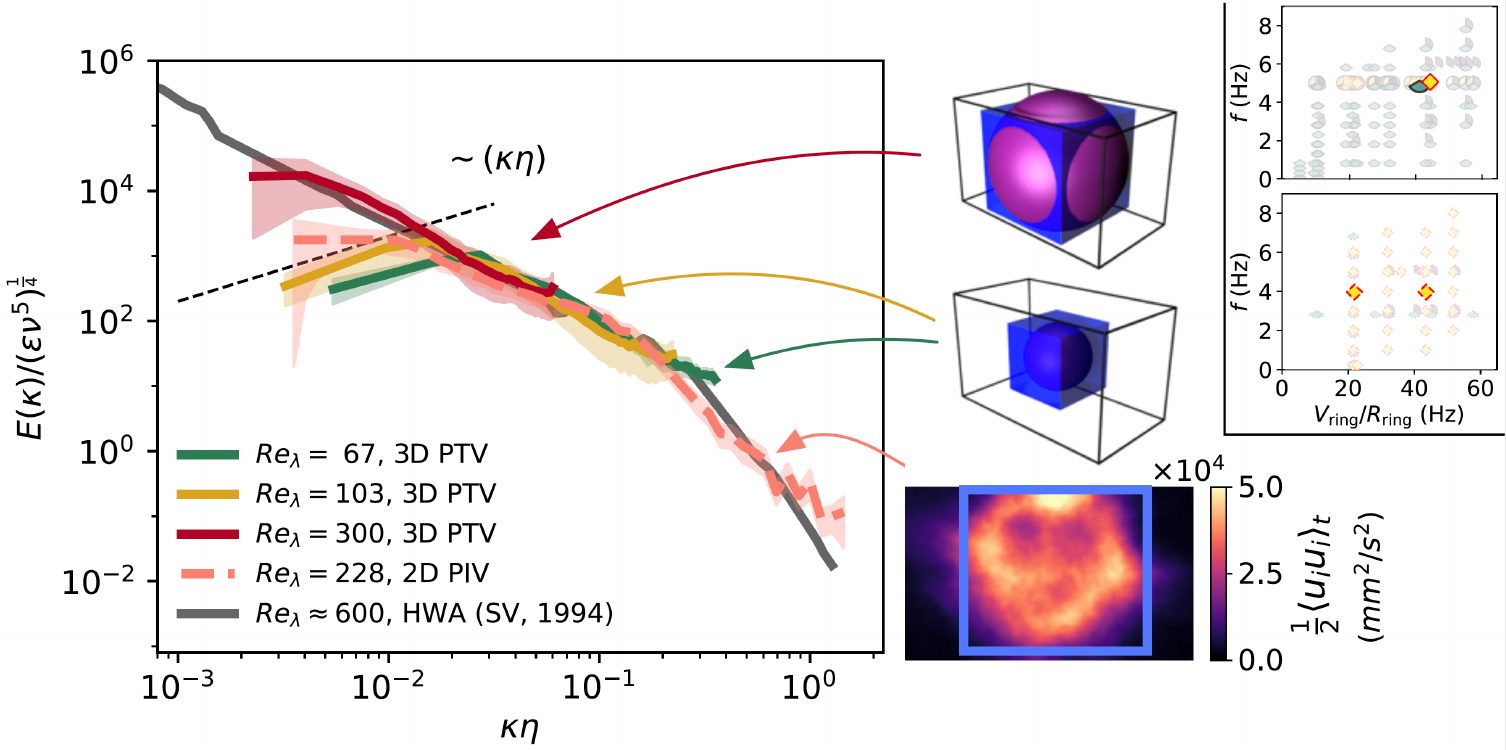}
\caption{\textbf{3D energy spectra of velocity fields obtained by 3D PTV and 2D PIV.} The blue cubes or square on the right represent the domain that the DFT is conducted (Eq.~\ref{eq:energy_spectrum_function_using_uk_dft}). The turbulent core ($r=R_{\rm blob}$) is represented by the purple sphere. }
\label{fig:energy_spec_3d}
\end{figure}

We report the energy spectrum of a fluctuating velocity field measured by 3D PTV. 
The extraction of all three components of velocity vectors enables the computation of the spectrum directly via Eq. \ref{eq:energy_spectrum_function_using_uk_dft}. 
Figure \ref{fig:energy_spec_3d} shows the energy spectrum of the central region of our turbulent blob.  
The temporal average of the spectrum is shown here as the turbulence inside the blob is statistically steady with weak dependence on the phase of forcing. 
We compare it to the corresponding spectrum from ~\cite{saddoughi_veeravalli_1994} by converting their 1D energy spectrum to $E(\kappa)$ via $E(\kappa)= \frac{1}{2} \kappa^3 \frac{d}{d\kappa} \left( \frac{1}{\kappa} \frac{d E_{11}(\kappa)}{d \kappa}\right)$. All energy spectra with three $Re_\lambda$ exhibit the Kolmogorov scaling in the inertial subrange. 
Furthermore, the rescaled spectra obtained by 3D PTV and 2D PIV are consistent with the universal function reported in~\cite{saddoughi_veeravalli_1994}. As discussed in the main text, both 3D PTV and 2D PIV are limited by the image resolution and the finite field of view. The size of the interrogation window determines the upper bound of the valid region of the spectrum ($\kappa \eta \leq \pi \eta/ W$). 
If the integral scale were greater than the field of view, the spectrum could not resolve the energy containing range. 
This can be seen with the curve of ($Re_{\lambda}=300$, 3D PTV) in  Figure \ref{fig:energy_spec_3d}. The same effect is observed for ($Re_{\lambda}=228$, 2D PIV) since the domain of analysis was restricted to the inner part of the energetic region($r\leq R_{\rm blob}$). 

We note that the geometry of the  velocity field obtained by 3D PTV is in general a cuboid and not a cube, reflecting the differences between sensor dimensions and resolution  and the maximum depth that 3D PTV accurately measures. A difference in the length of the data in each dimension affects the quality of DFT. This is  because three sinc functions with different frequencies are convoluted in  $\kappa$ space. 
If left uncorrected, this results in a wavy behavior of the energy spectrum  at high $\kappa$. 
To mitigate this effect, we pad the velocity field with  to make the shape of the array cubic before taking the DFT. Zero-padding simply extends the signal, and interpolates it in the wavenumber domain; however, cubing the array allows the spectral leakage to occur at the same frequencies along all the three directions, making the spectrum more readily interpretable.

\clearpage
\section{Energy balance}
\label{sect:energy_balance}
In this section we examine the energy balance of a turbulent blob in detail. We shall quantify how much dissipation occurs inside the blob, and identify its relation with the injected power.
The kinetic energy $ \mathscr{E}=\frac{1}{2} U_i U_i$ of a viscous flow obeys:

\begin{align}
    (\partial_t + U_i \partial_i) \mathscr{E} + \partial_i T_i = 2\nu S_{ij}S_{ij}
    \label{eq:energy_transport_eq}
\end{align}
where $S_{ij}=\frac{1}{2}(\partial_j U_i + \partial_i U_j)$ is the rate-of-strain tensor, and $T_i=U_i p /\rho - 2\nu U_j S_{ij}$ is the energy flux~\cite{pope_turbulent_2000}. 
The Reynolds decomposition $U_i=\langle {U}_i \rangle_t + u_i$ yields
\begin{align}
    (\partial_t + \langle U_i \rangle_t \partial_i)\langle \mathscr{E} \rangle_t + \partial_i (u_i \mathscr{E} + \langle T_i \rangle_t )  = - \overline{\epsilon} - \epsilon
    \label{eq:mean_energy_transport_eq}
\end{align}
where $\overline{\epsilon} = 2 \nu \langle \overline{S}_{ij}\overline{S}_{ij} \rangle_s$ with $\overline{S}_{ij}=\frac{1}{2}(\partial_j \langle U_i \rangle_t + \partial_i \langle U_j \rangle_t)$ is the mean rate of strain. Eq.~\ref{eq:mean_energy_transport_eq} reveals the decomposition of the energy dissipation in the mean and the fluctuating flows. In the case of a turbulent blob, the mean flow consists of eight vortex rings that repeatedly travel towards the center (See Figure 3a in the main text). On the $\sigma_h$ plane, the enstrophy measurements (Figure \ref{fig:fluctuation}f) indicate $\epsilon/\overline{\epsilon}=\omega^2/\overline{\Omega}^2=O(10^2)$. Below we assume $\overline{\epsilon} \ll \epsilon$ anywhere in the system. 

\subsection{Partition of dissipation in a turbulent blob}
The energy dissipation within a domain $\mathcal{V}$ that encloses a blob is 
\begin{align}
    \mathcal{D}(r/R_{\rm blob}) = \rho \int_{\mathcal{V}} \epsilon (\vec{r}) dV 
    \label{eq:energy_balance0}
\end{align}
where $\epsilon (\vec{r}) \equiv 2\nu \langle s_{ij}s_{ij} \rangle_t$ is a spatially varying dissipation rate. 
The radial profile of the local dissipation rate (or enstrophy density equivalently) shows that $\epsilon (\vec{r})$ depends only on the radial distance, is constant up to a radius $R_{\rm blob}$, and decays approximately as $\epsilon(r) \propto r^{-4}$ for $r\geq R_{\rm blob}$. 

\begin{align}
  \epsilon(r) =
        \begin{cases}
            \epsilon_0 & (r<R_{\rm blob}) \\
            \epsilon_0 (\frac{R_{\rm blob}}{r})^4 &  (r\geq R_{\rm blob}) 
        \end{cases}
    \label{eq:dissipation_model}
\end{align}

It is useful to compute the dissipation within a sphere of radius $r$ for a turbulent blob with this profile. The fraction of energy dissipation inside the sphere to the total is 
\begin{align}
   p(r/R_{\rm blob}) =
        \begin{cases}
             \frac{1}{4} \left( \frac{r}{R_{\rm blob}} \right)^3 & (r<R_{\rm blob})\\ \\
            1-\frac{3R_{\rm blob}}{4r} & (r\geq R_{\rm blob}).
        \end{cases}
    \label{eq:dissipation_fraction_model}
\end{align}
If we include a cutoff in the considered integral because the chamber size is finite, the expression becomes 
\begin{align}
   p(r/R_{\rm blob}, (L_{box}/2)/R_{\rm blob}) =
        \begin{cases}
             \frac{1}{4} \left( \frac{r}{R_{\rm blob}} \right)^3  (1-\frac{3R_{\rm blob}}{2L_{box}})^{-1} & (r<R_{\rm blob})\\ \\
            (1-\frac{3R_{\rm blob}}{4r}) (1-\frac{3R_{\rm blob}}{2L_{box}})^{-1} & ( R_{\rm blob} \leq r \leq \frac{L_{box}}{2}).
        \end{cases}
    \label{eq:dissipation_fraction_model_bound}
\end{align}
Note that fraction of energy inside the blob to the whole depends not only the blob radius but also its respective size to the system. The same function applies to the amount of energy and enstrophy inside the sphere with radius $r$. 

Figure \ref{fig:model_enclosed_energy}(a) shows the model function (Eq.~\ref{eq:dissipation_model}) and the scaled cummulative function $p$ (Eq. \ref{eq:dissipation_fraction_model_bound})for $R_{\rm blob}=\sqrt{6}R$. The value of $R_{\rm blob}$ with respect to the ring radius $R$ is taken from the experimental data (Figure 3b). There is a geometrical interpretation of $R_{\rm blob}=\sqrt{6}R_{\rm ring}$. The symmetrical configuration of the eight vortex rings defines an octahedron as explained in Sect{\ref{sect:gpe}. Since the vortex atmosphere is ellipsoidal, the majority of the energy is housed inside a sphere as large as the circumscribed sphere of the octahedron which has a radius of $\sqrt{6}R_{\rm ring}$. One can also imagine a polyhedron such that eight spheres with radius $R_{\rm ring}$ fused together on each face of an octahedron. The radius of such a circumscribed sphere is $(1+\sqrt{2})R_{\rm ring}\approx2.41R_{\rm ring}$, sufficiently close to $\sqrt{6}R_{\rm ring}\approx2.45R_{\rm ring}$.

\begin{figure}[htbp]
\centering
\includegraphics[width=\textwidth]{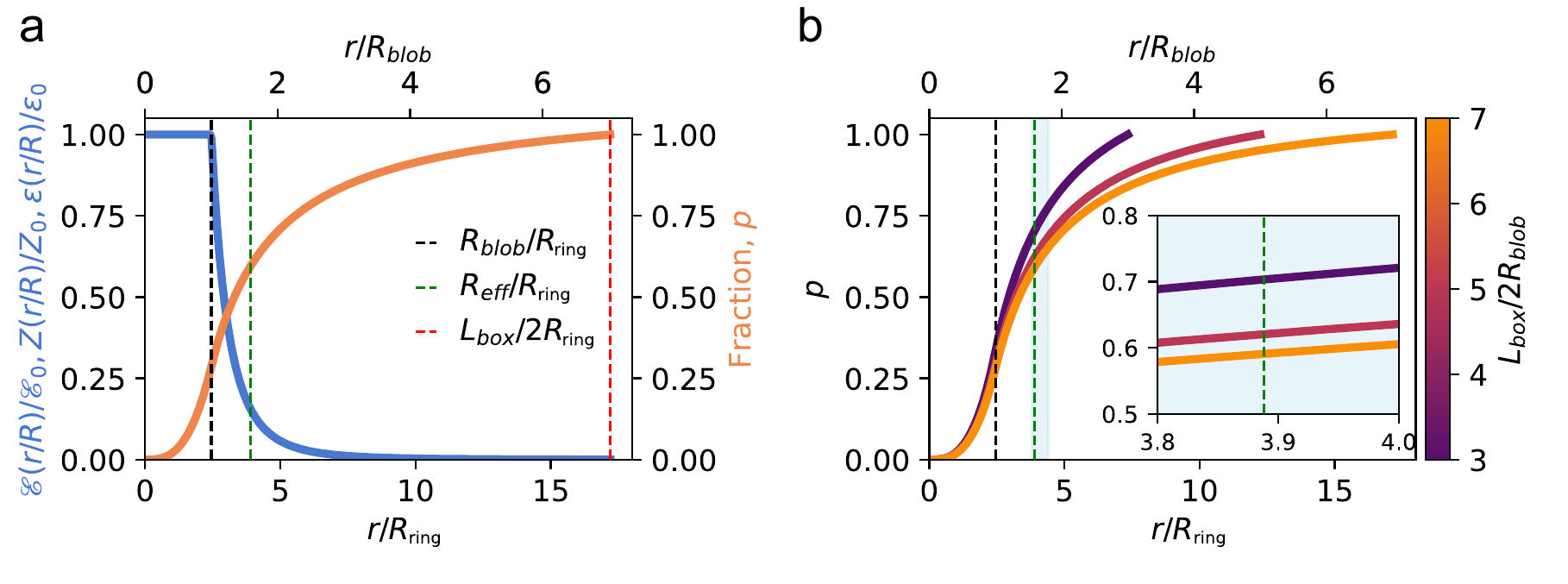}
\caption{\textbf{Model radial distribution of energy $\mathscr{E}$, enstrophy $Z$, and dissipation rate $\epsilon$.} (a) Radial distribution function (blue) and scaled cummulative function $p(r) $(orange) of energy/enstrophy/dissipation rate are shown for a case $R_{\rm blob}=\sqrt{6}R_{\rm ring}$ and $(L_{\rm box}/2)/R_{\rm blob}=7$. (b)Dependence of $p$ on the blob radius $R_{\rm blob}$ and the system size $L_{\rm box}$ is shown.}
\label{fig:model_enclosed_energy}
\end{figure}

The total dissipation is
\begin{align}
    \mathcal{D}_{tot} &= \frac{16}{3}\pi  \rho \epsilon_0 R_{\rm blob}^3 = \frac{4}{3}\pi  \rho \epsilon_0 R_{\rm eff}^3
    \label{eq:total_dissipation}
\end{align}
where $R_{\rm eff}=4^{\frac{1}{3}}R_{\rm blob}\approx1.59R_{\rm blob}$ is the effective radius of the blob. 
The dissipation within a sphere of radius $r$ is given by
\begin{align}
    \mathcal{D}(r/R_{\rm blob}) &= p(r/R_{\rm blob}) \mathcal{D}_{tot}.
    \label{eq:energy_balance_lhs_end}
\end{align}

\subsection{Estimation of dissipation rate}\label{sect:estimation of dissipation rate}
Measuring the dissipation rate robustly with PIV is a challenge~\cite{de_jong_dissipation_2008}. Multiple methods have been proposed, including direct computation from the rate-of-strain tensor to fitting to the structure function; however, they are often susceptible to the choice of interrogation window size and the random noise. ~\cite{de_jong_dissipation_2008} investigated five methods to estimate the dissipation rate from the PIV data of zero-mean, gaseous turbulence: (1) Direct method via the rate-of-strain tensor (2) Large eddy method based on the modeling in the subgrid scale (3) Method based on the scaling $\epsilon_0=A\frac{u'^3}{\mathcal{L}}$ (4) Fitting the energy spectrum to the Kolmogorov spectrum (5) Fitting the n-th order structure function to $(\epsilon_0 r)^{n/3}$. Their investigation on the sources of errors led to conclude that the method using the second-order structure function yields the most reliable estimate. Following their findings, we estimate the dissipation rate using the second-order structure function. To check the robustness of our measurements, we also computed the dissipation rate with the direct method and by fitting the energy spectrum to the universal curve. For the direct method, we compute
\begin{align}
    \epsilon_{s_{ij}}=6\nu \langle (\partial_1 u_1)^2 + (\partial_2 u_2)^2 + (\partial_2 u_1 ) (\partial_1 u_2) \rangle,
    \label{eq:dissipation_rate_direct_isotropic}
\end{align}
assuming isotropy ~\cite{hinze_turbulence_1975}. For the structure function method, we fit the longitudinal second-order structure function $D_{LL}(r)=\langle \delta u ^ 2\rangle  / \langle u_i ^2\rangle$ to $C_2 (\epsilon_{D_{LL}} r)^{2/3}$ where $\delta u(\vec{x}, r) = u_i(\vec{x}+r\hat{x}_i) - u_i(\vec{x})$, and $C_2=2.1$~\cite{saddoughi_veeravalli_1994}. For the spectral method, we fit the one-dimensional energy spectrum $E_{11}(\kappa_1)$ to $C_1 \epsilon^{2/3} \kappa_1^{-5/3}$ with $C_1=0.52$ ~\cite{sreenivasan1995universality}. Numerically, we look for the plateau of the compensated structure function $1/r (D_{LL}/ C_2)^{3/2}$ and the compensated energy spectrum $[E_{11}/( C_1 \kappa_1^{-5/3} ) ]^{3/2}$. This way, we avoid complications to identify the inertial subrange for each data set to fit. In addition, before applying the above explained methods, the PIV-extracted field is convoluted with a median filter (with a kernel size of 3x3 or 5x5 px$^2$) to remove spurious vectors.

\subsection{Partition of injected energy}
The energy balance over a domain $\mathcal{V}$ that encloses a blob is 
\begin{align}
  \mathcal{I} \approx 8K_{\rm ring}f
    \label{eq:injected_power}
\end{align}
because we expect only the energy within the vortex atmosphere is fed to sustain the blob. If there is only one vortex ring in a flow, the ratio of energy enclosed within the atmosphere to the total $c=K_{\rm ring}/K$ is completely determined by the shape of the vortex atmosphere (see Box 3), and remains nearly the same $c\approx 77\%$ for the ellipsoidal atmosphere. 
For the two canonical vortex ring models, the injected power $\mathcal{I}$ is given by
\begin{align}
    \mathcal{I} = 
    \begin{cases}
        4c \rho \Gamma_{\rm ring}^2 R_{\rm ring} f \left[ \ln{\frac{8R_{\rm ring}}{a} - \alpha}\right]& \text{(thin-cored vortex ring)} \\
        \frac{64\pi}{105} c \rho \Gamma_{\rm ring}^2R_{\rm ring} f & \text{(Hill's spherical vortex)}.
    \end{cases}
    \label{eq:energy_balance_rhs_end}
\end{align}
The coefficient represents the difference between the models. The generated vortex rings in this study has coefficients between these two models(Figure \ref{fig:vring_energy_comparison}); however, it depends on the experimental conditions ($L_*, v_{eff}, D_o, D_p$). Thus we directly measure the energy of the injected vortex rings from 2D PIV data shown in Figure \ref{fig:data_points_vring}.

\subsection{Balance of energy injection and dissipation}
We first turn our attention to the total energy balance ($\mathcal{D}_{\rm tot}$ vs $\mathcal{I}_{\rm tot}=\mathcal{I}/c$) in the chamber. Regardless of the methods to estimate the dissipation rate, the total dissipated power monotonically increases with the injected power over almost two decades (Figure \ref{fig:dissipated_vs_injected_energy_whole_space}e-g). The method using the rate-of-strain tensor tends to overestimate the dissipation, compared to the methods involving the energy spectrum and the structure function. This is consistent with ~\cite{de_jong_dissipation_2008} as the noise in the measured velocity field contributes to this overestimation. Removing the noise with a median filter (Figure \ref{fig:dissipated_vs_injected_energy_whole_space}a-c) leads the estimates by all three methods to agree with each other(Figure \ref{fig:dissipated_vs_injected_energy_whole_space}d-f). This gives us more confidence that the filtered velocity field gives the more accurate estimate of the dissipation rate than the raw field. There, the dissipated power is indeed proportional to the injected power. We find $\mathcal{D}_{\rm tot}/\mathcal{I}_{\rm tot}=1.07\pm 0.09$ (3x3(=$5.6\eta \times 5.6\eta$)), $0.64\pm 0.05$(5x5=$9.3\eta \times 9.3\eta$) for the median-filtered fields. The attenuation with a larger kernel is expected as it blurs the velocity field more.

As discussed in the main text, the far-field contribution in $\mathcal{I}_{tot}$ should be neglected to assess the energy balance of the blob. The ratio
\begin{align}
    \frac{\mathcal{D}(\frac{r}{R_{\rm blob}})}{\mathcal{I}} = \frac{\mathcal{D}_{\rm tot}}{\mathcal{I}_{\rm tot}} \frac{p(\frac{r}{R_{\rm blob}})}{c}
    \label{eq:dissipation_efficiency_in_blob}
\end{align}
represents how much injected power via vortex rings is dissipated as a function of a (scaled) radius. In a sense, this gives a conversion rate from coherent vortex motion to turbulence in terms of energy. Considering $\frac{\mathcal{D}_{\rm tot}}{\mathcal{I}_{\rm tot}}\approx 1$, the power associated to the vortex ring atmosphere is completely dissipated within a sphere of radius $r=\frac{3}{4(1-c)}R_{\rm blob}\approx3.3R_{\rm blob}$ with $c=0.77$. Of which, 33\% is dissipated up to $r=R_{\rm blob}$ and 68\% is dissipated up to $r=R_{\rm eff}=4^{\frac{1}{3}}R_{\rm blob}$.

\begin{figure}[htbp]
\centering
\includegraphics[width=1\textwidth]{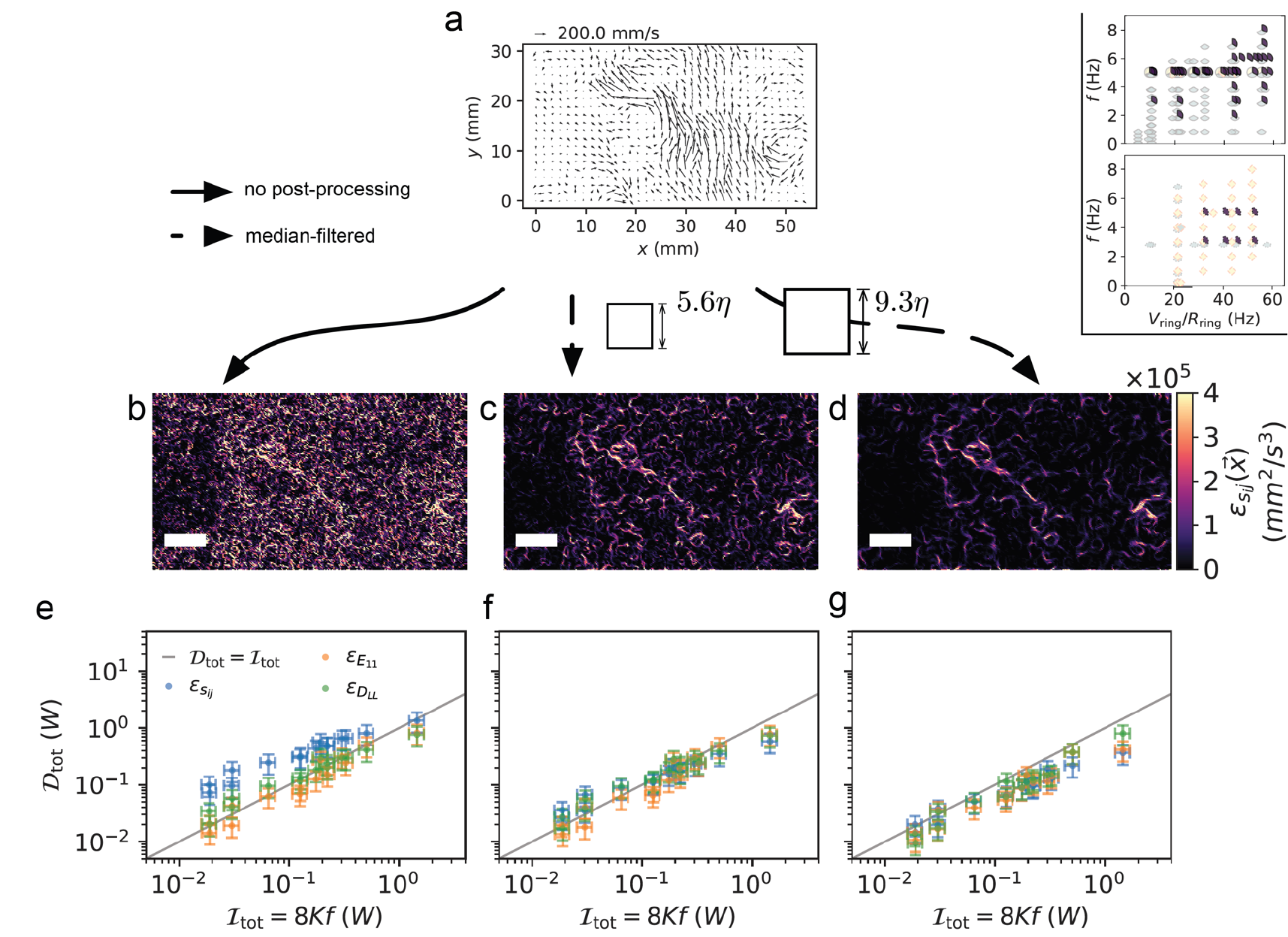}
\caption{\textbf{Dissipated vs injected power inside the chamber.} (a)To compute the dissipation rate, three velocity fields are used: a raw PIV-extracted field, and two median-filtered fields. (b-d) Median-filtering the velocity fields attenuates the local dissipation rate $\epsilon_{s_{ij}}(\vec{x})=2\nu s_{ij}(\vec{x})s_{ij}(\vec{x})$. (e) Dissipated vs injected power inside the chamber is plotted for the three methods to estimate the dissipation rate using the raw velocity field: $\epsilon_{sij}$ uses the rate-of-strain tensor. $\epsilon_{E_{11}}$ is obtained by fitting the energy spectrum to the Kolmogorov spectrum. $\epsilon_{D_{LL}}$ is obtained by fitting the second-order structure function to Kolmogorov's 2/3 law. (f) Same as (e) but with a median-filtered velocity field (Kernel size: $5.6\eta \times 5.6\eta$). (g) Same as (e) but with another medial-filtered velocity field (Kernel size: $9.3\eta \times 9.3\eta$).
}
\label{fig:dissipated_vs_injected_energy_whole_space}
\end{figure}

\subsection{Dissipation rate according to the energy balance}
Equating the dissipation $\mathcal{D}(r/R_{\rm blob})$ (Eq. \ref{eq:energy_balance_lhs_end}) with the injected power $\mathcal{I}$ (Eq. \ref{eq:energy_balance_rhs_end}) gives an expression about the dissipation rate inside the turbulent blob in terms of the properties of the vortex rings: $\epsilon_0 \propto \Gamma_{\rm ring}^2f / R_{\rm ring}^2 \propto V_{\rm ring}^2 f$ (main Figure 4b). With $R_{\rm blob}=\sqrt{6}R_{\rm ring}$, $\epsilon_0$ becomes
\begin{align}
    \epsilon_0=
    \begin{cases}
        \frac{c(\log{(8R_{\rm ring} / a)}-\alpha)}{8\sqrt{6}\pi p} \frac{\Gamma_{\rm ring}^2 f}{R_{\rm ring}^2}& \text{(thin-cored vortex ring)} \\
        \frac{2c}{105\sqrt{6}p} \frac{\Gamma_{\rm ring}^2 f}{R_{\rm ring}^2}& \text{(Hill's spherical vortex)}.
    \end{cases}
    \label{eq:dissipation_rate_from_energy_balance}
\end{align}
The prefactor of Eq.~\ref{eq:dissipation_rate_from_energy_balance} differs by the model of a vortex ring and $p(r/R_{\rm blob})$. Here is a quick calculation of the prefactor for $r/R_{\rm blob}=2\Leftrightarrow p=5/8$ (unbounded) and $c=0.77$. For the thin-cored model, we shall use $R_{\rm ring}/a=3$ as observed in the experiments, and $\alpha=2.04$~\cite{gharib_rambod_shariff_1998} for a viscous core. The prefactors of the thin-cored model and Hill's spherical vortex model are 0.02 and 0.01 respectively. This relationship is consistent with $\epsilon_0 = \alpha_1^2 V_{\rm ring}^2 f$ in the main Figure 4b. 

\clearpage
\section{Enstrophy flux}

As the vortex injection frequency is increased for fixed $V_{\rm ring}/R$ we observe a transition  from a reconnection-and-escape dynamic in which most of the enstrophy leaves with the outgoing vortices to a blob state in which little to none of the injected enstrophy escapes the blob. 
As illustrated in Figure~4 of the main text we examined this transition by plotting the time averaged enstrophy in a cross section perpendicular to the incoming rings, as well as by integrating the enstrophy outflux.

In this section we show the enstrophy flux data in greater detail and summarize a simple image-moment approach to quantifying the visual information contained in the enstrophy cross section images (not discussed in the main text).

The enstrophy balance is
\[
    \partial_t \int_\mathcal{V} \Omega^2 dV = 2 \int_\mathcal{V} \Omega_i (\partial_i U_j) \Omega_j dV - \oint_{\partial \mathcal{V}} \Omega^2 U_i n_i dS + \nu  \int_\mathcal{V}  \partial_j \partial_j \Omega^2 dV - 2 \nu \int_\mathcal{V}  (\partial_j \Omega_j)^2 dV \label{eq:enstrophy_transport}
\]
in a 3D incompressible, viscous fluid. The enstrophy is not conserved due to vortex stretching and viscous dissipation.
\begin{figure}[htbp]
\centering
\includegraphics[width=\textwidth]{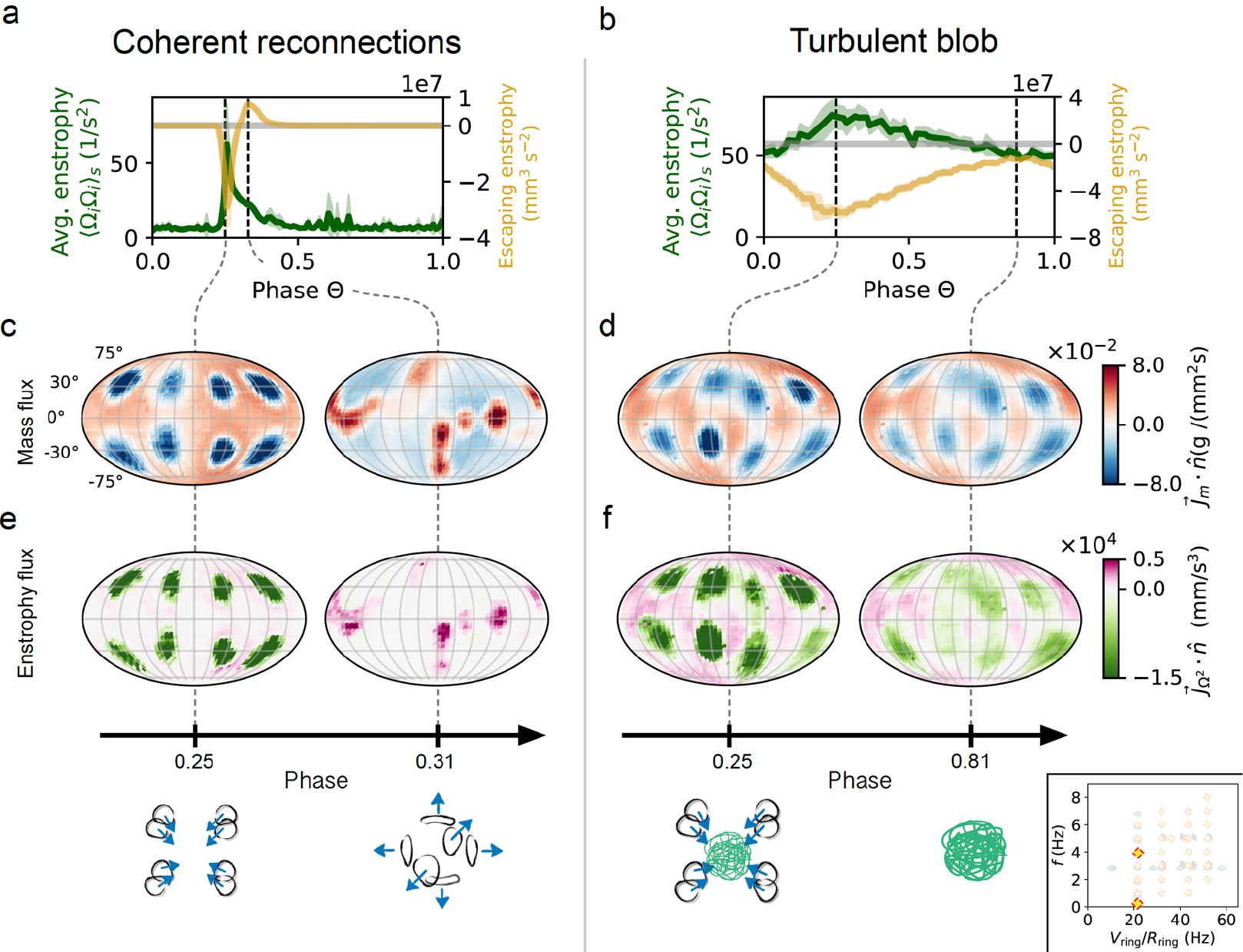}
\caption{\textbf{Enstrophy flux through a sphere with radius $R=1.2R_{\rm blob}$ reveals the influx always outweighs the outflux throughout a cycle when a turbulent blob is formed.} (a)Enstrophy averaged over a measured volume (Green) and the (net) enstrophy flux show the arrival of the vortex rings ($\Theta=0.25$) and the ejection event$\Theta=0.35$. The vortex rings are fired at $f=0.2$Hz. (b) The same as (a) but at $f=4Hz$ at which a turbulent blob is formed. }
\label{fig:enstrophy_n_mass_flux}
\end{figure}

To examine the confinement of enstrophy, we consider its normal flux $\vec{J}_{\Omega^2} \cdot \hat{n} = \Omega^2 \vec{U} \cdot \hat{n} $ through a sphere with radius $R>R_{\rm blob}$. In the case that coherent reconnections are dominant reactions ($f<f_c$), the enstrophy flux exhibits the inflow due to the eight incoming vortex rings, followed by outflow by the six outgoing vortex rings at a later phase (Figure \ref{fig:enstrophy_n_mass_flux}a and e). The mass flux also captures the inflow/outflow due to the mass transported by the incoming/outgoing vortex rings (Figure \ref{fig:enstrophy_n_mass_flux}c). The crucial difference is that the mass is always balanced, whereas the enstrophy is not. Integrating the enstrophy flux over the surface and cycles gives the net amount of (escaping) enstrophy at each phase as shown by a yellow curve in Figure \ref{fig:enstrophy_n_mass_flux}(a). The positive value corresponds to the enstrophy escaping from the considered sphere. On the other hand, the integrated enstrophy flux remains always negative throughout a cycle in the state of the turbulent blob (Figure \ref{fig:enstrophy_n_mass_flux}b). The enstrophy flux depicts that the inflow always outweighs the outflow, and the latter is more uniform than the case of the coherent reconnections(Figure \ref{fig:enstrophy_n_mass_flux}f). Because the mass outflow cannot be zero due to incompressibility (Figure \ref{fig:enstrophy_n_mass_flux}d), a small amount of enstrophy leaves the sphere. Nevertheless, the amount of enstrophy outflow remains smaller than the inflow. 

\subsection{Categorization of the flows using the time-averaged enstrophy distibution}
To translate the visible transition from reconnection-and-escape to confinement dynamics, we performed a simple image moment analysis as detailed below.

For a 2D image with intensity $I_{\rm img}(x, y)$, the (raw) moment of order ($i, j$) is defined as
\begin{align}
    m_{ij} = \sum_x \sum_y x^i y^j I_{\rm img}(x, y).
    \label{eq:imagemoments_raw}
\end{align}
The moments sequence for an image uniquely determines an image of finite size; however, they are not invariant under translation. On the other hand, the central moments 
\begin{align}
    \mu_{ij} = \sum_x \sum_y (x^i-\overline{x}) (y^j-\overline{y}) I_{\rm img}(x, y).
    \label{eq:imagemoments_central}
\end{align}
are translationally invariant by construction. The moments of inertia is analogous to the second-order image moments ($i=j=1$) when $I(x, y)$ is replaced by a mass distribution $\rho(x, y, z)$ of an object. Different images may have different second-order moments in the same way as the different objects may have different moments of inertia. To account for the total intensity, the normalized central moments are also used.
\begin{align}
    \nu_{ij} = \frac{\mu_{ij}}{\mu_{00}^{(1+(i+j)/2)}}
    \label{eq:imagemoments_normalizedCentral}
\end{align}
We characterized the distribution of the time-averaged enstrophy of various collision experiments by the sum of the second-order moments $\nu_{02}+\nu_{20}$. This quantity is invariant with respect to translation, scale, and rotation, and is also known as one of Hu invariants~\cite{hu_visual_1962}. Figure \ref{fig:blob_transition_imgMoments}a-c shows that the cross-like pattern takes a higher value than the blob pattern. This reflects the elongated distribution of the cross-like pattern created by the secondary rings. Figure \ref{fig:blob_transition_imgMoments}d plots our measurements on the parameter space, colored by $\nu_{02}+\nu_{20}$. Once the blob state is reached, the enstrophy distribution remains roughly the same. Hence, it tends to approach a fixed value if the blob radius were held constant. Strictly speaking, the normalized, second-order image moments take a higher value for a larger blob. As the blob size scales with the ring radius, Figure \ref{fig:blob_transition_imgMoments} was constructed by using only the rings with a certain radius $R_{\rm ring}=18\pm2$mm. This image analysis is consistent with the the linear trend.

\begin{figure}[htbp]
\centering
\includegraphics[width=\textwidth]{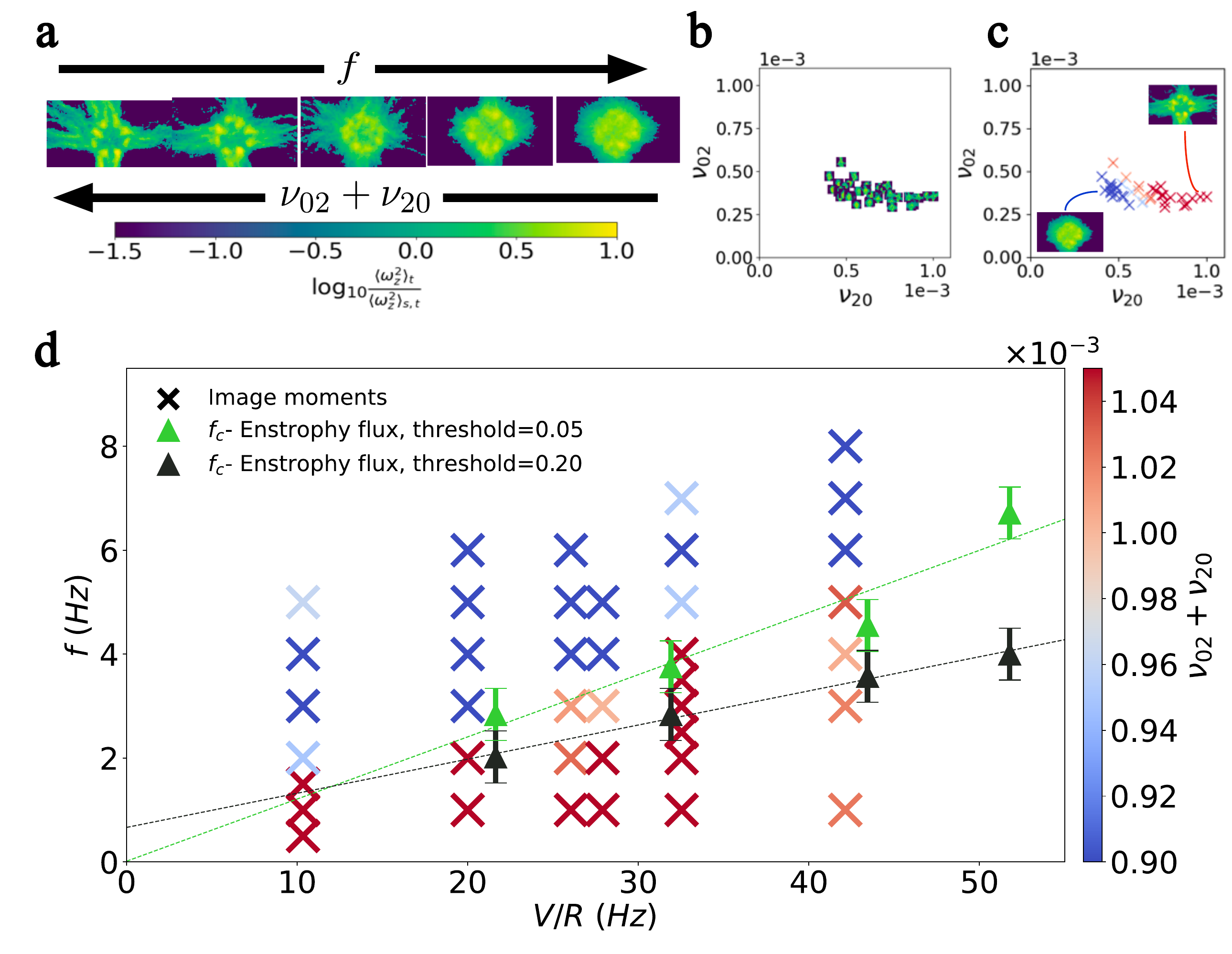}
\caption{\textbf{The second-order moments of the time-averaged enstrophy distribution categorizes the flows into the non-blob and blob states.} (a)The logarithmic, time-averaged enstrophy shows the formation of a turbulent blob at high forcing frequency. The corresponding second-order moments decrease as the ejection of enstrophy via secondary vortex structures becomes weak. (b)Images of thelogarithmic, time-averaged enstrophy are plotted on the $\nu_{02}-\nu_{20}$ plane. Blob patterns are clustered in the bottom left region. (c) The same plot as (b) but colored by $\nu_{02}+\nu_{20}$. (d) $\nu_{02}+\nu_{20}$ categorizes the flows into two groups. The separation is described by a linear relation $f\sim V_{\rm ring} / R_{\rm ring}$, and is consistent with the transitional frequencies obtained from the enstrophy flux.}
\label{fig:blob_transition_imgMoments}
\end{figure}

\clearpage
\newpage
\section{Gross-Pitaevskii Simulation} \label{sect:gpe}
In this section we report the result of our simulations of symmetric collisions of multiple quantum vortex rings using the Gross-Pitaevskii equation (GPE, Eq.~\ref{eq:gpe}). In all configurations we tested, we observe that reconnection events take place, and the secondary rings propagate away from the center of the collisions.

We simulate the dynamics of the quantum vortex rings through the Gross-Pitaevskii equation (GPE), a model equation that describes the time evolution of the superfluid wavefunction $\psi(x_i)=\sqrt{\rho(x_i)} \exp{[i\phi(x_i)]}$ where $\rho$ and $\phi$ are the spatially varying density and phase. The GPE is given by
\begin{align}
    \frac{d\psi}{dt} = -\frac{i}{2} \left(\Nabla^2 - |\psi|^2 \right) \psi.
    \label{eq:gpe}
\end{align}
The wavefunction can be mapped to classical hydrodynamic fluid velocity and density by the Madelung transform: $\vec{u}=\Nabla \phi, \rho=|\psi|^2$. The GPE is useful to study vortex dynamics as vortex reconnections occur without divergencies in physical quantities, and the topological dynamics was shown to be comparable to real viscous fluids~\cite{kleckner_creation_2013}.

All simulations were performed using a grid size of 0.5$\xi$ ($\xi$: healing length) and time step $\Delta t$=0.02 with periodic boundary conditions unless otherwise stated.

\subsection{Symmetric collisions of vortex rings}
We study collisions between vortex rings that are configured in a symmetrical fashion. For simplicity, we restrict ourselves to use only a circular ring with a fixed circulation and a radius. We associate a configuration to a polyhedron by the following rules. 1. Consider a plane $P_i$ that the $i$-th ring is 
embedded at an instant of time. 2. The bounded region by the $n$ planes is the polyhedron associated to the configuration of the rings. By this construction, the initial configuration of our experiment corresponds to an octahedron. 

We conducted GPE simulations of the symmetrical collisions that correspond to the platonic solids(a tetrahedron, a cube, an octahedron, a dodecahedron, and a icosahedron). 
They are the most symmetrical configurations for 4, 6, 8, 12, and 20 rings respectively. This is understood by counting the number of symmetry operations for the corresponding point groups. A tetrahedron belongs to $T_d$. A cube and an octahedron belong to $O_h$. A dodecahedron and an icosahedron belong to $I_h$ in the Schönflies notation. The result is summarized in Figure \ref{fig:gpe_platonic_solids}.

Two observations are made: (1)The secondary rings always travel away from the center of the collision. (2)The polyhedron corresponding to the configuration of the secondary rings is dual to the initial polyhedron. All adjacent vortex lines during the collision are anti-parallel so the reconnections occur at every possible site that correspond to the vertices of the initial configuration. A symmetric collision of four rings (a tetrahedron) results in four secondary rings (a flipped tetrahedron). A symmetric collision of six rings (a cube) result in eight secondary rings (an octahedron). The propagating directions of the secondary rings are given by the normal vectors of the faces (of the dual polyhedron). The pattern extends to all platonic solid configurations. In our experiment (the octahedral configuration), we observe six secondary rings (a cube) after the reconnection events via bubbles and Lagrangian trajectories (see SI Movie3).

\begin{figure}[htbp]
\centering
\includegraphics[width=\textwidth]{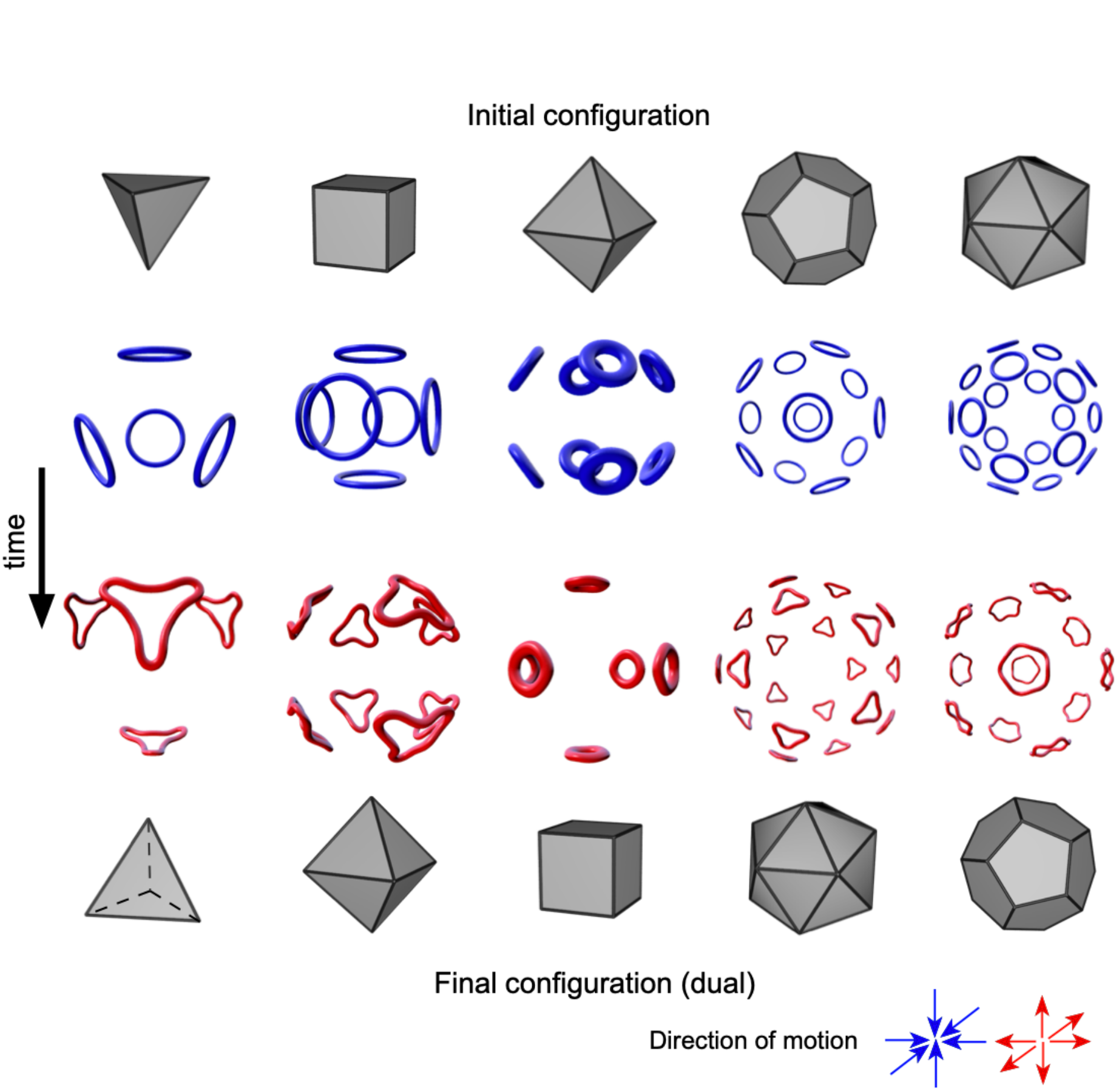}
\caption{\textbf{Gross-Pitaevskii simulation demonstrates the geometric rule of the symmetric vortex ring collisions.} When vortex rings are initially set on the faces of the platonic solids, the region bounded by the planes that the secondary rings live forms the dual solids, preserving the symmetry of the vortex structures before and after the collision. }
\label{fig:gpe_platonic_solids}
\end{figure}

\subsection{GPE simulation of a vortex ring train}
To generate Figure 1a, the rings with radius of $25\xi$ are initialized at a distance of 25\% away from the edge of the box in each axis, oriented to propagate towards the center of the box. We add noise ($\sim5\%$) to the amplitude of each ring in such a way that its center line remains smooth but not completely radially symmetric. This formation process repeats every 100 time steps to place a new set of rings. To enforce the wavefunction to be zero at the boundaries, we impose absorbing boundary conditions by adding a potential at the shell of the box with the thickness of 4$\xi$. 

\clearpage

\section{Supplementary movies}
\label{sect:supplementary videos}
\subsection*{Movie 1: Bubble visualization of a single-shot vortex ring collision}
This movie displays coherent vortex reconnections result in generation of six outgoing secondary rings when eight vortex rings collide. The bubbles, generated by hydrolysis, indicate the location of the cores of the primary and the secondary rings. The movie plays ten times slower than the real time.

\subsection*{Movie 2: Bubble visualization of a turbulent blob}
This movie displays the irregular motion of the bubbles when the vortex rings are fired at 5Hz, forming a turbulent blob. The first set of the collisions resembles a single shot (SI Movie 1). Yet, the bubbles at the center move more irregularly after a few cycles of vortex ring injection. 

\subsection*{Movie 3: Symmetric collision of vortex rings (Gross-Pitaevskii simulations)}
This movie demonstrates the geometrical rule of vortex ring collisions when they are initialized on the plane of the platonic solids: the configuration of the ejecting, secondary rings is dual to the configuration of the primary rings. 

\subsection*{Movie 4: Ejection pattern of a single-shot vortex ring collision}
The pathlines shows that a single-shot collision of eight vortex rings results in formation of six secondary rings, consistent with the GPE simulation (Movie 3). The trajectories are colored by the vortex rings of origins. Notice that each secondary vortex ring consists of four colors, reflecting its four `parent' vortex rings. 

\subsection*{Movie 5: Pathline visualization of a turbulent blob}
When a turbulent blob forms, the pathlines of particles that are transported by vortex rings become highly irregular due to turbulence within the blob. The ejecting pattern becomes more uniform than the case of coherent vortex reconnections. The colors are assigned by the vortex rings of origins. White trajectories are found in the turbulent blob, and are of unknown origins. 

\subsection*{Movie 6: Pathline visualization of a turbulent blob, colored by Lagrangian speeds}
This visualization of a turbulent blob shows the formation of a steady, isolated region with considerably higher energy than its surroundings. Here, all detected pathlines are shown with opacity weighted by the speed. The slow particles (purple) are dimmed more than the fast (yellow) ones. 

\subsection*{Movie 7: Mean flow structure and fluctuations of a turbulent blob}
This movie shows that the fluctuations dominate inside a turbulent blob (the blue cloud), whereas the time-averaged (mean) flow (the yellow cloud) is set up by the injected vortex rings outside the turbulent blob. The volume of measured field is 120mm$\times$102mm$\times$100mm with a voxel pitch of 2.2mm. 

\subsection*{Movie 8: Normal fluxes of mass and enstrophy in the two different phases: coherent vortex reconnections and a turbulent blob.}
These Mollweide plots show the normal component of each mass and enstrophy flux at the two phases. In the phase of coherent vortex reconnections (the top row), the mass and enstrophy fluxes indicate focused ejection. When a turbulent blob is formed, the ejection is more uniform, and less enstrophy leaves the central region on average (the bottom row). Here, a spherical surface with radius $r \approx 1.25 R_{\rm blob}=30$mm is considered.
The left column represents the normal mass flux $J_m = \rho \vec{U}\cdot \hat{n}$ through the surface, and the right column is the normal enstrophy flux $\vec{J}_{\Omega^2} = \Omega^2 \vec{U}\cdot \hat{n}$. The measurements correspond to ($V_{\rm ring}/R_{\rm ring}, f$) = (21 Hz, 0.2 and 4Hz).

\subsection*{Movie 9: Energy and enstrophy field during a single-shot collision($L/D, v_{\rm eff}, f$) = (3.5, 318mm/s, 1Hz); $V_{\rm ring} / R_{\rm ring}=26$Hz.}
This movie shows the energy and enstrophy fields of vortex ring collisions on the central slice, obtained by 2D PIV. The vortex reconnections generate secondary rings, four of which are visible in this movie. Time-averaging these fields provides equivalent pictures as Figure 2b-c. 

\subsection*{Movie 10: Energy and enstrophy fields of vortex ring train at ($L/D, v_{\rm eff}, f$) = (3.5, 318mm/s, 5Hz); $V_{\rm ring} / R_{\rm ring}=26$Hz.}
This movie shows the energy and enstrophy fields of vortex ring collisions in the state of a turbulent blob on the central slice, obtained by 2D PIV. Contrary to the single-shot collision (SI Movie 9), energy and enstrophy remains relatively steady within a spherical blob, and they leave the central region only intermittently. Time-averaging these fields provides equivalent figures as Figure 2f-g.

\bibliography{main}

\begin{thebibliography}{84}
\bibitem [{a()}]{a}%
  \BibitemOpen
  \bibinfo {note} {This choice of $R_{\protect \rm ring}$ is arbitrary but the
  other candidates are sufficiently close to the first moment of vorticity
  $0.75A_{\protect \rm sph}$. In the comoving frame, the velocity becomes zero
  at $1/\protect \sqrt {2}A_{\protect \rm sph}\approx 0.71A_{\protect \rm sph}$
  away from the center. The square-root of the second moment is equal to
  $\protect \sqrt {3/5}A_{\protect \rm sph}\approx 0.77A_{\protect \rm
  sph}$.}\BibitemShut {Stop}%
\end{thebibliography}

\end{document}